# The Science Vision

*for the*

# Stratospheric Observatory For Infrared Astronomy

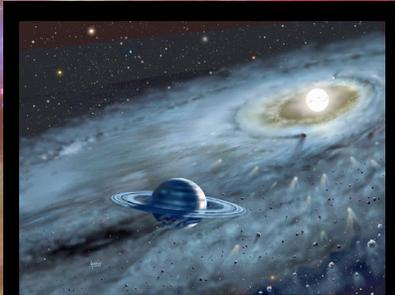
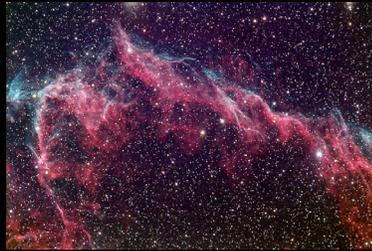
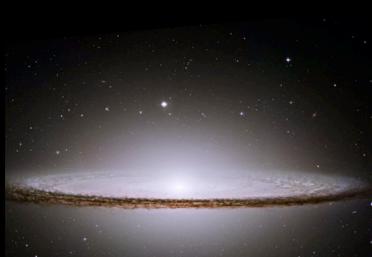
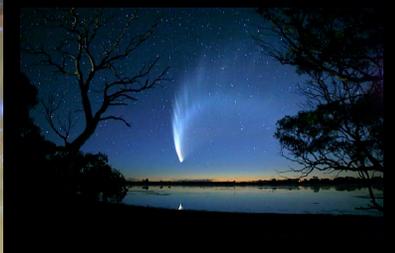

**Stratospheric Observatory for Infrared Astronomy**
NASA Ames Research Center
Moffett Field, CA

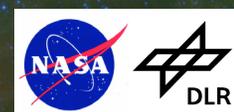

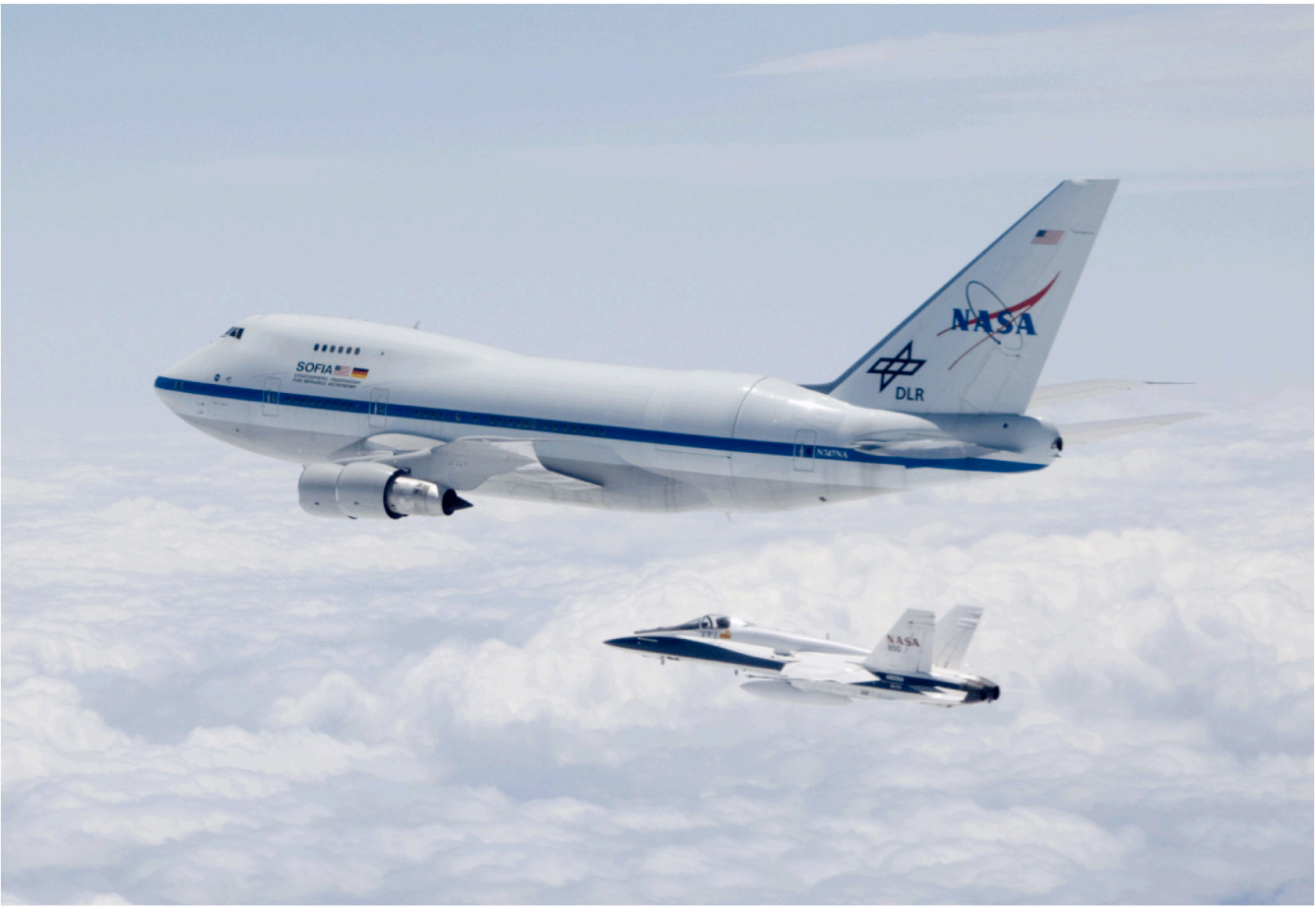

*The SOFIA flying observatory*

# Contents

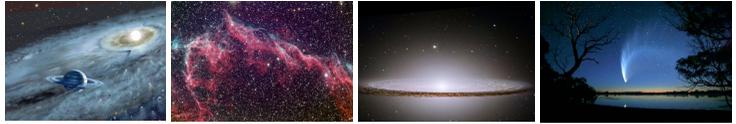

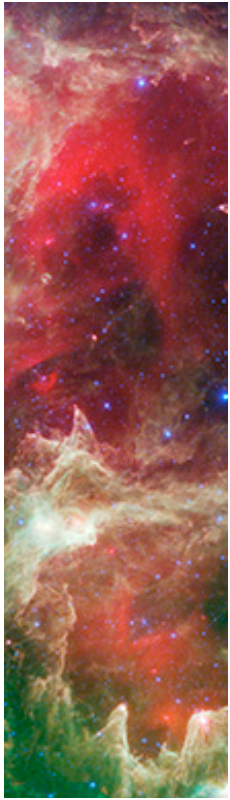

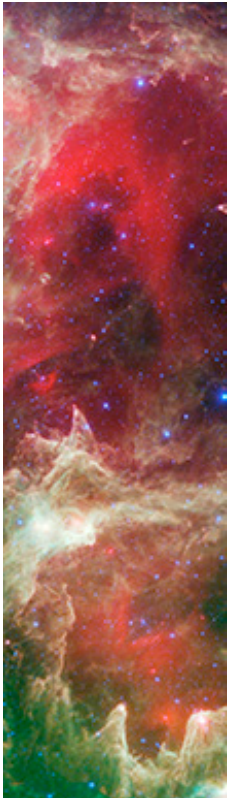

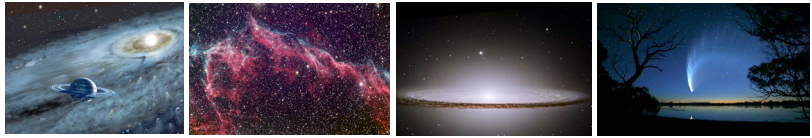

# Acknowledgements

| Name | Affiliation |
|---|---|
| **Lead Editors** | |
| *Introduction* | |
| Robert Gehrz | University of Minnesota |
| Daniel Lester | University of Texas |
| Thomas Roellig | NASA Ames Research Center |
| Eric Becklin | SOFIA-USRA |
| *The Formation of Stars and Planets* | |
| Neal Evans | University of Texas |
| James De Buizer | SOFIA-USRA |
| *The Interstellar Medium of the Milky Way* | |
| Margaret Meixner | Space Telescope Science Institute |
| Xander Tielens | Universiteit Leiden |
| *Galaxies and the Galactic Center* | |
| Gordon Stacey | Cornell University |
| William Vacca | SOFIA-USRA |
| *Planetary Science* | |
| Jeff Cuzzi | NASA Ames Research Center |
| Dana Backman | SOFIA-SETI Institute |
| **Contributors** | |
| Lou Allamandola | NASA Ames Research Center |
| B-G Andersson | SOFIA-USRA |
| Edward Austin | NASA Ames Research Center |
| Dana Backman | SOFIA-SETI Institute |
| John Bally | CASA/APS, University of Colorado |
| Eric Becklin | SOFIA-USRA |
| Dominic Benford | NASA Goddard Space Flight Center |
| Ted Bergin | University of Michigan |
| Michael Bicay | NASA Ames Research Center |
| Jesse Bregman | NASA Ames Research Center |

| Name | Affiliation |
|---|---|
| **SOFIA Science Council** ||
| Andrew Harris (Chair) | University of Maryland |
| Jacqueline Fisher | Naval Research Laboratory |
| Thomas Geballe | Gemini Observatory |
| George Helou | California Institute of Technology |
| Thomas Henning | Max-Planck-Institut für Astronomie |
| Adrian Russell | National Radio Astronomy Observatory |
| Gordon Stacey | Cornell University |
| Jürgen Stutzki | Universität zu Köln |
| Alycia Weinberger | Carnegie Institution of Washington |

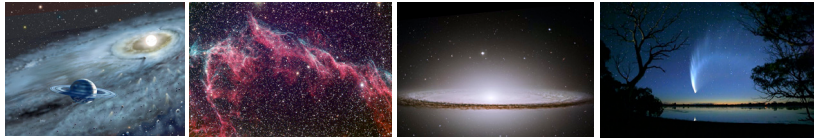

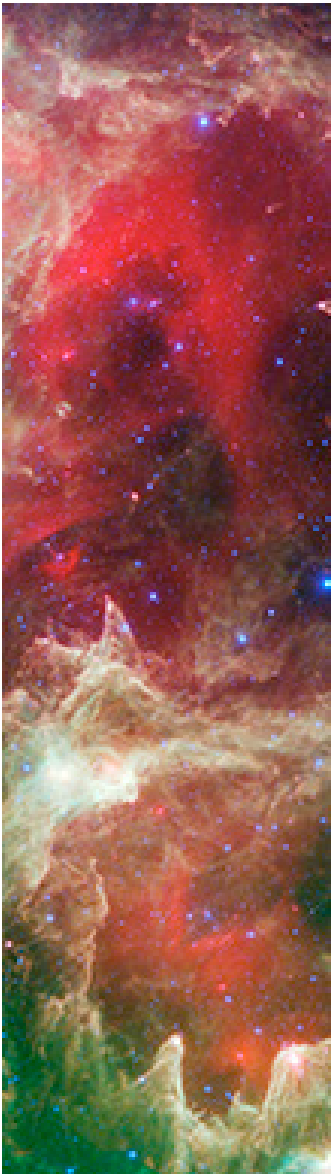

# *Executive Summary*

Humanity has fashioned tools to observe heavenly bodies at least since Galileo pointed his telescope at Jupiter and discovered its moons. This document introduces the science potential of a powerful new observatory, NASA/DLR's Stratospheric Observatory for Infrared Astronomy (SOFIA).

SOFIA consists of a German-built 2.7-meter telescope mounted in a modified Boeing 747-SP aircraft supplied by NASA. Operations costs and observing time will be shared by the United States (80%) and Germany (20%). Flying at altitudes up to 45,000-feet, SOFIA observes from above more than 99 percent of Earth's atmospheric water vapor, thereby opening windows to the universe not available from the ground. SOFIA will offer international science teams approximately 1000 cloud-free high-altitude science observing hours per year during its two decade design lifetime. More than 50 science proposals per year will be selected through a rigorous peer review process. Although the primary impact of SOFIA will be its science return, a small sample of which is outlined in this document, it will yield other returns as well. Compelling discoveries will follow the development of new technology — technology that can be demonstrated readily on SOFIA. Young scientists-in-training, educators, and journalists will also fly on SOFIA, making it a valuable training platform and public ambassador.

Nine first-generation science instruments are under development by institutions in both the US and Germany, including both imaging cameras and spectrographs. These instruments are a mixture of facility-class instruments with pipeline data products easily accessible to general observers and PI-class instruments employing the most advanced state-of-the-art technologies. SOFIA will observe at wavelengths from 0.3 $\mu$m to 1.6 mm. It will be capable of high-resolution spectroscopy (R > $10^4$) at wavelengths between 5 and 600 $\mu$m with its first-generation instruments. SOFIA's diffraction-limited imaging longward of 25 $\mu$m will produce the sharpest images of any current or planned IR telescope operating in the 30 to 60 µm region.

The SOFIA Observatory concept embodies a number of key advantages that make it a unique tool for astronomy in the coming decades:

- SOFIA is a near-space observatory that comes home after every flight. Its scientific instruments can be exchanged regularly for repairs, to accommodate changing science requirements, and to incorporate new technologies. These instruments do not need to be space qualified.

- SOFIA has unique capabilities for studying transient events. The observatory can operate on short notice from airbases worldwide, in both the northern and southern hemispheres, to respond to new scientific opportunities.

- SOFIA's diverse range of instrumentation will facilitate a coordinated program of analysis of specific targets and science questions. SOFIA's 20-year design lifetime will enable long-term studies and follow-up of work initiated by SOFIA itself and by other observatories, such as HST, Chandra, Spitzer, SMA, and ASTRO-F, as well as future facilities.

- SOFIA will present an ideal venue in which to educate students, where they can participate in hands-on, cutting-edge space technology developments.

- Because of its accessibility and ability to carry passengers, SOFIA will include a vigorous, highly visible Education and Public Outreach (E&PO) program designed to exploit the unique and inspirational attributes of airborne astronomy.

SOFIA, with its large suite of science instruments and broad wavelength coverage, will be capable of undertaking a huge breadth of different investigations. This document describes a sample of the exciting science programs that might be undertaken with SOFIA. It is important to recognize that the programs described in these sections are only representative of the science that SOFIA is capable of addressing — they are in no way a comprehensive listing of all of the investigations that might be conducted.

### The Formation of Stars and Planets

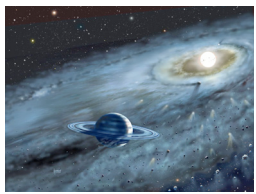

What physical, chemical, and dynamical processes are at work in the formation of stars and planets? While SOFIA will provide vital information on the formation of stars of all masses and on planet formation, we have focused here on three topics where SOFIA has unique capabilities.

*Massive Stars.* The process of massive star formation remains largely unknown despite recent progress in the observation and theory of low mass star formation. SOFIA will collect comprehensive data on hundreds of massive stars. These data will help understand how massive stars form in different environments by distinguishing physical, chemical, and dynamical differences between high and low mass star formation regions. SOFIA measurements of the broad spectral energy distribution of nascent massive stars, which are dark in the near- and sometimes mid-infrared, will constrain models of collapsing cores. SOFIA's high spatial resolution over a broad spectral range (25-300 $\mu$m) that encompasses the luminous output of massive stars will reveal structural details not seen by previous missions.

*Understanding Protoplanetary Disks.* To understand the origin of the Solar System, we must investigate the environment in which planets form: circumstellar disks around young stars. Over the last two decades, the study of circumstellar disks has focused on the shape of their spectral energy distributions and direct millimeter interferometric imaging. SOFIA data will be used to refine accretion and cooling models of circumstellar disks, and to study the kinematics, composition, and evolution of disks around low-mass young stellar objects.

*Astrochemistry in Star Forming Regions.* The study of exo-planetary systems and their formation is one of the fastest growing topics in astronomy, with far-reaching implications for understanding our place in the universe. Intimately related to this topic is the study of the chemical composition of the gaseous and solid-state material out of which new planets form and how it is modified in the dense protostellar and protoplanetary environments. With the ability to track the formation of complex hydrocarbons, SOFIA, in concert with ALMA, Herschel, and JWST, has an important role in tracing our chemical origins.

## The Interstellar Medium of the Milky Way

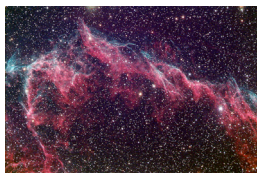

The interstellar medium (ISM) contains an elemental record of the generations of stars that have lived and died since the Galaxy's birth. SOFIA will gather spectra of bright sources and extended regions, probing the physics and chemistry of both. SOFIA will explore the physical processes governing how stars interact with their environments, the origin of dust, and the role of large, complex carbon molecules — notably polycyclic aromatic hydrocarbons (PAHs). PAH molecules are so ubiquitous and structurally revealing that they can be used to trace the chemistry of prebiotic molecules within star forming regions, even into the shrouded nurseries where stellar, and eventually planetary systems, form. SOFIA's telescope and excellent complement of spectroscopic instruments are well suited for spectral imaging of bright sources and extended regions of the ISM in the Milky Way and nearby galaxies. It can also do high resolution spectroscopy of particular regions that can resolve the narrow features of interstellar dust and kinematics of the ISM gas.

***The physical processes that regulate the interaction of massive stars and their environment.*** Stellar radiation from massive stars transforms the nearby environment, ionizing gaseous atomic hydrogen and driving the chemistry of star forming regions. SOFIA will read the infrared signatures of these massive-star-forming pockets and map their changing physical conditions, such as density, metallicity, and temperature.

***The origin of dust in the Milky Way and other galaxies.*** Dust is an important component of the ISM and the clouds from which stars form. Dust grains can provide effective shielding from harsh interstellar radiation for high-density clumps trying to become newborn stars, as well as an effective channel for cloud cooling. The chemistry and composition of the dust grains, as well as that of the ISM gas, determines how interstellar ices form. Thermal and radiative processing of these simple ices yields the complex chemistry that finds its way into planetary systems and eventually into living beings. Understanding the chemistry of interstellar dust and the environments where dust survives will thus provide important clues about the chemistry of life.

***The role of large and complex molecules, such as PAHs in the interstellar medium.*** The unique identification of large, possibly prebiotic carbon-based molecules in space has astrobiological significance. PAH molecules containing upwards of approximately 50 C-atoms are ubiquitous in the interstellar medium. Although the presence of these molecules in space is undisputed, identification of specific species has not yet been achieved. Far-infrared spectroscopy of the lowest lying vibrational modes — corresponding to 'drumhead' modes — offer the best hope to identify them. SOFIA is ideally suited to study this question with its complement of far-infrared spectrometers at critical wavelengths.

### Galaxies and the Galactic Center

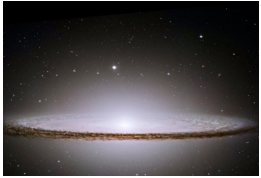

A key challenge of observational astrophysics is an understanding of the star-formation history of external galaxies. Investigation of the ISM in galaxies is essential to this understanding. Far-infrared fine structure lines are excellent probes of both the physical conditions of the ISM and the stellar radiation field properties within galaxies.

*The Galactic Center.* The center of our Galaxy is a few hundred times closer than the nearest galaxies with active nuclei enabling uniquely detailed studies of phenomena found only in galactic nuclei. Infrared observations, which first opened the window of visibility to our Galactic Center, are particularly valuable for investigating the many questions that arise in this complex environment. The central molecular zone represents the most massive concentration of dense gas in our Galaxy and provides important lessons about phenomenology of galactic nuclei in general. At a distance of only 8 kiloparsecs, the Galactic Center offers us critical, spatially resolved information about how the activity there is produced by the interactions of stars, powerful gas flows and stellar winds, strong magnetic fields, and the supermassive black hole. Because of the 20 – 30 magnitudes of visual extinction to the Galactic Center, its abundant energy emerges almost entirely in the infrared, and is well suited for study by SOFIA.

*The ISM and the Star Formation History of External Galaxies.* SOFIA provides a unique opportunity to study many of the properties of the warm ISM in nearby galaxies. These properties can be used to glean information about the star formation history of these galaxies. Arising in interstellar gas clouds heated by UV photons, far-infrared lines have distinct advantages over those available to optical and near-IR observers as relatively un-extincted probes of the extragalactic interstellar medium. With observations of these far-infrared lines, SOFIA's unrivaled mapping capabilities will allow investigations of the ISM in nearby galaxies at spatial resolution sufficient to resolve spiral arms and to distinguish large star formation regions.

### Planetary Science

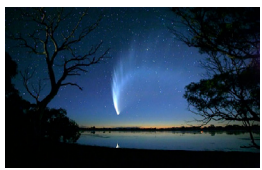

Infrared observations of comets, near-Earth asteroids, moons, and planets reveal evidence of their origin and, by extension, the origin of the Solar System. The existence of water and organic materials in primitive bodies offers clues as to where and how these bodies formed, and how those ingredients rained down on early Earth. SOFIA offers the unique capability, relative to space- or ground-based infrared observatories, to observe objects closer to the Sun than the Earth: for example, comets in their most active phases, and the planet Venus. SOFIA can point directly at bright planets and their inner moons, and observe spectral features of water and organic molecules at wavelengths blocked by Earth's atmosphere. SOFIA will observe stellar occultations and other transient events from optimum locations anywhere on Earth. Finally, SOFIA will be able to monitor seasonal and episodic changes in slow-orbiting outer planets over decade timescales.

*Primitive Bodies.* The Stardust and Deep Impact missions drew a complicated picture of comets and Solar System evolution. New models predict that dynamic mixing in the early Solar System led to great diversity among comets, more so than had been previously expected. SOFIA will build on these results and test new hypotheses by tracking the water, mineralogy, and organic content of comets, assembling a taxonomy of the more than 60 comets expected to be available for observation over its lifetime. Combined with dynamical modeling and data from other missions, this catalog will delineate the water and pre-biotic organics likely delivered to the early Earth. Water is the principal volatile in comets. Its isotopic composition — which SOFIA will investigate — is a key diagnostic criterion for understanding comets' origins.

Aside from comets, the tribes of primitive bodies include the asteroids of the inner Solar System, plus the outer Solar System dynamical families including Centaurs, Kuiper Belt Objects, and even more remote "scattered disk" objects. Scattered disk and Kuiper Belt Objects are collectively referred to as Trans-Neptunian Objects, and over 1000 such objects are now known. These classes of primitive bodies are notable for their diversity in size, density, and composition. Such diversity implies that a large number of objects must be observed to characterize them properly. SOFIA can measure the size of Trans-Neptunian Objects by flying anywhere on Earth to observe stellar occultations as the objects cross in front of background stars.

*Giant Planets.* The processes and properties of our Solar System's giant planets, including cloud formation and atmospheric dynamics, serve as ground truth for interpreting necessarily cruder observations of the ever-growing number of extrasolar planets. SOFIA can contribute to the vexing question of the H/He ratio in the giant planets, a problem that remains unsettled even after several spacecraft flybys and the Galileo entry probe into Jupiter's atmosphere. Departures from primordial ratios could be indicative of large-scale inhomogeneity and phase separation, in which He precipitates out of the outer atmosphere. SOFIA's high-resolution spectrometers will enable investigation of the global chemical inventory of all the gas giants, with especially good spatial resolution for Jupiter and Saturn. Neptune's spectral energy distribution, and by implication its atmospheric structure, have varied over the last two decades, perhaps as a seasonal response. Major episodic changes of Neptune's near-infrared spectrum have also been seen. Observations by SOFIA over the next two decades could help us understand these changes.

***Small Worlds of our Solar System: Venus and Titan.*** Comparable to the Earth in mass and probably initial volatile content, Venus may have lost nearly all its hydrogen to a fierce runaway greenhouse effect. Understanding its initial conditions would provide a crucial point in unraveling the formation of the Earth and other inner planets. Venus' slow rotation gives it unusual atmospheric dynamics, including a puzzling "super-rotating" middle atmosphere that transports substantial energy from the day-side to the night-side. By circumstance, Earth's sister planet Venus has never been explored thoroughly with broadband, high-resolution spectroscopy, but SOFIA will open the door to studies of many molecules predicted by theory, and the kinematics of Venusian winds. SOFIA can play the role of a Venus-focused spacecraft with the potential for discoveries in atmospheric chemistry and the ability to map dynamics well below the visible haze layer.

Titan has long been a target of central interest from the standpoint of organic chemical evolution. Its low abundance of atmospheric $H_2$ allows chemical pathways to proceed to great complexity that cannot be reached in the atmospheres of the gas giant planets. In spite of many new results obtained by the Cassini orbiter and Huygens probe, there are a number of missing key measurements that only SOFIA can make. SOFIA's broad wavelength range will capture the spectral signature of more molecular species in Titan's atmosphere and detect more lines per molecule, especially at short IR wavelengths, than Herschel, Spitzer, and ground-based telescopes. In addition, SOFIA's long mission lifetime will enable long-term seasonal studies of Titan's atmosphere.

### A Compelling Vision

SOFIA's vision will extend from Earth's sister planets to faraway galaxies, and will support a vast number of potential investigations over a two-decade lifespan. SOFIA is poised to address epic questions that drive NASA's space science program, providing a platform for collaboration, innovative technology, and scientific discovery. In doing so, SOFIA will inspire and train a new generation of young scientists.

# Chapter 1

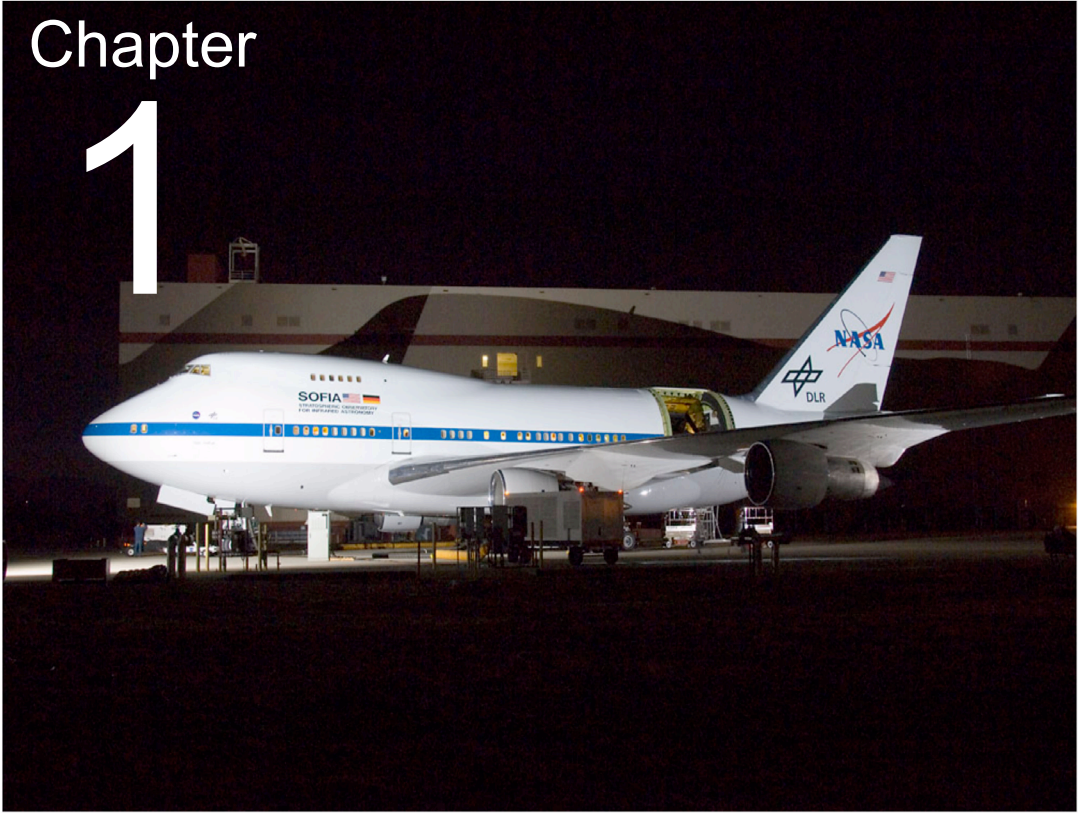

# Introduction

## 1.1 Overview

> Because of its accessibility, SOFIA will be an effective test bed for new technologies, and a training ground for a new generation of astronomical instrumentalists.

The Stratospheric Observatory For Infrared Astronomy (SOFIA, Figure 1-1) is a Boeing 747-SP aircraft housing a 2.5-meter gyrostabilized telescope designed to make sensitive measurements of a wide range of astronomical objects at wavelengths from 0.3 $\mu$m to 1.6 mm. This new observatory will be a key element in our research portfolio for chemical and dynamical studies of warm material in the universe, and for observations of deeply embedded sources and transient events. SOFIA is designed for at least two decades of operations and will join the Spitzer Space Telescope (Werner, et al. 2004, Gehrz, et al. 2007), Herschel Space Observatory (Pilbratt et al. 2003) and James Webb Space Telescope (JWST) (Gardner et al. 2006) as one of the premier facilities for panchromatic observations in thermal IR and submillimeter astronomy.

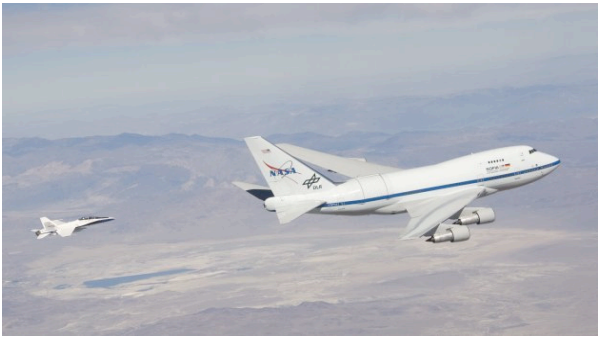

*Figure 1-1.* The SOFIA infrared observatory with chase plane during the first series of test flights to verify the performance of the modified Boeing 747SP. From the NASA Dryden Flight Research Center Photo Collection.

Furthermore, SOFIA will be a test bed for new technologies and a training ground for a new generation of instrumentalists and astronomers. SOFIA can be upgraded continually and can be used to evaluate state-of-the-art and high-risk technologies that could otherwise only be proven in space. SOFIA is a joint project of NASA and the Deutsches Zentrum für Luft- und Raumfahrt (DLR). The SOFIA telescope design and its evolving instrument complement build upon the successful heritage of NASA's Kuiper Airborne Observatory, a 0.9-meter infrared telescope that flew from 1971-1995. SOFIA, in full operation, will be capable of making more than about 100 8-10 hour scientific flights per year.

The SOFIA Science Center (SSC) is responsible for the science productivity of the mission, and is located at NASA Ames Research Center in Moffett Field, CA. Flight operations will be conducted out of NASA Dryden Flight Research Center's Aircraft Operations Facility (DAOF) in Palmdale, CA. The Universities Space Research Association (USRA) and the Deutsches SOFIA Institut (DSI) in Stuttgart, Germany manage science operations and mission planning for NASA Ames and DLR, while aircraft operations are handled by NASA Dryden. The SOFIA Program will support approximately 50 science teams per year, selected from peer-reviewed proposals. An on-going instrument development program will ensure that this easily serviceable facility remains state-of-the-art during its lifetime. The next call for new generation instruments will occur in 2011.

*Flying at altitudes where the precipitable water is a hundred times lower than at good terrestrial sites, SOFIA will be able to observe astronomical objects in a broad swath of wavelengths completely invisible from the ground. At a typical observing altitude of 45,000 feet, such column densities of less than 10 μm will allow observations with greater than 80% transmission from 0.3 μm to 1.6 mm.*

Science operations will start with a phased approach. Early Science (ES) with the Faint Object InfraRed Camera for the SOFIA Telescope (FORCAST), a mid-infrared imager, and the German Receiver for Astronomy at Terahertz Frequencies (GREAT), a high resolution far-infrared heterodyne spectrometer, will occur in 2010. The ES programs will study the chemistry of warm interstellar gas through spectroscopic observations of emission from molecules such as carbon monoxide (CO), and the detailed spatial morphology of the hot dust in bright star-forming regions. The first science flights will be predicated on science collaboration of the selected teams with the Principal Investigators (PIs) of both FORCAST and GREAT in order to ensure early science productivity. Routine observations will begin in 2012 in response to a General Investigator (GI) science proposal solicitation. There will be new science proposal solicitations every 12 months thereafter. About 20 GI science flights are planned annually at the start of science operations, with the rate ramping up steadily until approximately 100 flights each year are achieved in 2014.

### 1.2 SOFIA's Operational Envelope and Range

The column depth of precipitable atmospheric water above 45,000 feet (13.7 km) is typically less than 10 $\mu$m, a hundred times lower than at good terrestrial sites. At this altitude, atmospheric transmission averages 80% or better across SOFIA's wide wavelength range (see Figure 1-3). This includes large parts of the electromagnetic spectrum that are completely inaccessible from the ground.

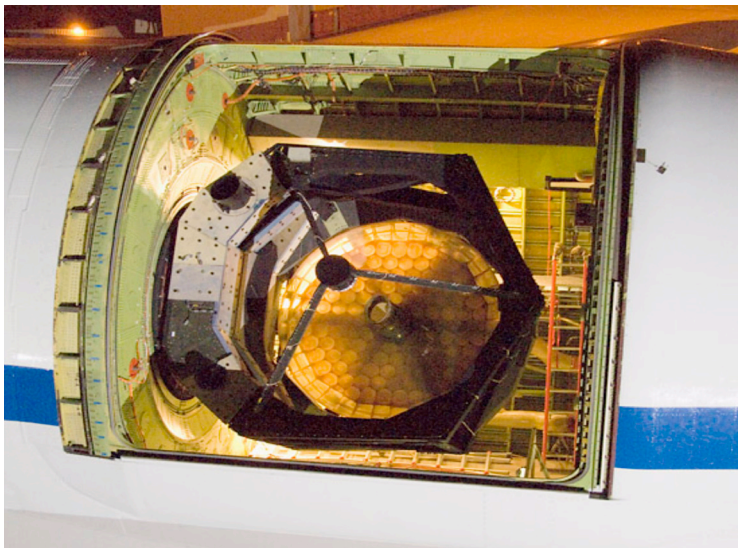

**Figure 1-2.** *The SOFIA telescope and primary mirror before the mirror was aluminized.*

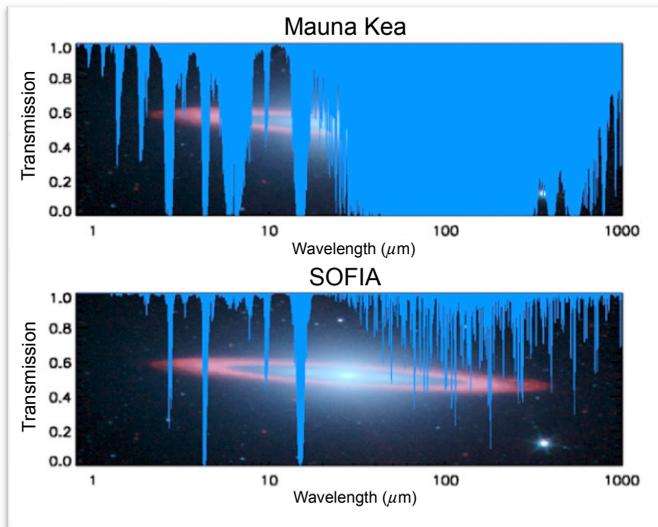

**Figure 1-3.** *The typical atmospheric transmission for SOFIA at an altitude of 45,000 feet as compared to the transmission on a good night at Mauna Kea (13,800 ft. MSL). From 1 to 1000 µm, the average transmission at SOFIA's altitude is ≥80% except in the center of absorption lines due to mostly telluric $H_2O$, $CO_2$, and $O_2$. Background image: IRAC false color image of the Sombrero Galaxy, courtesy of NASA/JPL-Caltech.*

Although some strong water absorption lines remain, spectroscopy between the water lines is possible and most of the flux makes it through enabling wide band photometry.

The SOFIA aircraft will ordinarily be staged in Palmdale, California, but will operate from other bases when necessary. Southern hemisphere bases will be used for observations of targets at extreme southerly declinations. SOFIA's deployment flexibility will allow measurements of transient events that are visible only at particular locations (e.g., stellar occultations).

### 1.3 The Telescope and Observatory

The SOFIA telescope (Figure 1-5, Figure 1-6, Table 1-1), was supplied by DLR as the major part of the German contribution to the SOFIA observatory. It is a bent Cassegrain design with a 2.7m (2.5m effective aperture) parabolic primary mirror and a 0.35m diameter hyperbolic secondary mirror. The telescope feeds two f/19.6 Nasmyth foci (the IR focus and a visible light focus for guiding), about 300 mm behind the instrument flange, using a gold coated dichroic and an aluminum coated flat. The secondary mirror provides chop amplitudes of up to ± 4 arcmin between 0 and 20 Hz. The visible beam is fed into the Focal Plane Imager (FPI) which is an optical focal plane guiding system. Independent of the FPI there are two other optical imaging and guiding cameras available: a Wide Field Imager (WFI) and Fine Field Imager (FFI). Both of these cameras are attached to the front ring of the telescope.

The telescope is mounted in an open cavity in the aft section of the aircraft (Figure 1-4) and views the sky through a port-side doorway. The telescope is articulated by magnetic torquers around a spherical bearing through which the Nasmyth beam is passed. The telescope has an unvignetted elevation range of 20-60 degrees. Since the cross-elevation travel is only a few degrees, the airplane must be steered to provide most of the azimuthal telescope movement required for tracking. Hence, the list of targets to be observed determines the flight plan. The focal plane instruments and the observers are on the pressurized side of the 21-foot diameter bulkhead on which the spherical bearing is mounted, allowing a shirt-sleeve working environment for the researchers and crew.

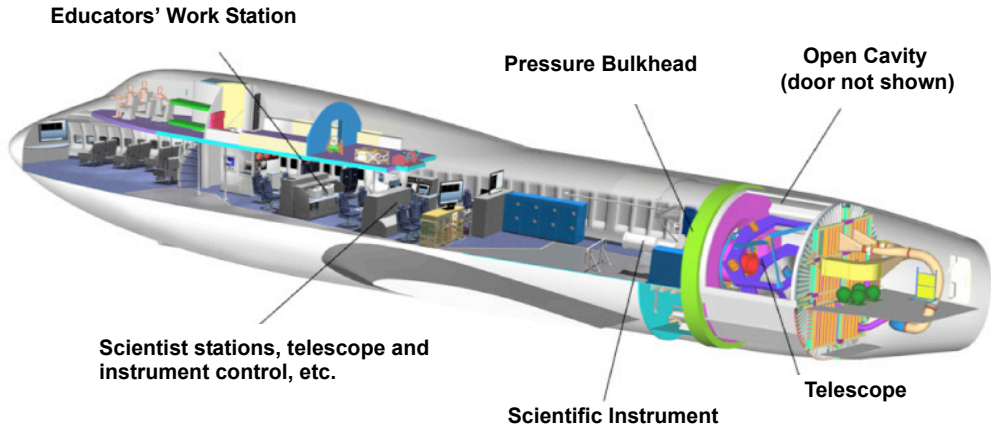

*Figure 1-4.* *A cutaway view of the SOFIA Observatory.*

**Table 1-1. SOFIA System Characteristics**

| | | | |
|---|---|---|---|
| **Nominal Operational Wavelength** | 0.3 to 1600 $\mu m$ | **Diffraction Limited Wavelengths** | $\geq 15\ \mu m$ |
| **Primary Mirror Diameter** | 2.7 meters | **Optical Configuration** | Bent Cassegrain with chopping secondary mirror and flat folding tertiary |
| **System Clear Aperture Diameter** | 2.5 meters | **Chopper Frequencies** | 1 to 20 Hz for 2-point square wave chop |
| **Nominal System f-ratio** | 19.6 | **Maximum Chop Throw on Sky** | +/- 4 arcmin (unvignetted) |
| **Primary Mirror f-ratio** | 1.28 | **Pointing Stability** | = 1.0" rms at first light<br>= 0.2" rms in operations |
| **Telescope's Unvignetted Elevation Range** | 20 to 60 degrees | **Pointing Accuracy** | = 0.5" with on-axis focal plane tracking |
| **Unvignetted Field-of-View Diameter** | 8 arcmin | **Total Emissivity of Telescope (Goal)** | 15% at 10 $\mu m$ with dichroic tertiary<br>10% at 10 $\mu m$ with aluminized tertiary |
| **Image Quality of Telescope Optics at 0.6 $\mu m$** | 1.5 arcsec on-axis (80% encircled energy) | **Recovery Air Temperature in Cavity (and Optics Temperature)** | = 240 K |
| **Diffraction Limited Image Size** | 0.1" • $\lambda\ \mu m$ FWHM | | |

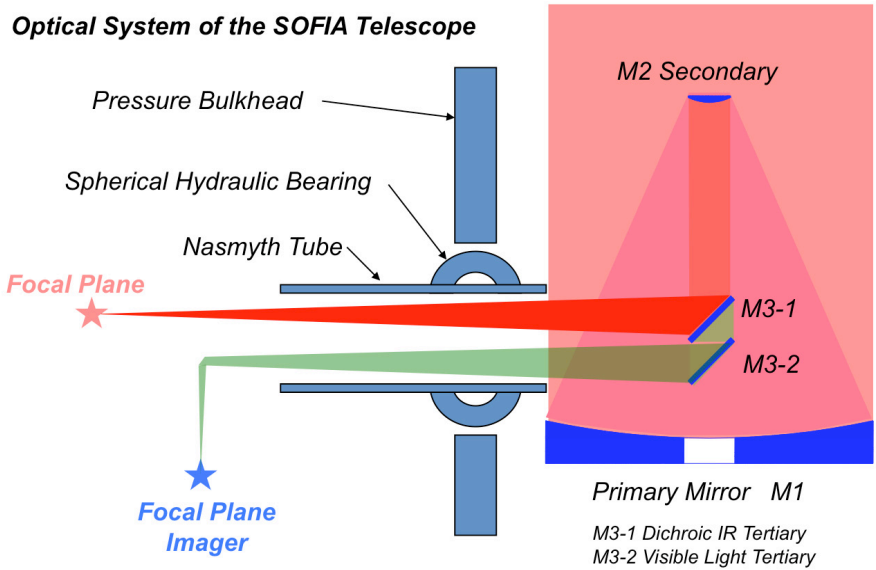

*Figure 1-5.* *The bent Cassegrain-Nasmyth optical configuration of the SOFIA 2.5-meter infrared telescope.*

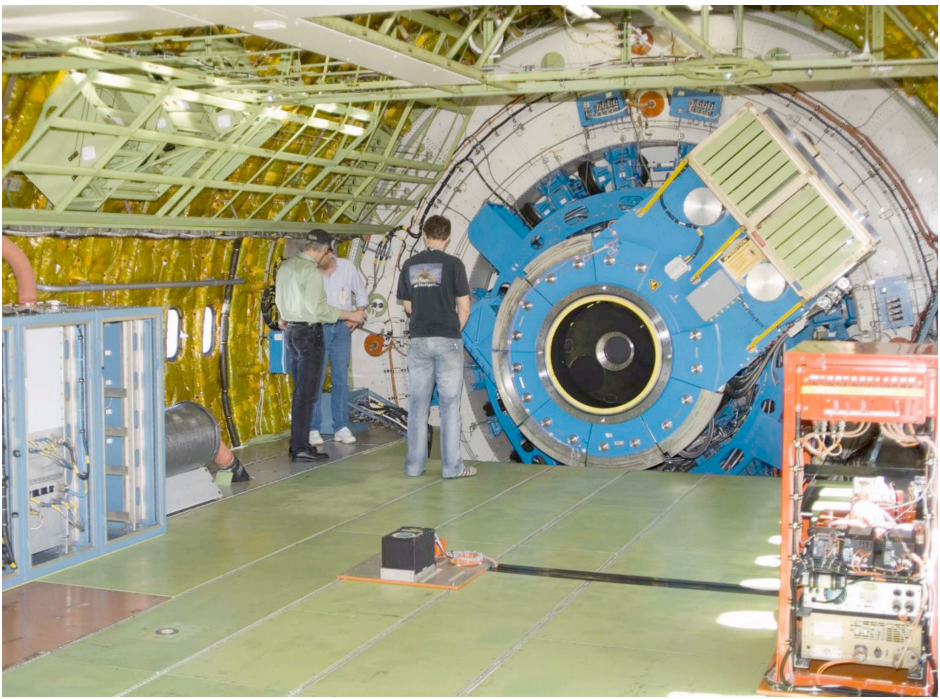

*Figure 1-6.* *Main cabin looking aft toward the pressure bulkhead and telescope assembly.*

### 1.4 SOFIA's First Generation of Instruments

Nine first generation Science Instruments (SIs) are under development for SOFIA (Table 2). They cover a much wider range of wavelengths and spectral resolutions than those of any other observatory (see Figure 1-7). These include three Facility Class Science Instruments (FSIs): the High-resolution Airborne Wideband Camera (HAWC), FORCAST, and the First Light Infrared Test Experiment CAMera (FLITECAM). The US FSIs will be maintained and operated by the Science Mission Operations (SMO) staff. FSI pipeline-reduced and flux calibrated data will be archived for general access by the astronomical community after a one year exclusive access (proprietary) period. In addition, there are six Principal Investigator (PI) class instruments maintained and operated by the PI teams. These instruments are designed to be less general in their potential applications than are the FSIs and are more likely to undergo upgrades between flight series, which has the advantage of keeping them more state-of-the-art at the expense of not having fixed capabilities. General investigators can propose to use these latter instruments in collaboration with the PI team that developed the instrument. In the present development plans, pipeline reduced data from the US PI instruments will also be added to the science archive, again after a one year exclusive access period. Two PI-class instruments are being developed in Germany, although the German PI instrument Field Imaging Far-Infrared Line Spectrometer (FIFI LS) will be available to the US science community as a Facility-like instrument under special arrangement with the FIFI LS team. The FIFI LS data will be pipeline-reduced and flux-calibrated before it is placed in the data archive. Further information about all the first-generation instruments can be found at:

*http://www.sofia.usra.edu/Science/instruments/index.html*

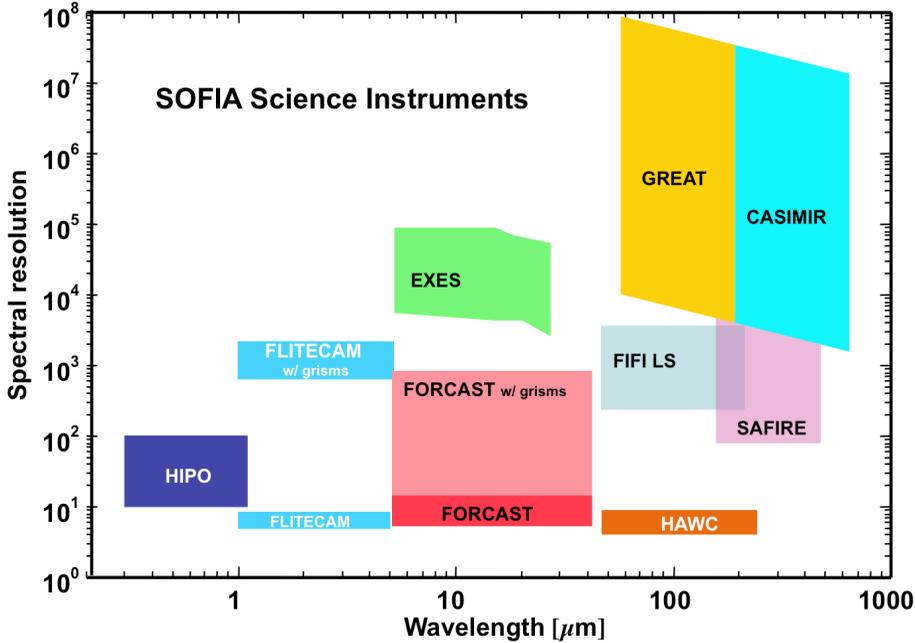

*Figure 1-7.* *SOFIA first generation instruments shown in a plot of log spectral resolution vs. log wavelength. FORCAST and GREAT are the first-light instruments.*

## Table 1-2. SOFIA Instrument Descriptions

| SOFIA Instrument | Description | Built by / PI | $\lambda$ range ($\mu$m) / spec res ($\lambda_c/\Delta\lambda$) | Field of View / Array Size | Available |
|---|---|---|---|---|---|
| **FORCAST** | Faint Object InfraRed CAmera for the SOFIA Telescope<br>Facility Instrument - Mid IR Camera and Grism Spectrometer | Cornell<br>T. Herter | 5 - 40<br>R ~ 200 | 3.2' x 3.2'<br>256 x 256 Si:As, Si:Sb | 2010 |
| **GREAT** | German Receiver for Astronomy at Terahertz Frequencies<br>PI Instrument - Heterodyne Spectrometer | MPIfR, KOSMA<br>DLR-WS<br>R. Güsten | 60 - 200<br>R = $10^6$ - $10^8$ | Diffraction Limited<br><br>Single pixel heterodyne | 2010 |
| **FIFI LS** | Field Imaging Far-Infrared Line Spectrometer<br>PI Instrument w/ facility-like capabilities - Imaging Grating Spectrometer | MPE, Garching<br>A. Poglitsch | 42 - 210<br>R = 1000 - 3750 | 30"x30" (Blue)<br>60"x60" (Red)<br>2 - 16x5x5 Ge:Ga | 2010 |
| **HIPO** | High-speed Imaging Photometer for Occultation<br>Special PI Instrument | Lowell Obs.<br>E. Dunham | .3 - 1.1 | 5.6' x 5.6'<br>1024x1024 CCD | 2012 |
| **FLITECAM** | First Light Infrared Test Experiment CAmera<br>Facility Instrument - Near IR Test Camera and Grism Spectrometer | UCLA<br>I. McLean | 1 - 5<br>R~2000 | 8.2' x 8.2'<br>1024x1024 InSb | 2012 |
| **CASIMIR** | CAltech Submillimeter Interstellar Medium Investigations Receiver<br>PI Instrument - Heterodyne Spectrometer | Caltech<br>J. Zmuidzinas | 200 - 600<br>R = $3 \times 10^4$ - $6 \times 10^6$ | Diffraction Limited<br>Single pixel heterodyne | 2012 |
| **HAWC** | High-resolution Airborne Wideband Camera<br>Facility Instrument - Far Infrared Bolometer Camera | Univ of Chicago<br>D. Harper | 50 - 240 | Diffraction Limited<br>12x32 Bolometer | 2013 |
| **EXES** | Echelon-Cross-Echelle Spectrograph<br>PI Instrument - Echelon Spectrometer | UT/UC Davis<br>NASA Ames<br>M. Richter | 5 - 28<br>R = $10^5$, $10^4$, or 3000 | 5" to 90" slit<br><br>1024x1024 Si:As | 2013 |
| **SAFIRE** | Submillimeter And Far InfraRed Experiment<br>PI Instrument - Bolometer array spectrometer | GSFC<br>H. Moseley | 145 - 450<br>R ~ 2000 | 160" x 320"<br>32x40 Bolometer | 2013 |

### 1.4.1 FORCAST

FORCAST is a facility class, mid-infrared diffraction-limited camera with selectable filters for simultaneous continuum imaging in two bands, within the 4-25 and 25-40 $\mu$m spectral regions. FORCAST will also conduct low-resolution grism spectroscopy in the 4-8, 16-25 and/or 25-40 $\mu$m regions. The high-sensitivity widefield imaging uses 256x256 Si:As and Si:Sb detector arrays which sample at 0.75 arcsec/pixel giving a 3.2 arcmin x 3.2 arcmin field-of-view. For small objects, chopping can be performed on the array to increase sensitivity.

### 1.4.2 GREAT

GREAT, a 2-channel heterodyne instrument, offers observations in three frequency bands with frequency resolution down to 45 kHz. The lower band, 1.4-1.9 THz, covers fine-structure lines of ionized nitrogen and carbon. The middle band is centered on the cosmologically relevant 1-0 transition of deuterated molecular hydrogen (HD) at 2.6 THz and the rotational ground-state transition of OH. A high-frequency band includes the 63 $\mu$m transition of [OI]. The receivers employ sensitive superconducting mixer elements, superconductor-insulator-superconductor (SIS) tunnel junctions and hot electron bolometers. A polarizing beam splitter allows simultaneous measurements of two channels at the same time.

### 1.4.3 FIFI LS

FIFI LS contains two medium resolution (R~1700) grating spectrometers with common foreoptics feeding two 16 x 25 pixel Ge:Ga detector arrays. A beamsplitter allows the two Littrow spectrometers to simultaneously observe an object in two spectral lines in the wavelength ranges 42 - 110 $\mu$m, and 110 - 210 $\mu$m, in 1st and 2nd order respectively. An image slicer in each spectrometer redistributes 5 x 5 pixel diffraction-limited spatial fields-of-view along the 25 entrance slits. FIFI LS will offer instantaneous coverage at 170 km/s resolution over a velocity range of ~1500 to 3000 km/s around selected lines for each of the 25 spatial pixels.

### 1.4.4 FLITECAM

FLITECAM will provide seeing-limited imaging from 1-3 $\mu$m and diffraction-limited imaging from 3 to 5.2 $\mu$m. Its objective is to cover near infrared science applications taking advantage of good atmospheric transmission and low thermal background. FLITECAM will also provide moderate resolution grism spectroscopy (R~2000).

### 1.4.5 HIPO

HIPO is a special-purpose instrument designed to provide simultaneous high-speed, time-resolved imaging photometry at two optical wavelengths. HIPO makes use of SOFIA's mobility, freedom from clouds, and near-absence of scintillation noise to provide data on transient events like stellar occultations and data acquisition at optical and near-IR wavelengths. HIPO and FLITECAM can be mounted simultaneously to enable data acquisition at two optical and one near-IR wavelengths. HIPO has a flexible optical system and numerous readout modes, allowing specialized observations.

> *SOFIA's optical performance will allow it to produce the sharpest images of any current or planned IR telescope at many scientifically valuable infrared wavelengths.*

### 1.4.6 HAWC

HAWC is a far-infrared camera designed to cover the 40-300 $\mu$m spectral range at diffraction-limited resolution. HAWC utilizes a 12x32 pixel array of bolometer detectors constructed using the silicon pop-up detector (SPUD) technology developed at Goddard Space Flight Center. The array will be cooled by an adiabatic demagnetization refrigerator and operated at a temperature of 0.2 K. HAWC may eventually be upgraded to perform far infrared polarimetry.

### 1.4.7 CASIMIR

CASIMIR is a submillimeter and far-infrared heterodyne receiver that uses sensitive superconducting mixers, including both tunnel junction superconductor-insulator-superconductor (SIS) and eventually hot electron bolometers (HEB). The local oscillators are continuously tunable. CASIMIR will cover the 500-2100 GHz frequency range in seven bands: SIS mixers in four bands up to 1200 GHz, and HEB mixers in three bands covering the rest. Four bands can be selected for use on a given flight. The receiver has an intermediate-frequency (IF) bandwidth of 4 GHz, processed by a high-resolution backend commercial Fast Fourier Transform (FFT) spectrometer with 250 kHz resolution.

### 1.4.8 EXES

The Echelon-Cross-Echelle Spectrograph (EXES) operates in three spectroscopic modes (R~$10^5$, $10^4$, and 3000) from 5-28 $\mu$m using a 1024 x 1024 Si:As blocked impurity band (BIB) detector. High dispersion is provided by a large echelon grating. This requires an echelle grating to cross-disperse the spectrum, resulting in continuous wavelength coverage of ~5 cm$^{-1}$ and a slit length of ~10" at R=$10^5$. The echelon can be bypassed so that either the echelle or low order grating can act as the sole dispersive element. This results in a single order spectrum with slit length of roughly 90" and R=$10^4$ or 3000, respectively. The low-resolution grating also serves as a slit positioning camera when used face on.

*1.4.9 SAFIRE*

SAFIRE is an imaging spectrometer covering the spectral region from 145 to 450 $\mu$m with a spectral resolving power of about 2000. The instrument uses a 32 x 40 array of superconducting bolometers to provide background limited performance for critical science applications. The nominal field of view is about 5.3 arcminutes.

### 1.5 SOFIA's Performance Specifications with its First Generation Instruments

SOFIA will be capable of high-resolution spectroscopy (R $\geq 10^4$) at wavelengths between 5 and 600 $\mu$m (see Figure 1-7). The 8 arcminute diameter field of view (FOV) allows use of very large format detector arrays. Despite the relatively large thermal IR background, the 2.5-meter aperture of the SOFIA telescope will enable measurements with about an order of magnitude better photometric sensitivity than IRAS and a factor of > 3 better linear spatial resolution than Spitzer, and will match or be more sensitive than the European Space Agency's Infrared Space Observatory (ISO) (Figure 1-8, left). SOFIA's capability for diffraction-limited imaging beyond 15 $\mu$m will produce the sharpest images of any current or planned IR telescope operating in the 30 to 60 $\mu$m region (Figure 1-8, right). SOFIA's performance for line flux measurements with various first generation instruments is shown in Figure 1-9. Each instrument will have an exposure time calculator on the SOFIA website to enable prospective observers to evaluate the feasibility of the programs they propose to conduct. See the followng URL: http://www.sofia.usra.edu

> SOFIA will be an ideal platform for the first or early deployment of new detector and instrumentation technology.

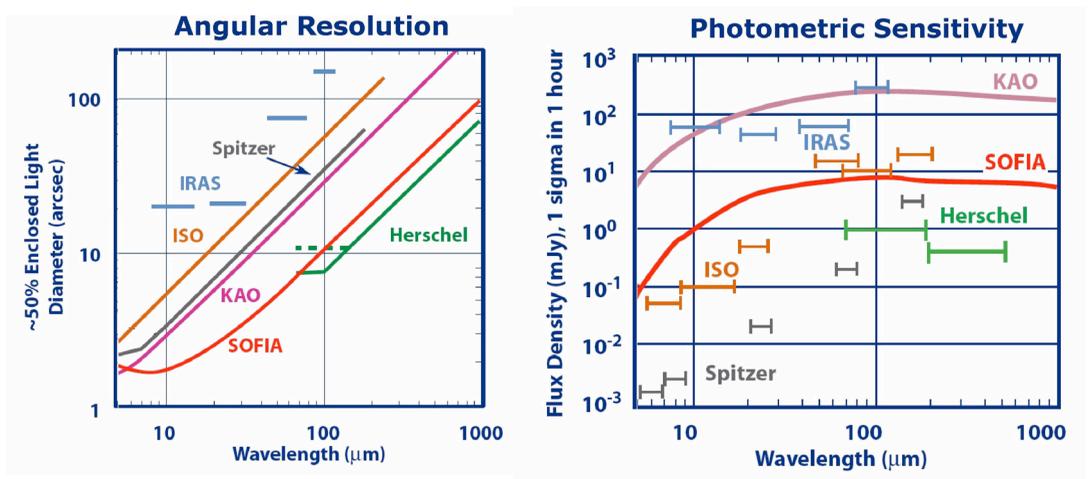

***Figure 1-8.** SOFIA's sensitivity will be comparable to that of ISO (right). It will form images three times smaller than those formed by the Spitzer Space Telescope (left).*

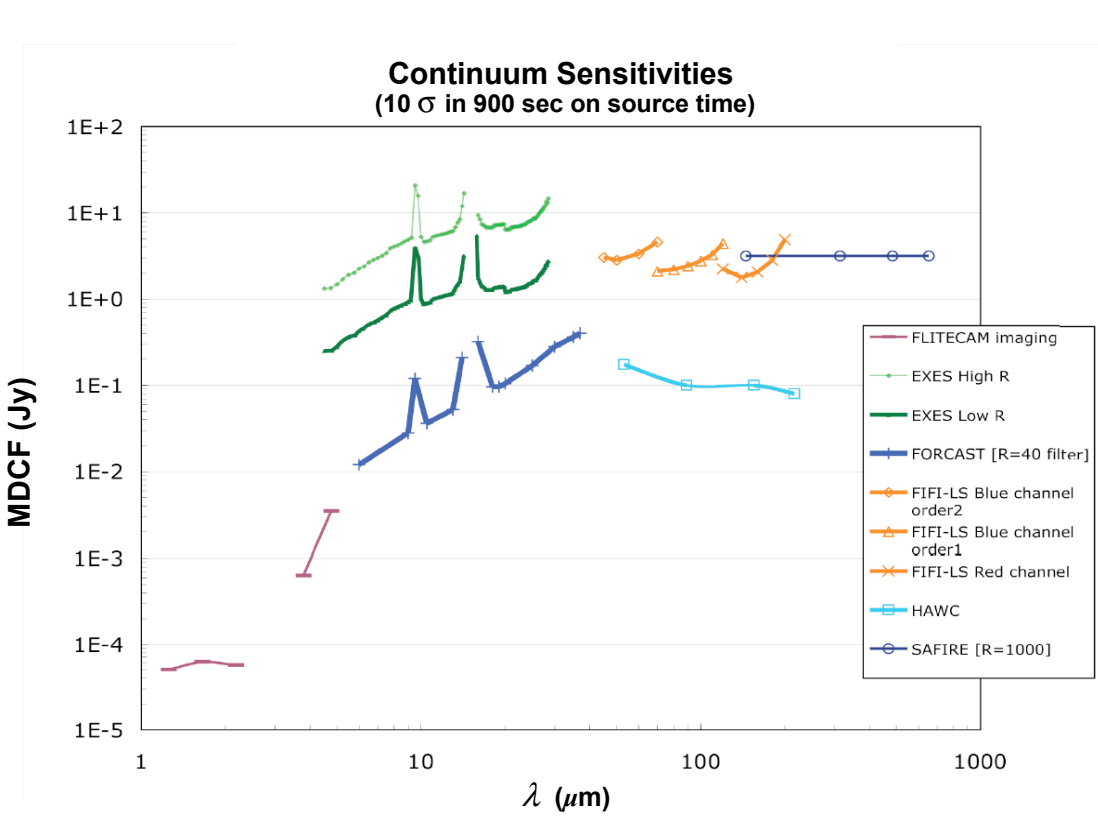

***Figure 1-9.*** *The continuum sensitivity of SOFIA instruments expected at the time of full operational capability. Shown is the 10 σ minimum detectable continuum point source flux densities (MDCF) in Janskys for 900 seconds of integration on source. Observing and chopper efficiency have not been included.*

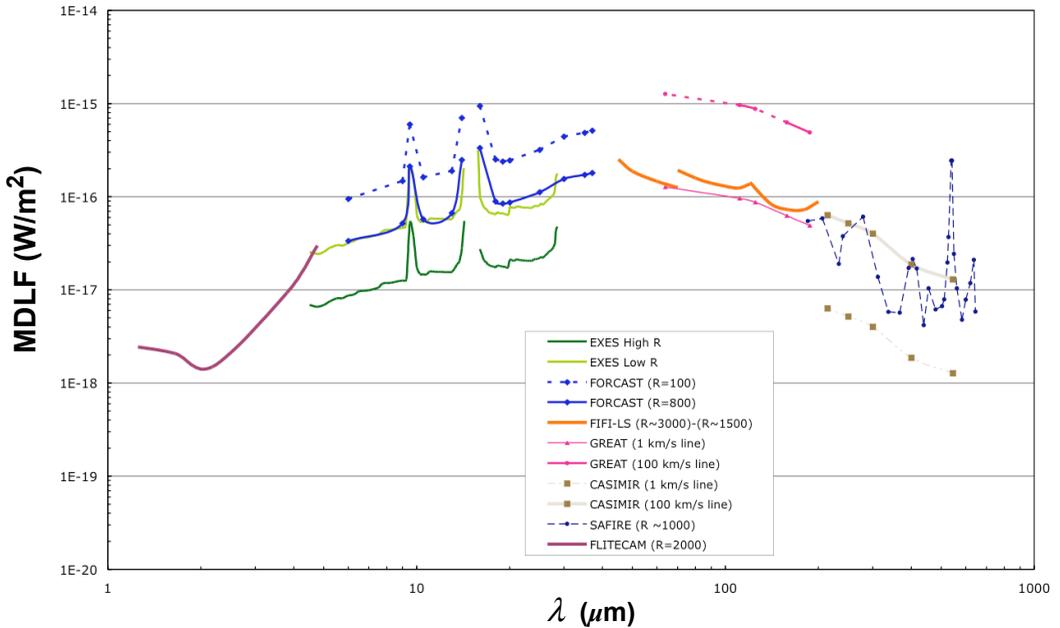

*Figure 1-10.* *The expected line sensitivity of SOFIA spectrometers at the time of full operational capability. Shown is the 10 σ minimum detectable line flux (MDLF) in watts per meter squared for 900 seconds of integration on source. Observing and chopper efficiency have not been included.*

### 1.6 Future Instrumentation

A major advantage of SOFIA over space-based missions is its ability to rapidly incorporate instrument improvements and to accommodate upgrades to respond to new technological developments. Focal plane technology is still expanding rapidly in the far-IR, and major advances in detector sensitivity and array size are anticipated. Those technologies expected to advance during SOFIA's lifetime were explored at a workshop entitled "SOFIA's 2020 Vision: Scientific and Technological Opportunities," held at Caltech December 6-8, 2007.

The conference website is located at:

http://www.sofia-vision.caltech.edu/

The conference papers may be viewed at:

http://www.sofia-vision.caltech.edu/program.html

The conference summary paper (Young et al. 2008) may be viewed at:

http://www.sofia.usra.edu/Science/07Dec_SOFIA_Vision/ SOFIA_2020Vision_white_paper_final.pdf

The next call for new instruments will be in FY 2011 and there will be additional calls every 3 years resulting in one new instrument or major upgrade to observatory instrumentation each year. Suggestions for future instrumentation already include:

- Expanded Heterodyne Wavelength coverage
- Arrays of Heterodyne Detectors
- Polarimeters
- R~2000, 5 to 60 $\mu$m Integral Field Unit Spectrometer
- R~200, 5 to 100 $\mu$m Spectrometer
- R~100,000, 28-100 $\mu$m Spectrometer
- R~100,000, 1-5 $\mu$m Spectrometer
- Kinetic Induction Detector (KID) Spectrometer

Recent evaluations of the scientific case for SOFIA have stressed the importance of imaging and spectroscopic polarization measurements for a number of key investigations. Strong consideration is therefore being given to adding polarimetric capability to HAWC within the next few years.

## 1.7 Unique Advantages of SOFIA

The SOFIA Observatory embodies a number of key advantages that will make it a unique tool for astronomy in the coming decades.

**SOFIA is a near-space observatory that comes home after every flight.** One of its great strengths is that its scientific instruments can be exchanged regularly to accommodate changing science requirements and new technologies that do not need to be space qualified. Furthermore, large, massive, complex and sophisticated instruments with substantial power and heat dissipation needs can be flown on SOFIA. Finally, simple repairs and adjustments can be performed on an instrument in flight, thereby increasing SOFIA's science productivity.

**SOFIA has unique capabilities for studying transient events.** The observatory can operate from airbases worldwide on short notice to respond to new discoveries in both the northern and southern hemispheres. SOFIA has the flexibility to respond to events such as supernovae and nova explosions, cometary impacts, comet apparitions, eclipses, occultations, near-Earth objects, activity in Active Galactic Nuclei, and activity in luminous variable stars.

**SOFIA's wide range of instruments will facilitate a coordinated science program through analysis of specific targets.** No other observatory operating in SOFIA's wavelength range can provide such a large variety of available instruments for

such a long period of time. A particular advantage of SOFIA is that it will be able to access events unavailable to many space observatories because of the viewing constraints imposed by their orbits. For example, SOFIA can observe astrophysical events which occur close to the sun, enabling temporal monitoring of supernovae, novae, and variable stars throughout the year. SOFIA will be the only infrared mission allowing observations of the inner planets and comets when they are brightest and most active.

**SOFIA's 20-year operational lifetime will enable long-term temporal studies and follow-up of work initiated by SOFIA itself and by other observatories.** Many space missions are relatively short compared with the critical cycle of observation, analysis, and further observation. The Herschel and James Webb Space Telescope (JWST) Observatories will raise scientific questions that will benefit from follow-up observations well after their missions have ended. On the basis of present plans, SOFIA will be the only facility operating between 25 and 350 $\mu$m following Herschel and JWST. SOFIA will keep the community engaged in fundamental science research until the next generation of missions is launched.

### 1.8 Training Students and Developing Technology with SOFIA

The continuous training of instrumentalists is a high priority for the United States and German science community. SOFIA will facilitate the training of students and faculty in instrument hardware and software development. It will present an ideal venue in which to educate students, where they can participate in hands-on, cutting-edge technology developments. These opportunities are generally not available to students working on satellite projects.

SOFIA will energize the next generation of young experimental astrophysicists and help to develop their talents in many different scientific and engineering areas. SOFIA graduate and post-doctoral students will form a rich reservoir of talent and will become the next generation of Principal Investigators and Instrument Scientists, as has been the legacy of the KAO.

SOFIA will also be an ideal platform for the first or early deployment of new detector and instrumentation technology. Unlike a typical space-borne observatory, SOFIA will always be able to utilize the latest state-of-the-art technology in terms of sensitivity, detector response time, observation technique, spectral resolution and more, by releasing instrument AO's on a regular and frequent (i.e., every 2 to 3 years) basis. SOFIA instruments can be much more complex and much larger in volume, weight and power consumption than those deployed on space-based observatories.

*Training of instrumentalists is a high priority for the United States and German science communities. As a hands-on observatory, SOFIA will offer training opportunities that space-based instruments cannot.*

### 1.9 Synergy and Complementarity between SOFIA and other Missions

The comparative performance of SOFIA with respect to other infrared missions has been noted in section 1.4 - "SOFIA's First Generation of Instruments" on page 1-7. SOFIA's science program will cover a long time period, so SOFIA observations will be complemented by data from present facilities such as HST, Chandra, Spitzer, SMA, and AKARI and future facilities such as WISE, Herschel, JWST, ALMA, SPICA, and SAFIR (see Figure 1-11). Specific examples of how the SOFIA science results will supplement and can be complemented with results from other missions can be found in the science examples given in the following sections.

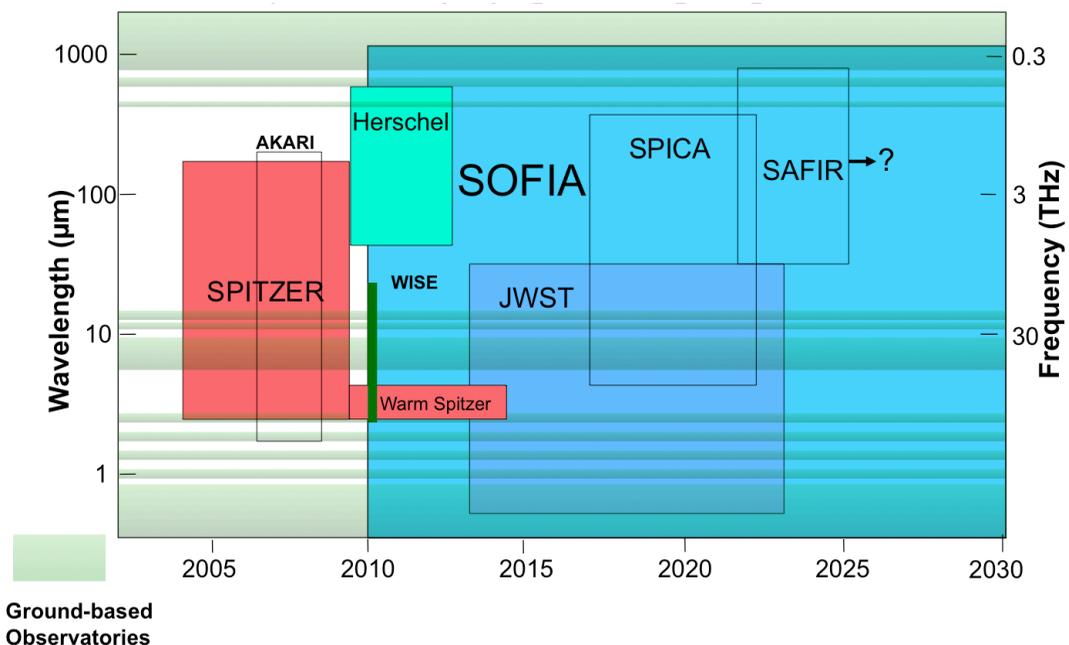

***Figure 1-11.*** *SOFIA's flight lifetime and time-frame will make it the premier facility for doing far-IR and submillimeter wave astronomy from 2010 until the mid 2030s. It will be the only facility available for wavelength coverage in the 28-1200 µm spectral region and for high resolution spectroscopy during much of that period. The SPICA and SAFIR missions have yet to be formally approved. The length of the SAFIR mission is undetermined at present.*

In general, SOFIA will excel at those observations that demand some combination of: good mid and/or far-infrared atmospheric transmission, reasonably high spatial resolution, very high spectral resolution, and/or the ability to rapidly deploy to a specific location on the Earth. In particular, SOFIA will be a powerful probe for understanding the physics and chemistry of a wide variety of astronomical objects, ranging from solar system objects out to distant galaxies.

### 1.10 SOFIA's Education and Public Outreach Program

Because of its accessibility and ability to carry passengers, SOFIA will include a vigorous, highly visible Education and Public Outreach (E&PO) program designed to exploit the unique attributes and programmatic excitement of airborne astronomy. SOFIA is the only major research observatory designed to bring non-scientists into routine close contact with scientists in a research environment. SOFIA is uniquely capable of giving the nation's science educators, such as Kindergarten through high school teachers, college faculty, science museum personnel, and amateur astronomers, hands-on experience with the processes of scientific inquiry. The objectives of the E&PO program are to a) enhance science, technology, engineering and math (STEM) education in communities across the U.S. and Germany, b) establish long-term relationships between educators, scientific researchers, and NASA and DLR, and, c) communicate the value of scientific research to the public. These goals will be achieved by having educators team with SOFIA researchers and engineers, and by having educators and journalists participate in the observatory's research flights. SOFIA's E&PO objectives are responsive to the NASA and DLR strategic educational goals of strengthening the future workforce, attracting and retaining STEM students, and engaging citizens in the mission of the space agencies.

*Because of its accessibility and ability to carry passengers, SOFIA will include a vigorous, highly visible Education and Public Outreach (E&PO) program designed to exploit the unique attributes and programmatic excitement of airborne astronomy.*

### 1.11 Scope of this Science Vision

This publication describes a number of exciting science programs that might be undertaken with SOFIA. These programs have been organized along four themes:

- The Formation of Stars and Planets — Chapter 2
- The Interstellar Medium of the Milky Way — Chapter 3
- Galaxies and the Galactic Center — Chapter 4
- Planetary Science — Chapter 5

It is import to recognize the programs described in these chapters are only representative of the science that SOFIA is capable of addressing — they are in no way a comprehensive listing of all of the investigations that might be conducted. Each chapter has a corresponding table in Appendix B listing a sample of the investigations, both in terms of the numbers of objects that are needed to obtain the needed data and the amount of SOFIA observing time that would be required. Finally, to aid the reader, we have listed many of the astronomical lines discussed in the text, with accurate wavelengths, as Appendix B.6. We also list the atmospheric transmission expected at typical observing conditions for the rest wavelengths.

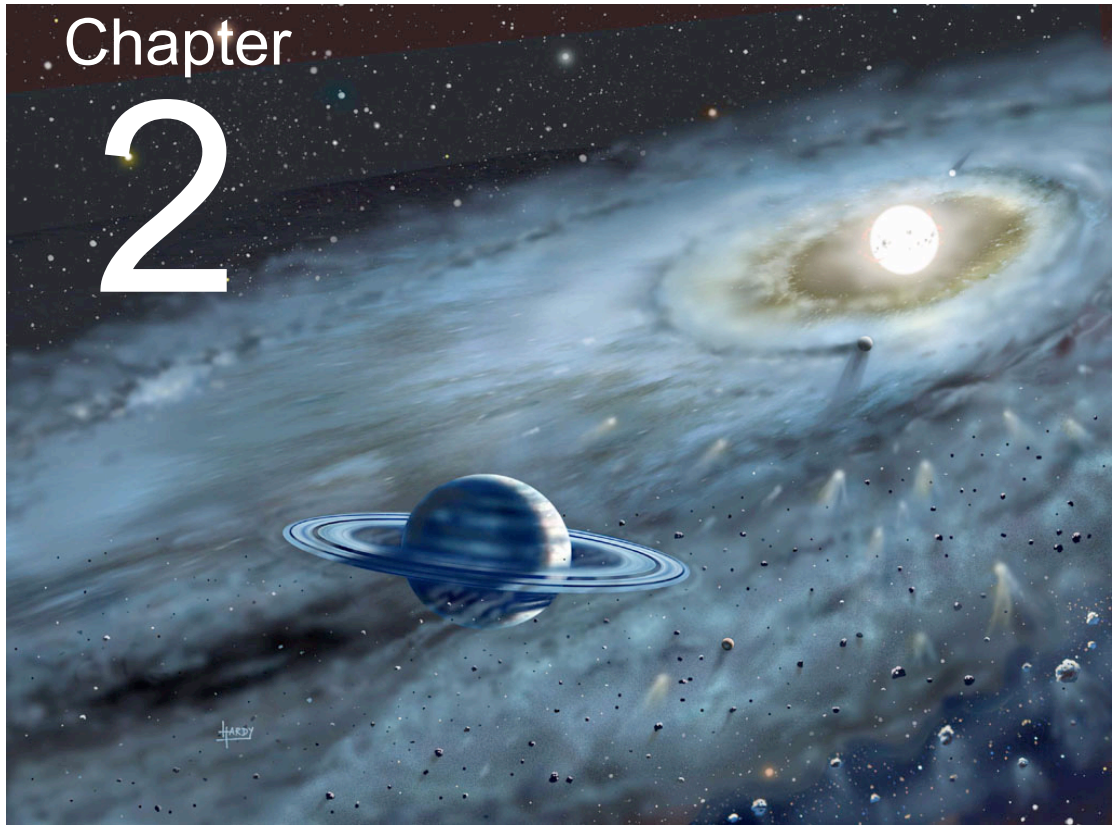

# The Formation of Stars and Planets

## 2.1 Overview

The formation of stars and planets is a central problem of astrophysics. On the largest scales, star formation is a key component of galaxy evolution. On smaller scales, the linked processes of star and planet formation provide the foundation for understanding both our own solar system and the wild diversity that we are finding among other planetary systems. The cycling of gas and dust from the interstellar medium into stars and planets and back into the interstellar medium occurs in dust-enshrouded environments and at relatively low temperatures. Thus, most emitted radiation emerges at the infrared wavelengths from 1 $\mu$m to about 1 mm.

While SOFIA will provide vital information on the formation of stars of all masses and will shed light on many aspects of star and planet formation, we have focused here on three topics where SOFIA has unique capabilities. These are the formation of massive stars, circumstellar disks, and astrochemistry. In the following subsections, we address each of these, noting interconnections among them and connections to the other topics in this document.

The major overarching question this chapter addresses is: *What physical, chemical, and dynamical processes result in the formation of stars and planets?* In recent years, theory and observations have made great progress on low-mass star formation, but the process of massive star formation remains a largely unknown process. SOFIA will be used to address physical, chemical, and dynamical differences as a function of mass to help understand this problem. The observations we are detailing in these sections can also be applied to numerous star forming regions throughout the Galaxy, allowing us to understand how star formation varies as a function of environment (i.e., Galactic Center, in spiral arms, between spiral arms, the outer Galaxy).

## 2.2 The Formation of Massive Stars

The overall star formation rate in a galaxy determines how rapidly the interstellar medium (ISM) is converted into stars, planets, and smaller solid bodies. The star formation rate in galaxies is traced by massive stars. Massive stars dominate the energy budget of young galaxies; their radiation, both direct and reprocessed by enshrouding dust, powers the optical/near-infrared and far-infrared peaks in the radiation field of the Universe. Understanding galaxy evolution will require a better understanding of massive star formation, as we can study the process in detail within our own Galaxy. Massive stars are also responsible for the creation and distribution of elements in the galaxy. They are therefore ultimately responsible for the chemical building blocks necessary for the origin of planets and life (see sections 2.3 and 2.4). Despite their overall importance in astronomy, the formation of massive stars is poorly understood. Here we highlight some of the important observations with SOFIA that can help to make significant progress in the field of massive star formation.

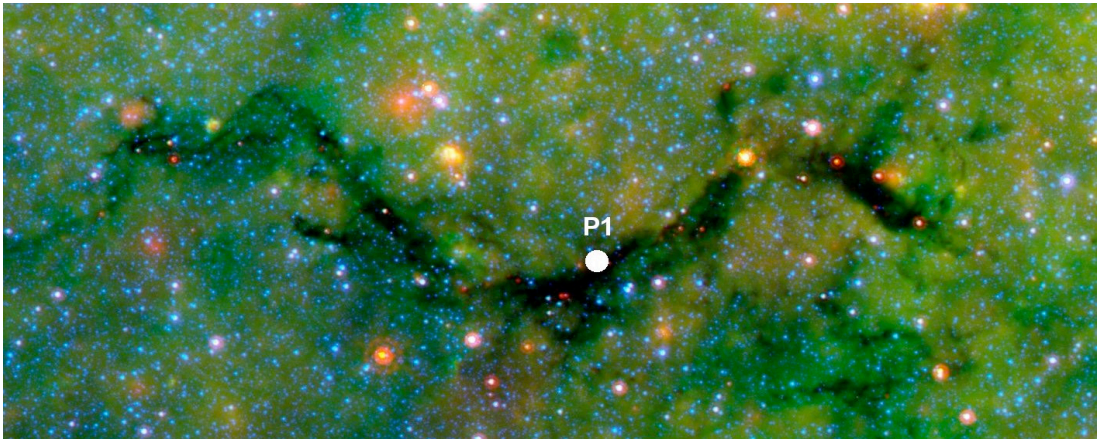

***Figure 2-1.*** *Infrared dark cloud G11.11-0.12 in Sagittarius as viewed by Spitzer shows a string of dense, dusty clumps extending 21 pc that appear dark against the bright infrared background of the Galactic plane. It is within dark clouds and dense millimeter cores like these that massive stars are thought to be born. This false-color image is a composite of infrared data taken by Spitzer's IRAC and MIPS instrument (blue represents 3.6 µm light; green shows light of 8 µm; and red is 24 µm light) and is 1 degree wide. The white dot represents the location of the sub-mm source P1 that is discussed in Figure 2-2. Credit: NASA/JPL-Caltech/S. Carey (SSC/Caltech).*

### 2.2.1 Disentangling the Luminosity

There are two key questions for understanding star formation and its role in galaxy evolution: *What controls the initial mass function and what physical conditions and processes lead to the formation of the most massive stars*? We can answer these questions in our own Galaxy if we can separate the individual sources of luminosity in regions forming massive stars, but we need to study regions that retain enough of their dense gas so that we can tie initial conditions to stellar outcomes.

One of the major road blocks in the study of massive star formation is that the highest mass stars are generally born in the densest and most obscured cloud cores deep within giant molecular clouds. As a consequence, these young massive stellar objects can only be studied at relatively long wavelengths due to high extinction. A new generation of mm and sub-mm Galactic Plane surveys (i.e., Bolocam Galactic Plane Survey, SCUBA2 All Sky Survey, ATLASGAL) are, or will be, producing catalogs of thousands of such cloud cores.

These surveys complement catalogs (Egan et al. 1998; Simon et al. 2006) of infrared-dark clouds (IRDCs), so called because they appear dark at near-infrared (and often even mid-infrared) wavelengths against the bright infrared background of the Galactic Plane (see Figure 2-1). It is within such cold and dense cloud cores and IRDCs that massive star formation is believed to take place. Complementary surveys with Spitzer (GLIMPSE and MIPSGAL) are providing near- and mid-IR images of massive star forming regions in the Galactic Plane (Figure 2-1). These surveys have uncovered many potential sites of high-mass star formation that now demand a more subtle analysis.

The study of young massive stars is further hampered by the gregarious nature of their formation. Unlike low-mass stars, massive stars almost never form in isolation, with recent observations showing that more than 95% of O stars form in clusters (de Wit et al. 2004, 2005), necessitating observations with high spatial resolution.

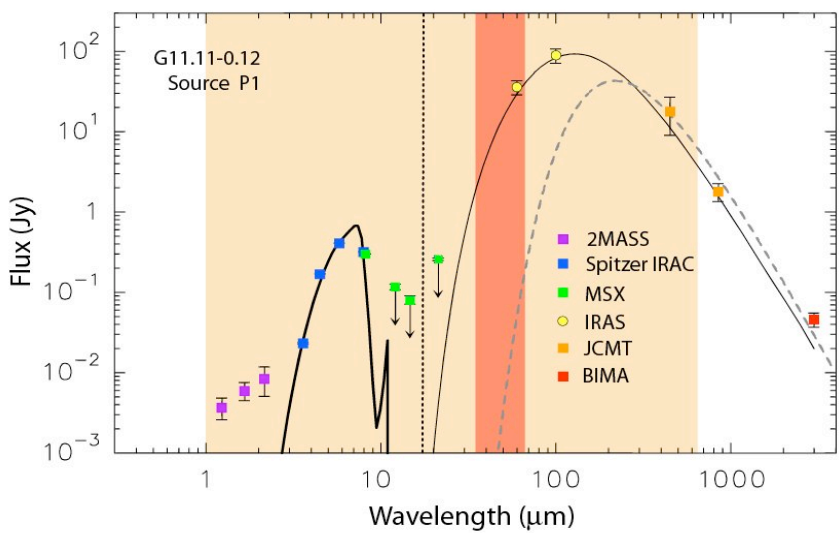

***Figure 2-2.*** *Spectral energy distribution data for sub-mm source P1 in the IRDC shown in Figure 2-1. The thin black solid line at long wavelengths corresponds to the model of a central star of $L=1200\ L_{sun}$ that is surrounded by an envelope. These are fit through the IRAS data points, which are likely upper limits due to contamination by flux from several nearby sources. A second possible model of the long wavelength points is shown as the dashed grey line, which fits the sub-mm/mm data better, but has significantly different luminosity. However, neither model fits the shorter wavelength data. A separate model (thick black line) of a 550K blackbody can fit the mid-IR data only, but not the far-IR or near-IR. Better spatial resolution at far-IR wavelengths and better spectral coverage throughout the SED is needed to disentangle and model the luminosity in this area. The light yellow area shows the wavelength coverage of SOFIA's instruments, and the dark orange area is where SOFIA's instruments are unique. Data points shown are fluxes due to dust emission at 1.2, 1.65, and 2.2 µm from 2MASS (purple); 3.6, 4.5, 5.8 and 8.0 µm from Spitzer IRAC (blue), MSX detection at 8.2 µm and upper limits at 12, 15 and 21 µm (green); 60 and 100 µm from IRAS (yellow); the 450 and 850 µm JCMT (orange), and 3 mm from BIMA (red). Figure adapted from Pillai et al. 2006.*

However, the bulk of the energy emerges in the far-infrared, where previous missions (KAO, ISO, IRAS) have had very poor spatial resolution. Furthermore, many of the young massive sources are so bright at far-IR wavelengths that they satu-rate the detectors of sensitive space satellites (e.g., Spitzer). With the improved spatial resolution of SOFIA and the dynamic range of its instruments, we will be able to observe these sources without saturation problems and separate emission of different sources in crowded fields for detailed analysis. With this information in hand, we address the question: *What are the different sources of luminosity in massive star forming cores?*

Far-IR observations with the improved spectral and spatial sampling available with FORCAST and HAWC are needed to constrain the spectral energy distributions (SEDs) of individual young massive cores. Combined with high spatial resolution images from ground-based interferometers (like CARMA, SMA, and ALMA) one can model the different contributions to the luminosity of the source. Figure 2-2 shows the SED of a sub-mm core in the IRDC G11.11-0.12 that actually displays emission at wavelengths shorter than 20 $\mu$m. For some very young massive stars displaying broad wavelength emission (as shown in Figure 2-2), a single model cannot properly explain both the shorter wavelength (2 - 25 $\mu$m) and longer wavelength emission (350 $\mu$m - 3 mm). The higher spatial resolution and broad spectral coverage in the 25 to 300 $\mu$m region afforded by SOFIA can aid modeling by filling in much needed data throughout the SED. Herschel will be able to image only at wavelengths longer than 60 $\mu$m where the diffuse interstellar cirrus is brighter and may cause confusion. However, it should be pointed out that most high-mass star forming cores are totally opaque at wavelengths less than 20 $\mu$m, and thus typical SED models and data look like the part of Figure 2-2 that lies to the right of the vertical dotted line. Sub-mm and mm points alone cannot constrain the luminosity of the source because there is no information on where the SED turns over (see dashed versus solid line fits in Figure 2-2). IRAS 60 and 100 $\mu$m data provide only upper limits to the flux at these wavelengths because of contamination from the crowded fields they lie in and the very coarse spatial resolution (~2-5') of IRAS. Furthermore, most of these sources are saturated with MIPS on Spitzer. Again, SOFIA will be key to filling in the information throughout the SED for such sources and accurately yielding estimates of bolometric luminosity which is a fundamental measurement for understanding the stellar content of cold, opaque cloud cores. Given the large mid-IR to far-IR brightnesses of these sources (typically tens of Jy at 38 $\mu$m and hundreds of Jy at 100 $\mu$m), one could easily measure a well-sampled SED with a signal-to-noise ~10 with FORCAST and HAWC for 20 sources in about 15 hours.

### 2.2.2 Probing the Interiors of Massive Star Forming Cores

Material falling onto stars transitions from infall to rotation, encounters an accretion shock, and settles into a disk. In clumps forming clusters and massive stars,

> *SOFIA has a unique combination of broad spectral coverage from optical to millimeter wavelengths and high spatial resolution in the far-IR free of saturation, which will provide critical multi-wavelength data to aid massive star formation modeling.*

the dynamics are further complicated by supersonic turbulence. While few detailed models of line profiles expected from collapsing cores have been published, the "turbulent cores" model (e.g., McKee & Tan 2003) would predict very dense infalling gas around forming massive stars, but the model of competitive accretion (e.g., Bonnell et al. 1997) would not. Although copious observations of molecular lines at mm/sub-mm wavelengths exist, it remains very difficult to test theories. This gives rise to an important question: *What are the chemistry and dynamics of the core interiors?*

Owing to a combination of extreme abundance variations and sensitivity to density and temperature, different lines arise primarily in different parts of the massive star forming core. ALMA will probe dynamics using rotational lines from relatively low energy levels, probing temperatures and densities up to about 100 K and $10^6$ cm$^{-3}$ (Table 1 in Evans 1999). Lines that require even higher densities and/or temperatures probe the innermost regions, and these lie in the far-infrared (pure rotational) or mid-infrared (ro-vibrational). The ro-vibrational lines have revealed gas as hot as 1000 K with densities exceeding $10^7$ cm$^{-3}$ (e.g., Carr et al. 1995). This gas has high abundances of some molecules and probes regions within hundreds of AU of the central star.

High spectral resolving power ($R > 10^5$) between 5 and 150 $\mu$m is needed to study the dynamics, and this is not available on Spitzer, Herschel, or JWST. SOFIA has unique capabilities in this area from 5-28 $\mu$m (EXES), as well as from 60 to 150 $\mu$m (GREAT), and additional capabilities from 150 to 600 $\mu$m (GREAT, CASIMIR). These instruments provide new opportunities to probe the dynamics and chemistry closer to the forming massive star than has been possible with mm/sub-mm lines.

> *SOFIA has extremely high spectral resolution capabilities from the mid-IR to sub-mm, providing unique opportunities to probe the dynamics and chemistry closer to forming massive stars than has been previously possible.*

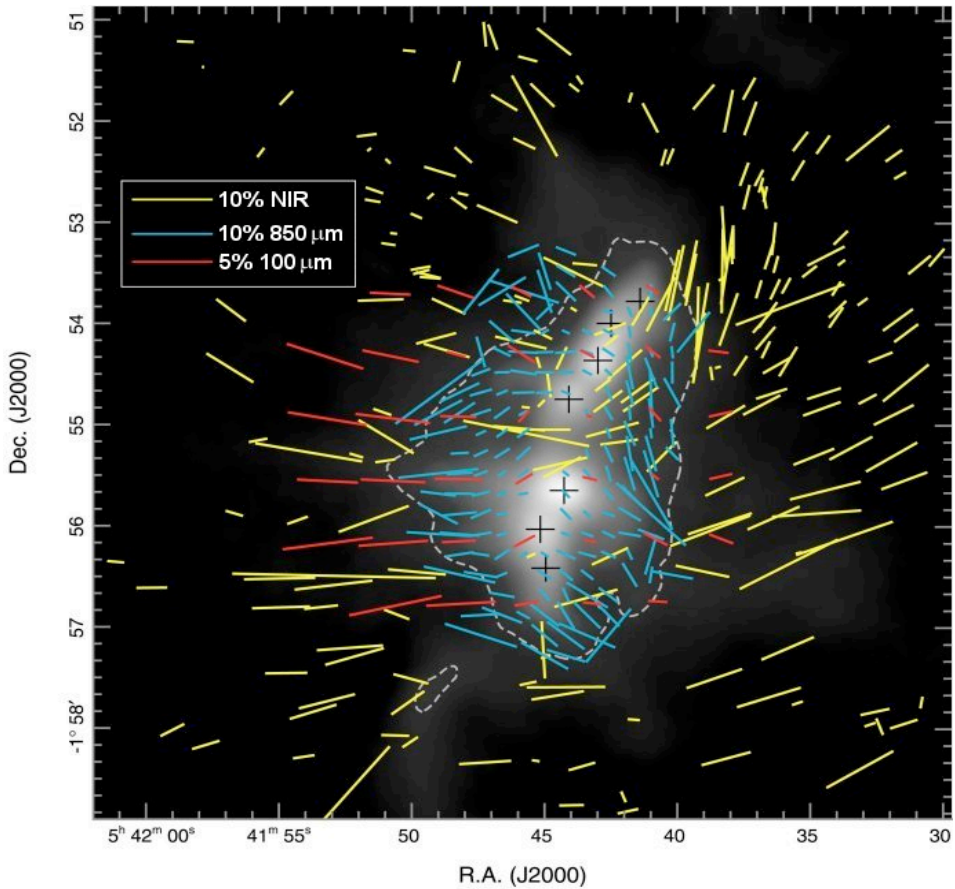

***Figure 2-3.*** *A comparison of near-IR (1.6 μm, yellow), far-IR (100 μm, red) and sub-mm (850 μm, blue) polarimetry of the NGC2024 star forming region with a background image of the 850 μm dust continuum intensity map (from Kandori et al. 2007). All are indicators of magnetic fields in this region. All data show a good agreement in the outer regions, but the correlation significantly breaks down in the inner dense region (within the dashed curve) where the near-IR polarization appears very different from the longer wavelength polarization, signifying a different magnetic geometry in the interior. To complete the 3-D magnetic picture, mid-IR polarization data at a similar scale to ground-based sub-mm observations is required. In order to tackle this, and other similar problems, a next-generation SOFIA based mid/far-IR polarimeter is crucial. This mid/far-IR instrumentation and technique are especially important for understanding magnetic fields in the interiors of massive star formation regions, where near-IR polarization measurements are often not possible.*

In particular for massive young stellar objects, the [OI] line (63 μm) is strong in the accretion shock, but it is also likely to be photo-excited at cloud surfaces creating photo-dissociated regions. Radial velocities from high-spectral resolution observations are needed to distinguish this emission from that in outflows and that from photo-dissociated regions, where ISO showed it to be ubiquitous. As has been shown by its ground-based counterpart, TEXES, the high spectral resolution capability of EXES will probe material within a few hundred AU of the forming star, providing data on hundreds of ro-vibrational lines simultaneously (Knez et al. 2003). Combining these unique probes with spectral imaging with ALMA will provide strong constraints on the dynamics in regions where massive stars form.

### 2.2.3 Magnetic Fields in Massive Star Forming Regions

Star formation theories disagree on the importance of magnetic fields. Current observations cannot resolve this disagreement, in part due to the limited observational platforms to make the necessary polarimetric observations. Polarization observations with SOFIA will allow us to address the critical question: *What are the magnetic field properties in massive star forming regions?*

Recent debate has centered on whether magnetic pressure (Mouschovias & Tassis 2008) or turbulence (Crutcher et al. 2008; Troland & Crutcher 2008) is the primary regulator in slowing the gravitational collapse of a star forming cloud core. The magnetic fields in regions of star formation are traced through observations of the polarization of dust particles that align themselves with magnetic field lines. Observations of polarization at different wavelengths probe different depths within star forming regions. Therefore, observations at sub-mm to radio wavelengths only yield part of the picture, tracing the cool dust in magnetic fields far from the stellar source. As one goes to the shorter wavelengths afforded by SOFIA, one probes further into the star forming region (Figure 2-3). Hence, polarization measurements at all wavelengths where dust emits radiation will give a 3-dimensional picture of the magnetic fields in a star forming region. How well ordered the magnetic fields are in massive star forming regions can tell us how important turbulence is compared to magnetic pressure and gravity. SOFIA is especially important for polarization measurements of massive star forming cores, where often light at wavelengths shorter than $5\,\mu m$ cannot escape. The magnetic field information in the inner regions of massive cores is therefore missing, and mid/far-IR polarization measurements from a next-generation SOFIA instrument are needed to complement polarization measurements at longer wavelengths (i.e., SMA, ALMA).

> *Mid/far-IR polarization measurements from a next-generation SOFIA instrument will be a unique tool that will, among other things, discern the magnetic field geometries in the interiors of massive star forming cores.*

## 2.3 Understanding Proto-planetary Disks

To understand the origin of our Solar System, we need to investigate the environment in which planets form within circumstellar disks around young stars. Over the last two decades, the study of circumstellar disks has focused on the shape of their spectral energy distributions (in a similar fashion discussed in the massive star formation section) and direct millimeter interferometric imaging (Hillenbrand et al. 1992, Mannings & Sargent 1997, Natta et al. 2004). However, the detailed evolution and structure of these disks is poorly understood. The chemi-cal evolution of the disks determines the ability to form planets, and perhaps even life. Water ice is key, as it affects the disk's ability to form massive planets (Ciesla & Cuzzi 2006) and provides the essential solvent for life. We will discuss its importance here as well as in section 2.4 on astrochemistry.

### 2.3.1 Deriving the Physical Properties of Disks

We know that low-mass stars like our Sun form via accretion from a disk, and that the disk forms from a collapsing, rotating envelope. However, we have little detailed understanding of these processes, especially the formation of the disk. A key question thus emerges: *What are the physical properties of circumstellar disks and how do these properties change as a function of stellar mass?* SOFIA is poised to play a key role in answering this question by studying disks around low-mass young stellar objects to allow us to refine the models of accretion disks and better understand the formation of our own Solar System. Furthermore, unlike massive stars, many low-mass stars are formed in relative isolation, removing the environmental ambiguities to the application of such models.

> *The 20-100 μm wavelength range is crucial in constraining physical parameters and the evolutionary stage of the disk. This is the region (28 to 60 μm) where SOFIA provides unique continuum imaging capabilities.*

Recent studies (i.e., Robitaille et al. 2006, 2007; Crapsi et al. 2008) have shown the ability to associate a physical structure and evolutionary stage with an observed SED is much better when the SED is well sampled over a broad wavelength range. Derived physical parameters, such as envelope mass, disk mass, and disk geometry are in better agreement with values ascertained via other techniques when the SED is well sampled. It was determined in those modeling studies that the addition of more data points in the range 20-100 $\mu$m is crucial in constraining parameters and the evolutionary stage of a source. In particular, this wavelength range probes the region where the envelope feeds the disk, the least-understood aspect of the process. This is precisely the region where SOFIA provides its unique continuum capability (28 to 60 $\mu$m) that cannot be observed from the ground, with Herschel, or even with JWST. Furthermore, instruments like FORCAST and HAWC provide narrower filters (~10% bandwidth) for more accurate and denser spectral sampling than previous (Spitzer's IRAC and MIPS) or future imagers (Herschel). Using the full range of SOFIA wavelengths will therefore provide well-sampled data throughout the entire SED of these sources, which can remove many ambiguities.

SOFIA will allow us to study the more massive analogs to Sun-like stars, the Herbig Ae/Be stars (2-10$M_{sun}$), with higher spatial resolution and better spectral sampling of their SEDs than previously achieved. Many Herbig stars were too bright for Spitzer to observe without saturation. These sources provide an important link between the formation processes of low-mass stars and the high-mass stars addressed previously. Through extensive observations of Herbig Ae/Be stars one will begin to understand at what mass limit certain physical properties related to star formation change in behavior and appearance, thus helping to understand the differences between low and high-mass star formation.

Studies with ISO found very interesting variations in composition in disks around Herbig Ae/Be stars (e.g., Meeus et al. 2001). The broad spectral features reveal grain growth and crystallization, processes that tie disk observations to the growth of planets. SOFIA science instruments FORCAST, FIFI LS, HAWC, and SAFIRE could be used to make well-sampled SEDs from 5 to 655 $\mu$m for a sample of Herbig Ae/Be stars. In order to achieve ~10 $\sigma$ detection per source with these four instruments, a total observing time of 60 hours would be needed to observe a statistically significant sample of 40 sources.

### 2.3.2 The Evolution and Chemistry of the Disk

Making SED observations towards a large sample of low-mass and Herbig Ae/Be stars will allow us to compare their disk properties as a function of stellar mass. However, physical properties of the disks can also be compared as a function of age, which may lead to a better understanding of grain growth, planet formation, chemical evolution of the disk, and disk clearing.

Planets are expected to form within the inner 20 AU radius of protostellar disks (Boss 2008) where emission is mostly between 10 and 60 $\mu$m. Again SOFIA will uniquely probe this critical region of the disk and will address one of the major questions in the study of star formation: *How are the building blocks of planets formed in these disks?* Grain growth in the circumstellar disks is the first step toward planet formation. In the disk mid-plane, micron-size grains can grow very quickly into kilometer size structures. Multi-wavelength SED observations can probe this growth, with several studies of YSOs showing evidence of grain growth in their disks (e.g., Natta et al. 2004, Bouwman et al. 2008). Associating the grain growth with other changes and with stellar ages will provide input to models of planet formation.

High spectral resolution observations with EXES and GREAT on SOFIA will probe disk kinematics and composition to address the question: *What is the chemical state of the disk?* As we will discuss in more detail in the astrochemistry section, SOFIA can follow the water trail from clouds to the planet-forming zones of disks. It is generally believed that water plays an essential role in the formation and early evolution of planetary systems, since water ice condensation will dominate the mass budget of newly-formed planetesimals. The 45 $\mu$m water ice band was observed towards several types of objects with ISO (e.g., van den Ancker et al. 2000). SOFIA, with an even higher spatial resolution compared to ISO, will be the only observatory that can now observe this transition. Also available to SOFIA is the 6 $\mu$m water feature, which cannot be observed from the ground (see sidebar in Figure 2-7 on page 17). Observing the strengths of these transitions towards circumstellar disks will provide direct measurements of temperature and water column density.

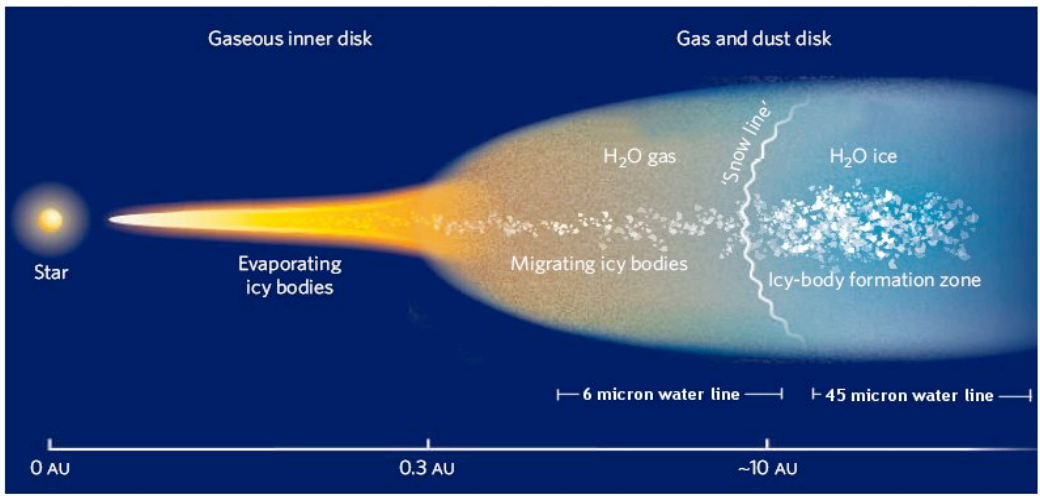

***Figure 2-4.*** *A schematic diagram showing the location of water in a circumstellar disk. NIR instruments can probe the steam water features within 0.3AU of stars from the ground. SOFIA can probe the gas and snow line by observing the 6 µm feature, as well as the colder water ice found in the outer disk through the 45 µm feature. SOFIA can also probe ice grain growth in star forming regions through these transitions to learn how planets form in their earliest stages. (Figure modified from van Boekel 2007)*

> *SOFIA can uniquely measure important spectral lines within circumstellar disks with extremely high spectral resolution. Velocity-resolved line profiles will probe the chemistry in circumstellar disks.*

Water condensing into ice beyond the "snow line" (Figure 2-4) in disks plays a crucial role in the formation of the cores of giant planets. This outer region of the disk can be probed by the 45 $\mu$m water ice vibration band with SOFIA. This spectral feature is the only direct probe of the amount of ices in disks that does not depend on a special viewing geometry. The origin and distribution of water in the inner proto-planetary disk is crucial to our understanding of the abundance of water on terrestrial planets in the habitable zones around stars. SOFIA can probe from the "snow-line" inward using EXES observations of the gas-phase $H_2O$ 6 $\mu$m vibration-rotation lines (Figure 2-4), especially in disks for which the grains have already grown to sizes of a few microns. Indeed, the 6 $\mu$m feature can also be used to study grain growth related to planet formation in the disk (Meijerink et al., 2008). Spitzer detections of gas phase $H_2O$ lines around 16 $\mu$m, combined with linewidths less than 35 km/s from near-infrared lines (Carr & Najita 2008, Salyk et al. 2008), suggest peak emission line intensities of ~1.0x $10^{-17}$ W/m² if velocity resolved. EXES can detect lines this bright in 900 sec (10 $\sigma$) with 8 km/s resolution, yielding important information about the chemical state of these very active regions of proto-planetary disks.

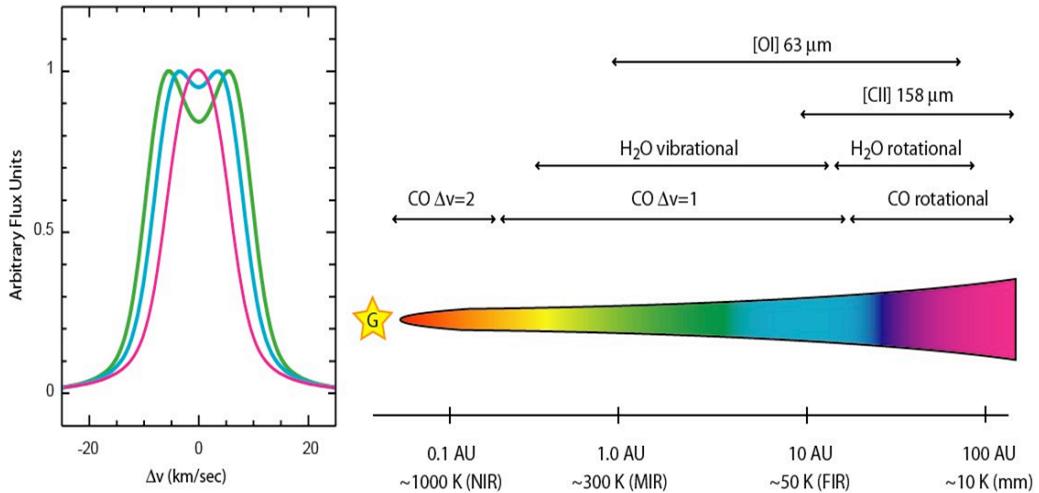

***Figure 2-5.*** *The plot on the left shows models of velocity-resolved lines in a disk, as would be observed with 4 km/s spectral resolution. The resultant double-peaked shape of the line is because it is the combination of the red and blue Doppler shifted line profiles (with respect to the systemic velocity of the star) coming from the emitting material on either side of the disk. From the separation of the peaks, one can distinguish the location within the disk of the chemical being probed. The colors of the lines on the left translate to the regions of the same colors in the simplified cartoon disk on the right. Determining the locations of molecules and atoms in disks could be done for the lines of several indicators, like [OI] at 63 μm, [CII] at 158 μm, and H2O to get a chemical inventory of the disk as a function of location. The estimated locations indicated are based in part on the chemical models of Kamp & Dullemond (2004).*

While SOFIA will not spatially resolve the region of the snow line, velocity-resolved line profiles in combination with Kepler's law can yield the distribution of chemicals in the emitting layers of circumstellar disks (Figure 2-5). For example, SOFIA is the only observatory that will have the capability to observe the 63 $\mu$m [OI] line and the 119 $\mu$m OH ground state line with high velocity resolution and hence will be able to pin down the location where these specific spectral features are produced (Figure 2-5). The location of other important emission features such as $H_2O$ and [CII] at 158 $\mu$m, and [OI] at 145 $\mu$m (with an upgrade to the GREAT instrument) can also be determined. Furthermore, there are some features which can only be measured by SOFIA: for example, the pyroxene feature at 40.5 $\mu$m which can be used to trace the iron content of protostellar disks. The state of iron is a very important tool for deriving the disk opacities and temperature. Disk models predict strong [OI] lines at 63 $\mu$m that can be used to probe the warm (T~200 K), dense (n~$10^6$ cm$^{-3}$) layers of the disk. ISO has detected the line from some Herbig Ae and T-Tauri disks with fluxes as high as $10^{-15}$ W/m$^2$, but it is crucial to separate the [OI] emission from the disk from that arising in the surrounding molecular cloud. SOFIA can map and, even more importantly, velocity-resolve the spectra to kinematically distinguish line emission from the disk from that of the surrounding cloud.

Another important topic is the lifetime of gas in protoplanetary disks, which is a vital constraint on models of planet formation. The frequency of giant planets may be determined by the timing of dissipation of gas, evaporation of ices, and removal of dust. If grain growth renders the disk optically thin in the dust continuum emission while a substantial reservoir of gas remains in the disk, the gas will affect the interactions and evolution of protoplanets and planetesimals. Model calculations show that the [OI] 63 $\mu$m line may be one of the most robust tracers of the presence of any residual gas in disks. SOFIA should be able to detect this line at least down to a total gas mass of $10^{-4}\,M_{sun}$ (0.1$M_{Jup}$) based on X-ray irradiated disk models. In contrast with Herschel, the SOFIA searches will not be limited by the line/continuum ratio because of the much higher spectral resolution of SOFIA.

A future MIR/FIR spectropolarimeter on SOFIA would also help in further understanding disk chemistry. Spectropolarimetry can yield a more accurate and complete diagnosis of the composition of the material on the disk surface. For instance, measurements of the flatness of 10 $\mu$m silicate feature can be due to the presence of larger grains (>2 $\mu$m, van Boekel et al. 2005) or due to small porous grains of size $\leq$ 0.3 $\mu$m (Voshchinnikov et al. 2006). The polarization signatures of these two types of particles can be distinguished by spectropolarimetry. While the 10 $\mu$m silicate feature can be studied in detail from the ground with present spectropolarimeters (Michelle on Gemini, and soon CanariCam on the Gran Telescopio Canarias), spectropolarimetric observations of the relatively unexplored 20 $\mu$m silicate feature with SOFIA would be unique and aid in the understanding of our incomplete picture of the chemistry of circumstellar disks.

### 2.4 Astrochemistry in Star Forming Regions

The study of exo-planetary systems and their formation is one of the fastest growing topics in astronomy, with far-reaching implications for understanding our place in the universe (see Chapter 5, Planetary Science). Intimately related to this topic is the study of the chemical composition of the gaseous and solid-state material out of which new planets form and how it is modified in the dense protostellar and protoplanetary environments. SOFIA, in concert with ALMA, Herschel, and JWST, has an important role in tracing our chemical origins.

### 2.4.1 The Oxygen Deficit

For the three most crucial elements to life as we know it — carbon, nitrogen, and oxygen — we have a good understanding of the location only for the carbon (about 2/3 in dust grains and 1/3 in CO in molecular clouds). Nitrogen is less well understood because it lacks key spectroscopic transitions. Oxygen, however, stands out as a puzzle that SOFIA can uniquely address.

> *SOFIA is the only mission that can provide spectrally resolved data on the [OI] lines at 63 and 145 µm to understand the oxygen deficit in star-forming clouds and disks.*

A key question for SOFIA is *Where is the oxygen in star-forming clouds and disks?* The main reservoirs of oxygen — O, OH, and $H_2O$ — all have their principal transitions at mid-infrared and far-infrared wavelengths. As stated in the disk section, SOFIA is the only mission that can provide spectrally resolved data on the [OI] lines at 63 and 145 $\mu$m, and the OH ground-state line at 119 $\mu$m, as well as probe H2O and OH in spectral windows that Herschel does not cover to answer this question.

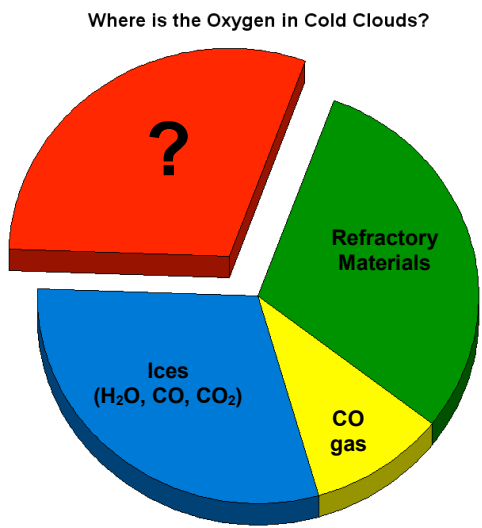

**Figure 2-6.** *A pie chart showing the oxygen budget in cold clouds. Almost 1/3 of the oxygen is unaccounted for.*

In fact, the total oxygen budget in clouds is still a puzzle. In diffuse clouds, optical and UV absorption line data have established that about 30% of the total (gas plus solid) elemental O abundance of $N(O)/N(H) = 4.6 \times 10^{-4}$ is contained in refractory material such as silicates, with the remainder in gaseous atomic oxygen (Meyer et al. 1998). Assuming that the refractory budget stays the same in dense clouds, one can make an inventory of the remaining volatile components. Ices (mostly $H_2O$, plus some CO and $CO_2$) contain about 25-30% and gaseous CO up to 10% of the oxygen, with precise values varying from cloud to cloud (e.g., van der Tak 2005). That leaves a significant fraction of the budget, about 1/3, unaccounted for (Figure 2-6). SWAS and ODIN have shown that the $O_2$ abundance in dense clouds is surprisingly low, $N(O_2)/N(H_2) < 10^{-7}$, as is gaseous $H_2O$ in the cold gas (Goldsmith et al. 2000). Thus, gaseous atomic O is the most likely reservoir, but this would be in conflict with models to explain the low gaseous $O_2$ and $H_2O$ abundances, which require most of the oxygen to be frozen out.

Limited ISO-LWS data on [OI] 63 and 145 $\mu$m emission or absorption from cold clouds exist but are difficult to interpret because of the high optical depth of the lines coupled with the fact that they are spectrally unresolved (Li et al. 2002). SOFIA is the only planned facility that will have heterodyne capabilities at these wavelengths to properly resolve the lines and derive oxygen abundances. Assuming a linewidth of 1 km/s, typical of cold clouds, the detections at 63 and 145 $\mu$m with ISO would correspond to line fluxes of $3.3\times10^{-16}$ Wm$^{-2}$ and $1.3\times10^{-16}$ Wm$^{-2}$, respectively. GREAT could observe both lines for 20 dark cloud sources in just under 10 hours.

### 2.4.2 Following the Water

Water is obviously one of the most important molecules to study in protostellar and protoplanetary environments, because it is a significant reservoir of oxygen. Like CO, it thus controls the chemistry of many other species. Water also plays an important role in the energy balance as a strong gas coolant, regulating the temperature in dense shocks and allowing clouds to collapse to higher temperatures (Doty & Neufeld 1997). In cold clouds, water is the most abundant molecule in icy mantles, and as we have discussed in the section on disks, its presence may help the coagulation process that ultimately produces planets. Asteroids and comets containing ice have likely delivered most of the water to our oceans on Earth, where water is directly associated with the emergence of life. Most of this water is actually formed in the clouds from which stars and planets form. Thus, the distribution of water vapor and ice during the entire star and planet formation process, and not just in disks, is a fundamental problem relevant to our own origins.

***Water in Absorption.*** Interpretation of lines in absorption is more straightforward than that for emission lines. If the observations can measure true continuum and zero level, the optical depth of a weakly excited absorption line is easy to determine. Assuming that the absorption seen is confined to a column defined by the angular size of the background source, it is possible to compare with other observations, even if made using other beam sizes, with a reasonable accuracy. EXES can provide a view of water that is complementary to that offered by Herschel. Herschel has several programs to detect water in absorption from extended cold gas components (and follow-up can be anticipated using CASIMIR). However, EXES will be sensitive to the warmer, more abundant water that exists in close proximity to deeply embedded young stars (see sidebar in Figure 2-7 on page 17). In this regard, the capabilities of EXES are a significant

> *SOFIA has the unique ability to spectrally resolve water vapor lines in the mid-IR which will allow it to probe and quantify the creation of water various star forming environments.*

improvement when compared to ISO detections of water vapor in absorption at 6 $\mu$m. With EXES one can observe essentially any rotational level of the main isotope, including the ground rotational state. A single wavelength setting of EXES will provide information on the water abundance and gas temperature in warm gas local to young massive stars. The spectral resolution offered by EXES is also significantly better than the more sensitive JWST/MIRI and therefore probes a different range of parameter space. This will prove to be a useful counterpoint to the studies of water in emission with Herschel/HIFI, providing a powerful tool to examine the line of sight water abundance structure in several key sources (e.g., Boonman et al. 2003). For example, bright massive protostars can be observed in the 6 $\mu$m $H_2O$ lines with EXES with reasonable exposure times (signal-to-noise of 50 in 15 minutes for a $2.0 \times 10^{-17}$ W/m$^2$ absorption line), allowing surveys with decent statistics to be performed.

***Water in Shocks.*** Also related to the production of water in the interstellar medium near protostars are shocks. It is thought that shocks are "water factories" because the high temperatures of the shocks (up to a few 1000 K) can allow reactions with energy barriers to proceed. The most important is the reaction of $O + H_2 \rightarrow OH + H$, followed by $OH + H_2 \rightarrow H_2O + H$, which should drive most of the atomic oxygen into water. Despite the detection of water in numerous shocks (Neufeld et al. 2000; Nisini 2003), there are lingering questions regarding the true water vapor abundance inside shocks and the spatial distribution when compared to other shock tracers, such as vibrationally excited $H_2$. This is primarily due to the low spatial resolution of instruments capable of observing cold post-shock water (e.g., SWAS and Odin) and the low spectral resolution of instruments capable of detecting hot shocked water (e.g., ISO and Spitzer).

SOFIA offers new observational insight into the unresolved question: *How much water is created in interstellar shocks?* Herschel has a strong effort directed towards the study of water in shocks using both a heterodyne (HIFI) instrument and direct detection (PACS). However, as has been stated, the EXES instrument offers the ability to spectrally resolve water emission lines, and this is not available with Herschel or JWST. Studies of SWAS emission suggest a sharp (2 orders of magnitude) increase in water abundance as a function of velocity in the line wings (Franklin et al. 2008).

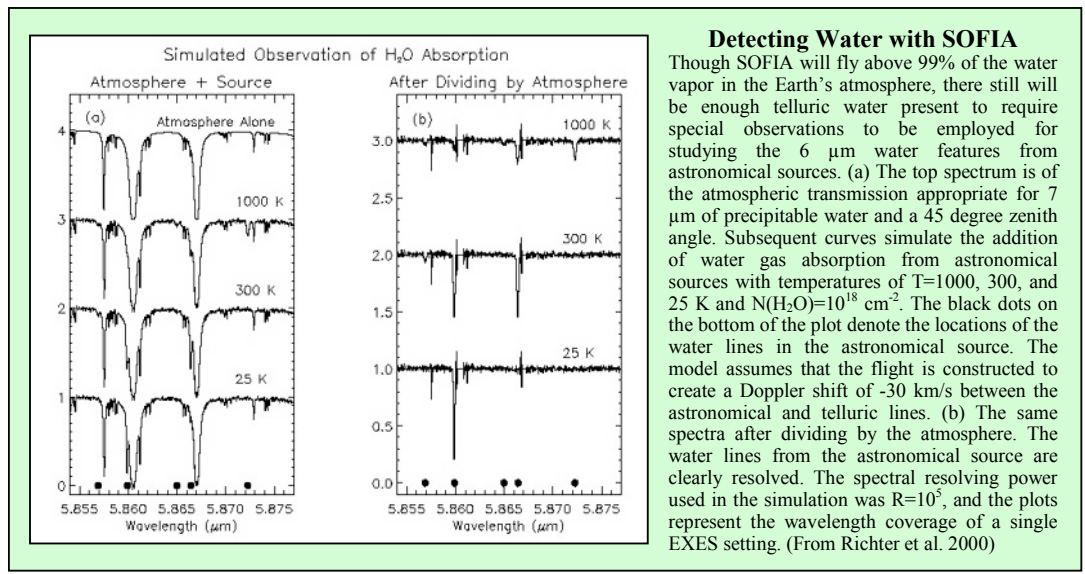

**Detecting Water with SOFIA**
Though SOFIA will fly above 99% of the water vapor in the Earth's atmosphere, there still will be enough telluric water present to require special observations to be employed for studying the 6 μm water features from astronomical sources. (a) The top spectrum is of the atmospheric transmission appropriate for 7 μm of precipitable water and a 45 degree zenith angle. Subsequent curves simulate the addition of water gas absorption from astronomical sources with temperatures of T=1000, 300, and 25 K and $N(H_2O)=10^{18}$ cm$^{-2}$. The black dots on the bottom of the plot denote the locations of the water lines in the astronomical source. The model assumes that the flight is constructed to create a Doppler shift of -30 km/s between the astronomical and telluric lines. (b) The same spectra after dividing by the atmosphere. The water lines from the astronomical source are clearly resolved. The spectral resolving power used in the simulation was $R=10^5$, and the plots represent the wavelength coverage of a single EXES setting. (From Richter et al. 2000)

*Figure 2-7. Detecting water with SOFIA — simulated observation of water absorption.*

EXES will have the capability to distinguish between different water emitting clumps along the line of sight. EXES also has the ability to detect the primary gas constituent, $H_2$. Any study of water chemistry requires knowledge of the gas temperature, density, and total column depth. The gas temperature/density can be estimated via observation of many water transitions. However, the total column requires observation of either $H_2$ or CO. With the EXES instrument one can spectrally resolve the emission of molecular hydrogen in shocks and provide a direct measurement of the total gas column in the shock as a function of velocity. In the sub-mm, SOFIA can also observe the optically thin $H_2^{18}O$ lines which are shifted from the telluric lines and which are also helpful for deriving water abundances. Given the limited lifetime of Herschel, it can do only a small fraction of young stellar objects. With continued developments to improve receiver noise at THz frequencies, SOFIA can be highly effective in this area. For example, the lowest 994 and 745 GHz lines of p-$H_2^{18}O$ are particularly well suited to constrain both the abundance and excitation and predicted line brightnesses are about $2.7 \times 10^{-17}$ W/m$^2$.

**Water Crystallinity.** Near-infrared spectroscopy has shown that $H_2O$ ice on dust grains in molecular clouds are typically amorphous (Smith et al. 1989). On the other hand, $H_2O$ ices in the comets are crystalline (Grun et al. 2001), although the amorphous phase was reported in comet Hale-Bopp (Davies et al. 1997). Some-how, in the process of stellar formation from the molecular cloud to a planet-bearing disk, this change occurs. This change from amorphous to crystalline state must happen as a part of the star formation process as the dust grains are carried along from molecular cloud to planet-bearing disk, but some star forming regions show crystalline ice while others do not (Boogert et al. 2000). The polarization signature of material is quite different for crystalline versus amorphous material, and consequently a future SOFIA-based MIR spectropolarimeter operating over the 6 $\mu$m and 12 $\mu$m features studying objects in various stages of the star formation process will be valuable to trace this processing of the water during stellar birth.

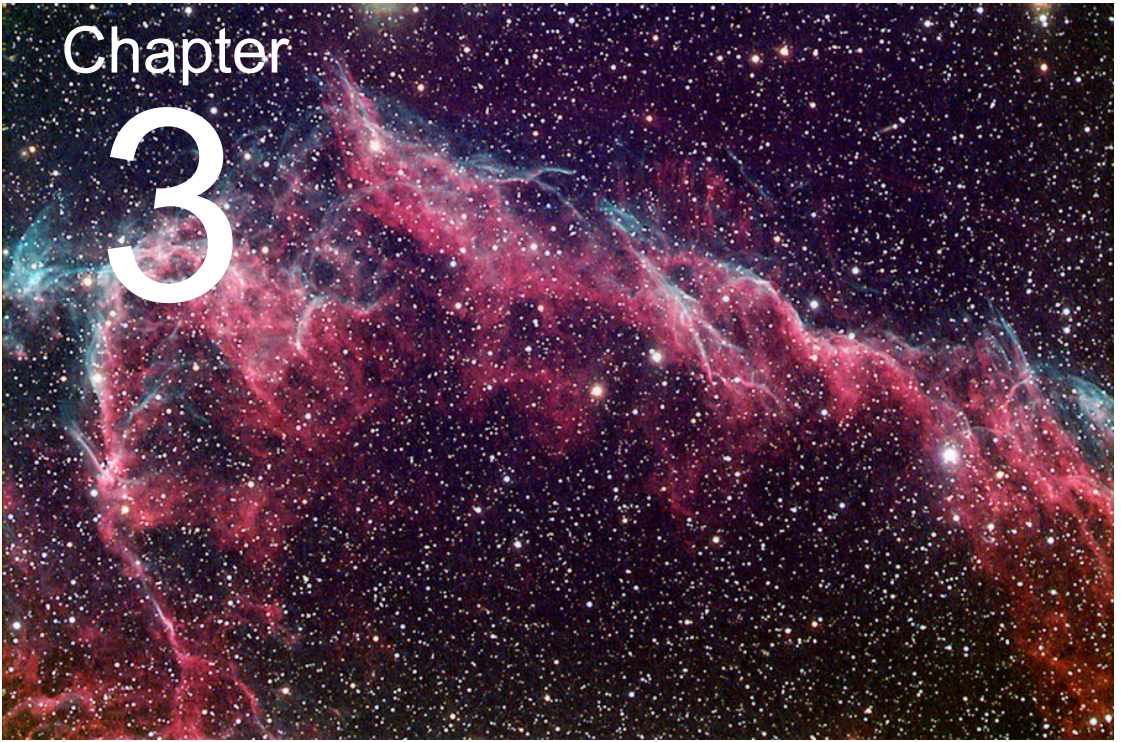

# Chapter 3

# The Interstellar Medium of the Milky Way

### 3.1 Overview

The interstellar medium (ISM) plays a central role in the evolution of galaxies as the birthsite of new stars and the repository of old stellar ejecta (Figure 3-1). The formation of new stars slowly consumes the ISM, locking it up for millions to billions of years in the form of stars. As these stars age, the winds from low-mass, asymptotic giant branch (AGB) stars and high mass, red supergiants (RSGs), and supernova explosions inject nucleosynthetic products of stellar interiors into the ISM, slowly increasing its metallicity. This constant recycling and associated enrichment drives the evolution of a galaxy's baryonic matter and changes its emission characteristics. The physical conditions and metallicity of the ISM are determined in large part by stellar feedback. SOFIA investigations are a key to understanding the ISM and the stellar feedback that controls it.

The spectral energy distribution (SED) in Figure 3-2 shows the main components that make up a galaxy such as the Large Magellanic Cloud. SOFIA will be the only observatory in the next decade that is sensitive to the entire far-infrared (far-IR) bump in the SED, which is dominated by emission from the ISM. The dust features from the smallest, molecular sized dust grains, polycyclic aromatic hydrocarbons (PAHs), are prominent in the mid-IR. The dust continuum from large dust grains dominates the far-IR and submillimeter wavelengths. The main cooling lines of the neutral ISM are spread across the far-IR peak. SOFIA's rapid mapping capability and excellent spectroscopic instrument complement are well suited for spectral imaging of bright sources and extended regions of ISM in the Milky Way and nearby galaxies. SOFIA's high resolution spectroscopy can resolve the narrow features of ISM dust and kinematics of the ISM gas. Moreover, SOFIA's new instrumentation program can address new scientific questions about the ISM.

In this section, we focus on SOFIA's capability to address several key questions about the ISM in galaxies, which will be a significant, broad and vigorous area of research with SOFIA.

- What are the physical processes that regulate the interaction of massive stars and their environment? In particular, what is the interrelationship of massive stars, HII regions, and their associated Photodissociation Regions (PDRs)?
- What are the origins and characteristics of dust in the Milky Way and other galaxies? In particular, how do the dust characteristics depend on the types of stars that make up a galaxy?
- What is the role of large and complex molecules, such as polycyclic aromatic hydrocarbons (PAHs) in the ISM? In particular, is there a dominant species of PAHs that can be uniquely identified?

- What is the rate of deuterium depletion in the ISM caused by the cycling of ISM into stars? In particular, what is the abundance of HD in the ISM and how does it vary with local conditions and chemistry?

SOFIA investigations of the ISM physics, chemistry and evolution in the Milky Way and nearby galaxies will provide critical understanding necessary to interpret the galaxy evolution studies of the high redshift Universe that will be investigated by JWST and ALMA.

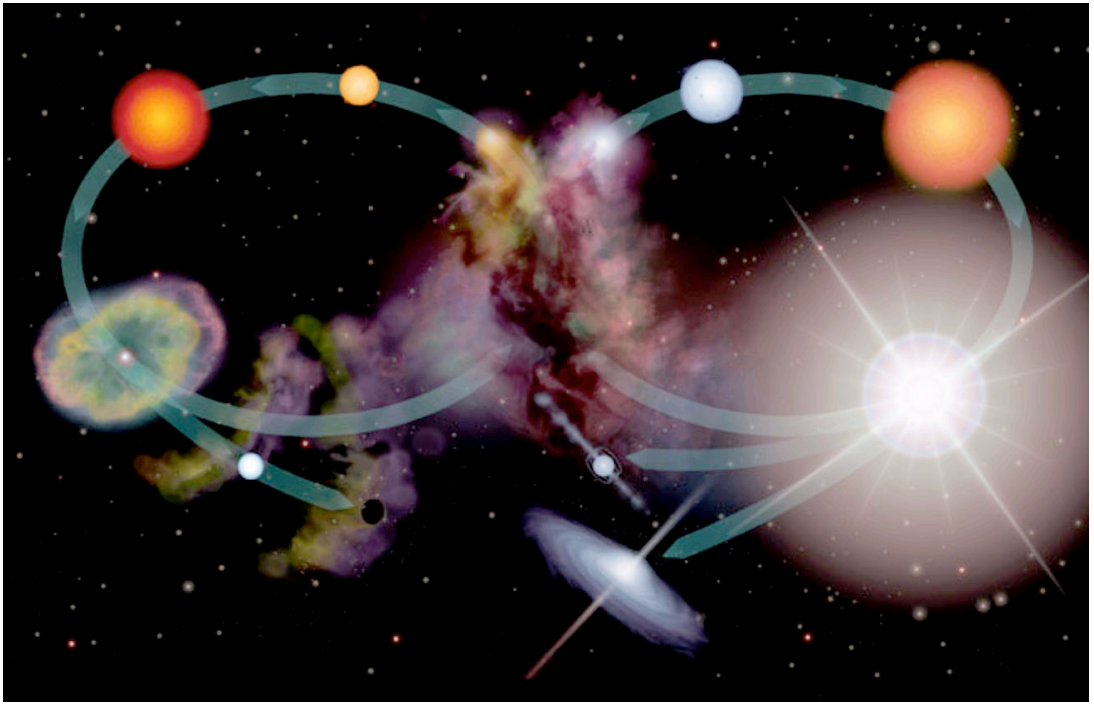

**Figure 3-1.** *Artist's conception that captures the lifecycle of stars and the ISM. On the left is the cycle for a solar mass star that ends as a white dwarf. On the right is the cycle of a massive star that forms a supernova and ends as a neutron star or black hole. Image courtesy: http://hea-www.cfa.harvard.edu/CHAMP/EDUCATION/PUBLIC/ICONS/*

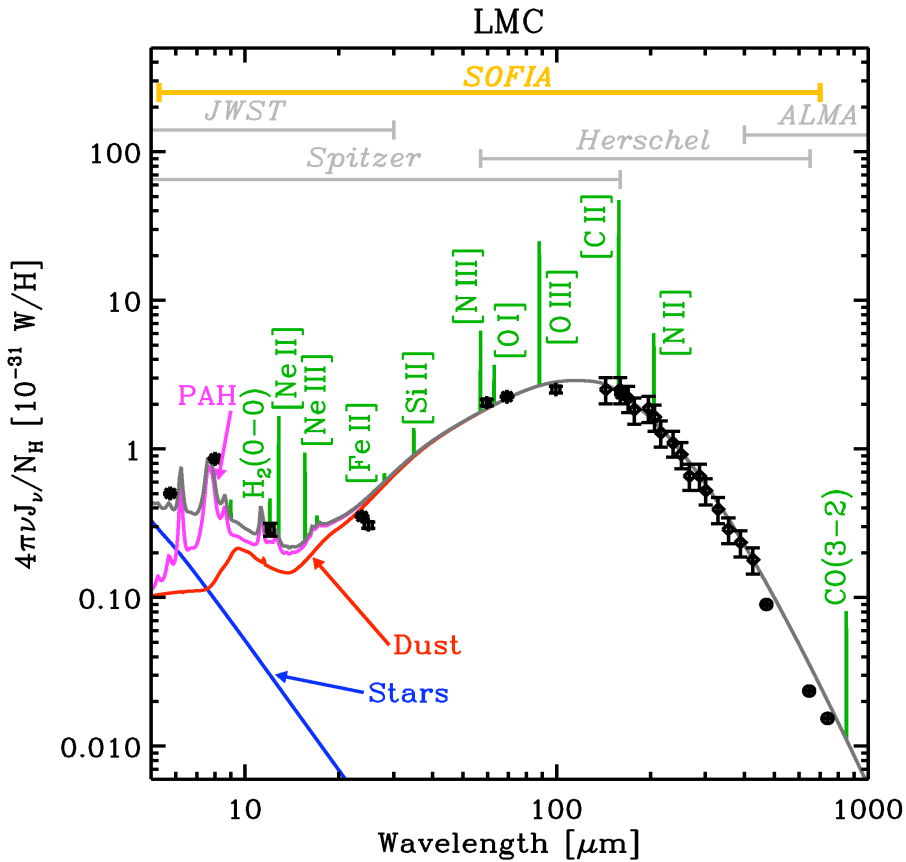

***Figure 3-2.*** *The spectral energy distribution of the entire LMC, based on data from Spitzer, IRAS and FIRAS (Bernard et al. 2008). SEDS are fitted with the dusty PDR model of Galliano et al. (2008). Spitzer has and Herschel will provide good photometric coverage of a galaxy's spectral energy distribution (SED) over a portion of the wavelengths. SOFIA will provide excellent wavelength coverage and spectroscopic capability across the entire SED. In the future, JWST and ALMA will provide complementary wavelength coverage and work on nearby galaxies and the most distant Universe. Figure courtesy of Galliano.*

### 3.2 Massive Stars and the ISM: Photodissociation Regions (PDRs)

The intense radiation from massive stars dominates the energetics and chemistry of its surrounding molecular clouds creating HII and photodissociation regions (PDRs). By observing these regions, we can understand how massive stars interact with their environment and, in particular, address the following questions:

- What is the interrelationship of massive stars, HII regions, and their associated PDRs?

- How do the infrared spectroscopic signatures of PDRs depend on the physical conditions, including density, stellar radiation field, and metallicity?

- What does this tell us about the physical conditions in regions of massive star formation and the star formation process?

Photodissociation regions (PDRs) are very complex to model because they link chemistry, gas and dust energetics, and radiative physics (e.g., Wolfire et al. 1990; Kaufman et al. 1999; Hollenbach and Tielens 1999). For that reason, they require a number of multi-wavelength observations to fully understand the physical processes and to constrain the models properly. Because PDRs are in dynamic and thermal equilibrium, the heating of the gas caused by photoelectric heating from dust grains is balanced by cooling from line emission that is dominated by the far-infrared atomic fine structure lines of [OI] 63 and 146 $\mu$m and [CII] 158 $\mu$m.

Figure 3-3 illustrates theoretical expectations for PDRs of differing densities and incident radiation fields. The far infrared intensity ($I_{FIR}$) of PDRs results from thermal radiation of large dust grains in radiative equilibrium with the incident $F_{UV}$ stellar radiation field, $G_0$, and can be used to measure it. The ratio of the dominant cooling lines of [OI] 63 $\mu$m and [CII] 158 $\mu$m to the far infrared intensity represents the photo-electric heating efficiency, and is shown on the x-axis of Figure 3-3.

The ratio of the [OI] and [CII] cooling lines provides a diagnostic of the gas density because of their different critical densities, and is shown on the y-axis of Figure 3-3. In addition to these cooling lines, the pure rotational lines of $H_2$, CO and other simple molecular species, provide constraints on the temperature and density of the molecular gas in PDRs.

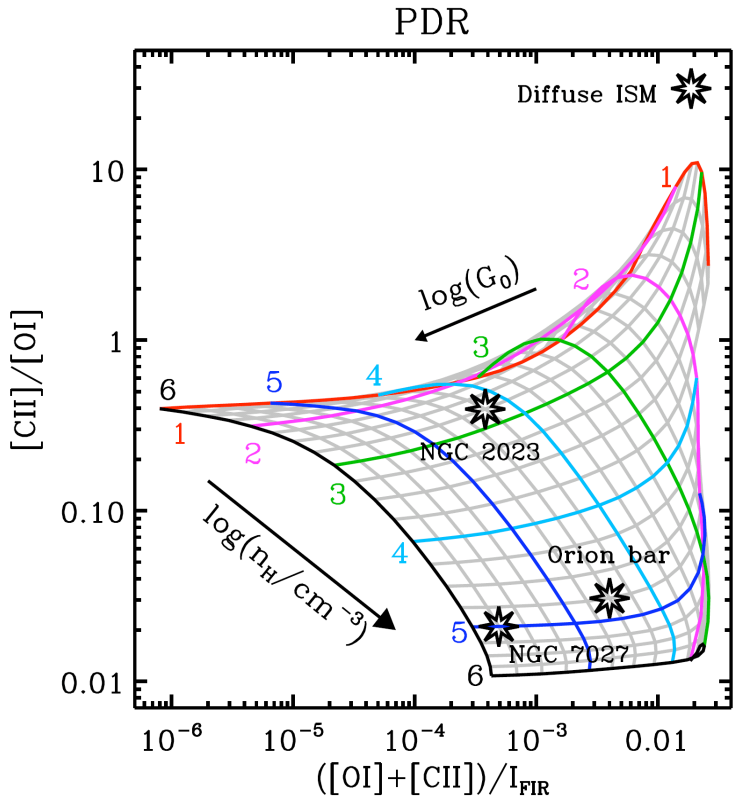

*Figure 3-3.* The wide range of gas density, $n_H$, and incident UV radiation, $G_0$ (in units of local interstellar radiation field strength), of PDRs can be effectively probed using measurements of the dominant cooling lines, [CII] 158 $\mu$m and [OI] 63 $\mu$m, and a measurement of the total far-infrared intensity, $I_{FIR}$. The stars mark the results from some well observed PDRs. Figure is adapted from Kaufman et al. (1999).

The photoelectric heating efficiency in PDRs increases for smaller dust grains. PAH molecules, the smallest of dust grains, are considered very important in the heating of PDRs (Bakes & Tielens 1994). Indeed, the mid-IR spectra of PDRs in galactic and extragalactic sources are dominated by emission features at 3.3, 6.2, 7.7, and 11.2 $\mu$m due to PAHs perched on broad emission bands as well as a steeply rising continuum. If properly calibrated, the PAH features, which are also observable in high red-shift galaxies, could be used to measure the physical conditions of PDR gas and star formation regions in the distant Universe where the cooling lines may be difficult to measure (Figure 3-4).

The relative strength and characteristics of PAH features vary from source to source and within sources reflecting local physical conditions (Hony et al. 2001; Peeters et al. 2002; Rapacioli et al. 2005; Flagey et al. 2006; Galliano et al. 2008;

Smith et al. 2007). For example, Figure 3-4 compares the strength of CH stretching and bending modes (11.3 $\mu$m) relative to the CC stretching modes (6.2 and 7.7 $\mu$m) for various regions in the nuclear starburst galaxy M82. Extensive laboratory and experimental studies have revealed that these PAH feature ratios measure the relative importance of ionized to neutral species in the emitting PAH population. The observed variations reflect the variations in the PAH ionization balance due to variations in the local density, temperature and/or radiation field. For a few well-studied regions, these variations in the PAH spectral characteristics have been linked to variations in the physical conditions (Figure 3-4; Galliano et al. 2008). Thus it appears promising to calibrate the PAH features and link their ratios to PDR physical conditions. In order to calibrate the PAH-PDR relation shown in Figure 3-4, it is important to measure the PAH emission, dominant FIR cooling lines, and molecular gas line emissions for a significant sample of PDRs that cover the full range of physical conditions in the Milky Way ISM.

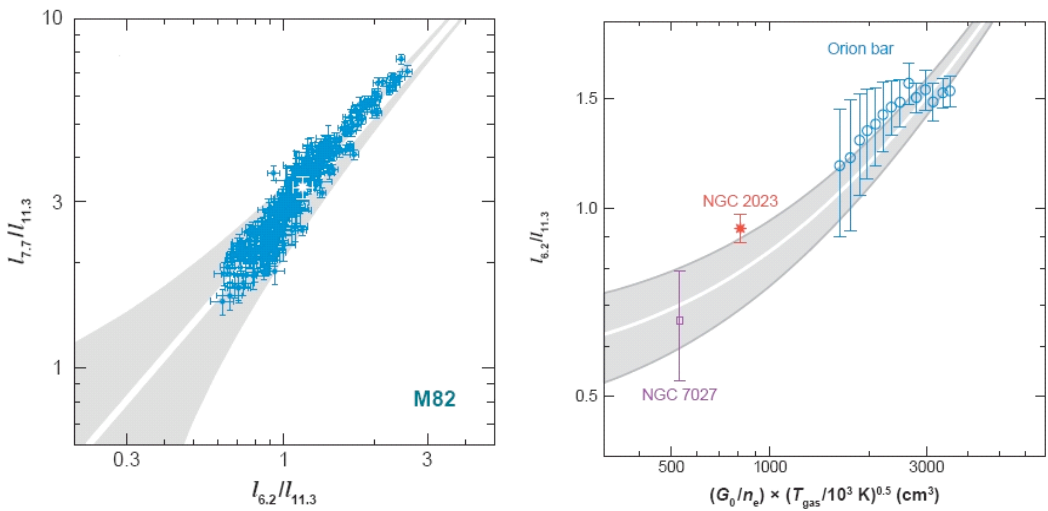

*Figure 3-4.* Left panel: Variations in the relative strength of the 6.2, 7.7, and 11.3 $\mu$m PAH bands observed in the starburst galaxy, M82 (Galliano et al., 2008). As these data illustrate, the 6.2 and 7.7 $\mu$m bands – due to PAH cations – vary relative to the 11.3 $\mu$m band – due to neutral PAHs – by a factor of 4 in this data set. Similar variations are observed in many other sources. Right panel: The observed ratio of the 6.2 to 11.3 $\mu$m band – a measure of the ionization balance of PAHs is related to the physical conditions in a few well-studied PDRs through the ionization parameter, $G_o T^{1/2}/n_e$, a measure of the ionization rate over the recombination rate (Galliano et al. 2008).

SOFIA is optimally suited to studying the diagnostic relations of PDRs because it can observe all of the important infrared spectral signatures of PDRs. The main cooling lines of [CII] 158 $\mu$m, [OI] 63 and 146 $\mu$m can be spectrally mapped with FIFI LS and can be spectroscopically resolved by GREAT high-res spectroscopy to investigate the kinematics of the PDR gas. The far infrared (FIR) intensity can be

*SOFIA observations of PDRs will reveal how massive stars interact and shape their surrounding environments from which the next generation of stars are born. The program will also result in well calibrated diagnostics that can be used for studying the high red-shift Universe.*

mapped at high angular resolution with HAWC. PAHs, which are main sources of photo-electric heating, can be spectrally mapped by FORCAST. One can then relate all of these aspects for carefully selected regions that probe the range of density, temperature, metallicity, and UV fields. A sample of PDRs will be selected to test the full range of densities ($10^2 < n < 10^6$ cm$^{-3}$), temperatures ($100 < T < 1000$ K) and radiation fields ($10 < G_0 < 10^6$) outlined in Figure 3-3, and, in addition, investigate the dependencies on metallicity and stellar populations. This sample will include giant HII region complexes powered by clusters of stars as well as reflection nebulae around cooler stars.

SOFIA will provide a major step forward in comprehensive observations of Milky Way PDRs because of its wide spectroscopic instrument complement and facile mapping capability. The program outlined in the ISM observing plan table in Appendix B demonstrates the feasibility of this SOFIA study of PDRs. The results of this program will reveal how massive stars interact and shape their surrounding environments from which the next generation of stars are born. The program will also result in well calibrated diagnostics that can be used for studying the high red-shift Universe.

## 3.3 The Diversity and Origins of Dust in the ISM: Evolved Star Contributions

Spitzer has revealed that dust characteristics can change with different extragalactic environments (Marwick-Kemper et al. 2007; Armus et al. 2007; Spoon et al. 2006). The stardust enrichment budget may drive these differences. However, the links between the composition and quantity of dust injected by the evolved star populations will require a comprehensive study of Milky Way sources with SOFIA. Such a study will establish the framework necessary to interpret the infrared spectra of more distant galaxies by Spitzer and JWST.

Dust is an effective tracer of the stellar mass injection into the ISM. As the stellar ejecta cool upon expansion, small dust grains condense and then absorb the stellar radiation, "down-converting" it to infrared emission. The observed IR spectrum of such objects becomes thereby a measure of the mass returned to the interstellar medium. Quantifying this relationship between stellar mass-loss rate and the spectral energy distribution of the ejecta requires, however, knowledge of the properties of the dust since much of the IR emission occurs in strong vibrational modes characteristic for the emitting material.

In order to link the observed spectral signatures of dust to their sources and their contribution to the lifecycle of the ISM, we will need to address a number of questions:

- What are the characteristics of dust injected by different stellar sources?
- What is the contribution of low-mass versus massive stars to the ISM budget?
- What does this imply for the lifecycle of dust and gas in galaxies?

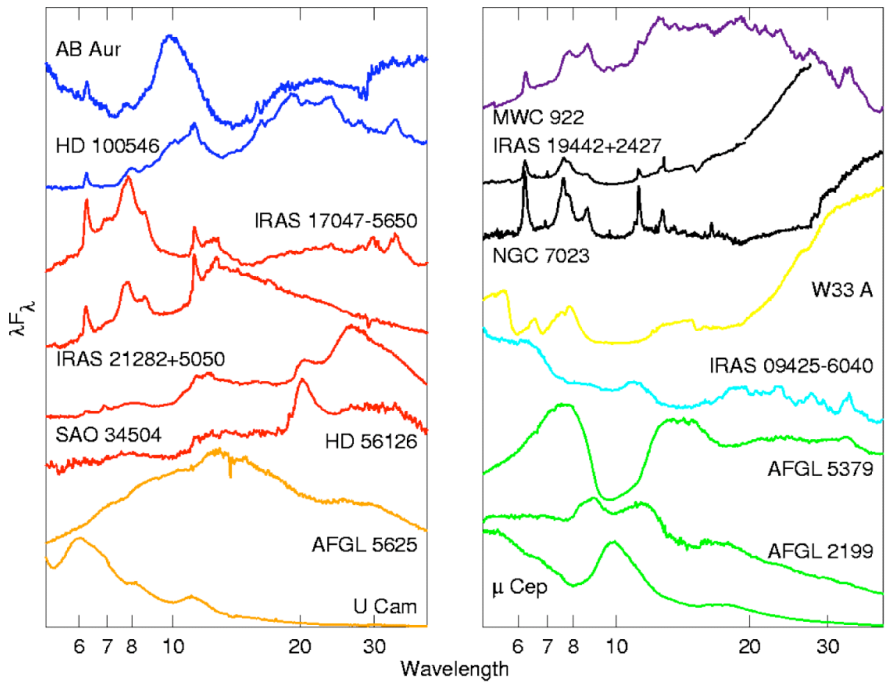

***Figure 3-5.*** *The rich and spectrally diverse stardust revealed by ISO and Spitzer requires systematic study. A selection of ISO SWS spectra of a variety of objects shown here illustrates the rich spectral diversity of the dusty Universe. Key: dark blue: Herbig AeBe; red: post-AGB and PNe; orange: C-rich AGB; green: O-rich AGB; light blue: mixed chemistry AGB; yellow: deeply embedded YSO; black: HII region/reflection nebula; purple: mixed chemistry post-AGB.*

Stardust is formed in the ejecta of old, dying stars — including low-mass (<8M$_\odot$), Solar-type descendants such as asymptotic-giant-branch (AGB) stars and post asymptotic-giant-branch (post-AGB) stars and novae. Massive stars (>8M$_\odot$), such as red supergiants, luminous blue variables, Wolf Rayet stars, and supernovae also form dust in their ejecta. Figure 3-5 illustrates the spectral richness of interstellar dust in the 3 $\mu$m to 40 $\mu$m range and demonstrates the compositional variations between different sources. The IR spectra of oxygen-rich AGB and post-AGB stars with high mass-loss rates are dominated by amorphous and crystalline silicates (e.g., mainly forsterite (Mg$_2$SiO$_4$) and some enstatite (MgSiO$_3$)) while those with low mass-loss rates show predominantly oxides (e.g., corundum (Al$_2$O$_3$), wuerstite (MgO), spinel (MgAl$_2$O$_4$)) (Sylvester et al. 1999, Molster et al., 2002a; Cami 2001). Carbon-rich AGB and post-AGB objects show predominantly silicon carbide (SiC), hydrogenated amorphous carbon, magnesium sulfide (MgS) and polycyclic aromatic hydrocarbon species (Hony et al. 2002; Peeters et al. 2002). The far-IR is less well studied but additional features occur at 45 and 60 $\mu$m due to $H_2O$ ice and a key forsterite band at 69 $\mu$m (Sylvester et al. 1999; Molster et al. 2002b). It is clear that the mid-infrared region carries great diagnostic potential for the determination of dust composition and mass-loss rates from observed spectral energy distributions. Hence, mid-IR spectra for Milky Way dusty stars which cover the full range, shown in Figure 3-6, are necessary for a proper assessment of the contribution of different classes of stellar objects to the mass budget of galaxies.

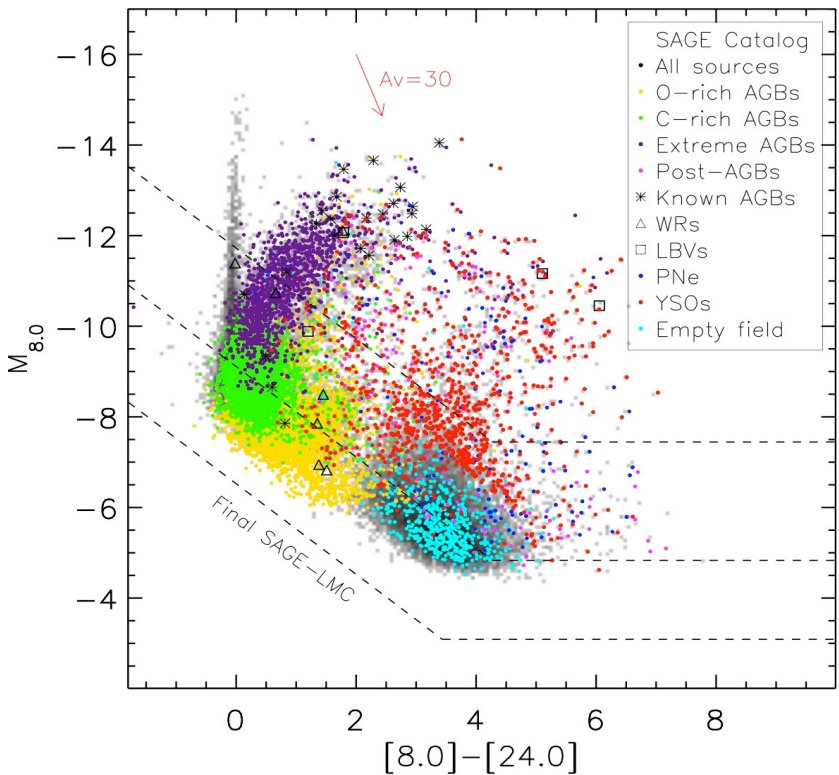

***Figure 3-6.*** *Infrared color-magnitude diagram of stars in the nearby galaxy, the Large Magellanic Cloud, obtained by the SAGE Spitzer Legacy Program (Blum et al. 2006; Meixner et al. 2006). Different classes of objects – indicated by different color symbols – segregate into different parts of this diagram due to differences in dust composition and spectral characteristics, while their distribution reflects intrinsic variations in mass-loss rates. The GAIA mission will provide accurate distances for stars in the Milky Way and this will enable similar color-magnitude diagrams for our own galaxy. The dashed lines indicate the SAGE photometry limits for the LMC and the predicted sensitivity limits (10 $\sigma$ in 900 sec) for FORCAST in the GRISM mode for these types of objects at 3 and 10 kpc, respectively, in the Milky Way. Figure is courtesy of M. Sewilo & SAGE team.*

ISO has made a first foray into this area, limited to bright, well-known galactic objects in these classes (essentially a flux-limited sample). Through the SAGE leg-

> *Over the lifetime of SOFIA, we can initiate a comprehensive program to determine the characteristics of stardust injected by all stellar sources into the interstellar medium of the Milky Way.*

acy program, Spitzer has focused on addressing the questions of dust origin and evolution on a galactic scale (Figure 3-6) through spectroscopy of objects in the low metallicity environments of the LMC and the SMC. JWST will be able to probe the issues around dust sources in all galaxies in the local group. With SOFIA, we can initiate a comprehensive program to determine the characteristics of stardust injected by all stellar sources into the interstellar medium of the Milky Way. A SOFIA study can uniquely include the faintest (~1 Jy) and most numerous stars that are too faint in nearby galaxy studies. In addition, GAIA will provide for the first time accurate distances and that is a prerequisite for accurate mass-loss rates. In particular, Hipparchos has derived accurate distances ($\sigma <$ 15%) for 8 AGB stars (Whitelock et al. 2008). In the next decade, the GAIA mission — which will measure parallaxes ~100 times more accurate than Hipparchos — can be expected to provide accurate distances to approximately 10,000-100,000 AGB stars. Thus, for the first time, determination of reliable mass-loss rates from evolved stars and an assessment of the mass balance issues on a Milky Way wide scale will be possible. Moreover, we can compare the characteristics of stardust injected into the local ISM with those of interstellar dust derived from UV/optical extinction or IR emission measurements.

This SOFIA program would consist of infrared spectroscopy from 3 to 100 $\mu$m, with R ~200-1000, of a volume-limited sample of AGB objects selected through their infrared colors. Rare objects — including post-AGB objects, luminous blue variables, and Wolf-Rayet stars — will require a large galactic volume to be sampled. Current or planned optical and infrared surveys of the Milky Way will yield large samples of single and binary stars in the various brief evolutionary phases that occur near the end-points of their lives. Ground-based optical surveys include the nearly complete INT Photometric H$\alpha$ Survey of the northern galactic plane (IPHAS; Drew et al. 2005) and the complementary VST Photometric H$\alpha$ Survey (VPHAS+) of the southern galactic plane that will begin in 2009. Ground-based near-IR surveys include the UKIRT Infrared Deep Sky Survey (UKIDSS; Lawrence et al. 2007) in the north, and the VISTA sky surveys in the south. The *Spitzer* GLIMPSE (Benjamin et al. 2003) and MIPSGAL (Carey et al. 2005) mid-IR surveys will be complemented by the Herschel 70-500 $\mu$m Hi-GAL Survey of the same Milky Way regions. These surveys will enable many stars in different advanced evolutionary stages to be identified, from their characteristic optical/IR spectral energy distributions (SEDs). The program outlined in the ISM observing plan table (Appendix B) demonstrates the feasibility of this evolved star spectroscopic survey with SOFIA. This SOFIA program will provide the spectroscopic signatures of dust formed in different stellar birth sites and thus enable us to link the dust composition and mass-loss rate to the type of evolved stars. This unique database will trace the origin and lifecycle of dust in the Milky Way. Moreover, it will guide interpretation of the dust composition in galaxies near and far in the Universe.

### 3.4 The Role of Large, Complex Molecules in the ISM: Identification of PAHs

Polycyclic aromatic hydrocarbon (PAH) molecules containing upwards of approximately 50 C-atoms are ubiquitous in the interstellar medium (Figure 3-7). As a class, these species are the most complex molecules so far identified in space comprising ~10% of the elemental carbon in the local Universe. Interstellar PAHs are thought to range from individual molecules and molecular clusters to amorphous carbon nanoparticles which are a conglomerate of PAHs (Figure 3-8). PAH molecules are likely formed in the outflows from C-rich AGB stars as byproducts or molecular intermediaries of the carbon-soot formation process (Frenklach and Feigelson 1989; Cherchneff et al. 1992). Once injected into the interstellar medium, these species are processed, weeding out the weaker species by interaction with UV and X-ray photons and cosmic rays as well as by sputtering in supernova shocks (Jochims et al., 1994; Le Page et al. 2001; Micelotta et al. 2009). In addition, grain-grain collisions may shatter carbon nanoparticles and release embedded PAHs into the gas phase (Jones et al. 1996).

*SOFIA's unique far-infrared spectroscopic capabilities will offer new insights on PAH species through the molecule specific drumhead modes.*

Condensation of PAHs in ice mantles in dense molecular clouds may be followed by UV photolysis producing aromatic ketones, alcohols, ethers, amines, N-heterocycles, and $H_n$-PAHs (PAHs with excess H atoms added; Bernstein et al. 2002). Significantly, and perhaps not coincidentally, such compounds are common in living systems, performing a number of important biochemical functions. Evaporation of these ices near a nascent protostar and its surrounding planetary disk will release these complex chemicals again into the gas phase. Hence, PAH chemistry that occurs in and around dense clouds may be at the base of the reservoir of prebiotic organic compounds available to evolving planetary systems. Key questions surrounding PAHs are:

- What is the census of PAH molecules in the ISM?
- What processes affect the PAH population in the ISM?
- What is the complex molecular inventory in regions of star formation particularly in the terrestrial habital zone?
- Are portions of the aromatic population converted into species of astrobiological significance?

Although the presence of PAH molecules in space is undisputed, identification of specific species has remained challenging. Investigations of PAHs so far have been driven by sensitivity issues that promoted a focus on the strong mid-IR fundamental modes, involving nearest neighbor atoms that are not sensitive to the larger molecular structure. For example, the top of Figure 3-7 notes how the wavelengths and features in the mid-IR correspond to the various types of vibrational modes in all of the PAHs. ISO observations of the fundamental mid-IR modes have revealed large spectral variations between different regions and within individual regions. These variations reflect variations in the molecular structure and chemical composition of the emitting interstellar PAH family (Hony et al., 2001; Peeters et al., 2002; Joblin et al., 2008; Sloan et al. 2007; Flagey et al. 2006). In particular, there is increasing evidence (Goto et al. 2002; Sloan et al. 2007) that the change in observed profiles reflects a transition from relatively unprocessed mixture of aliphatic and aromatic compounds in post-AGB objects to almost pure aromatic structures in planetary nebulae as the radiation field hardens (Pino et al. 2008). A similar chemical change is observed for PAHs associated with planet-forming disks around protostars (Sloan et al. 2007; Boersma et al. 2008). The increased sensitivity of the IRS on Spitzer has opened up the 15-20 $\mu$m spectral range, which probes (next-nearest neighbor) deformation modes of the C-skeleton (Sellgren et al. 2007; Smith et al. 2007). These data have not been fully analyzed, but these modes have still only limited diagnostic value when similar molecular structures are present in the PAH family.

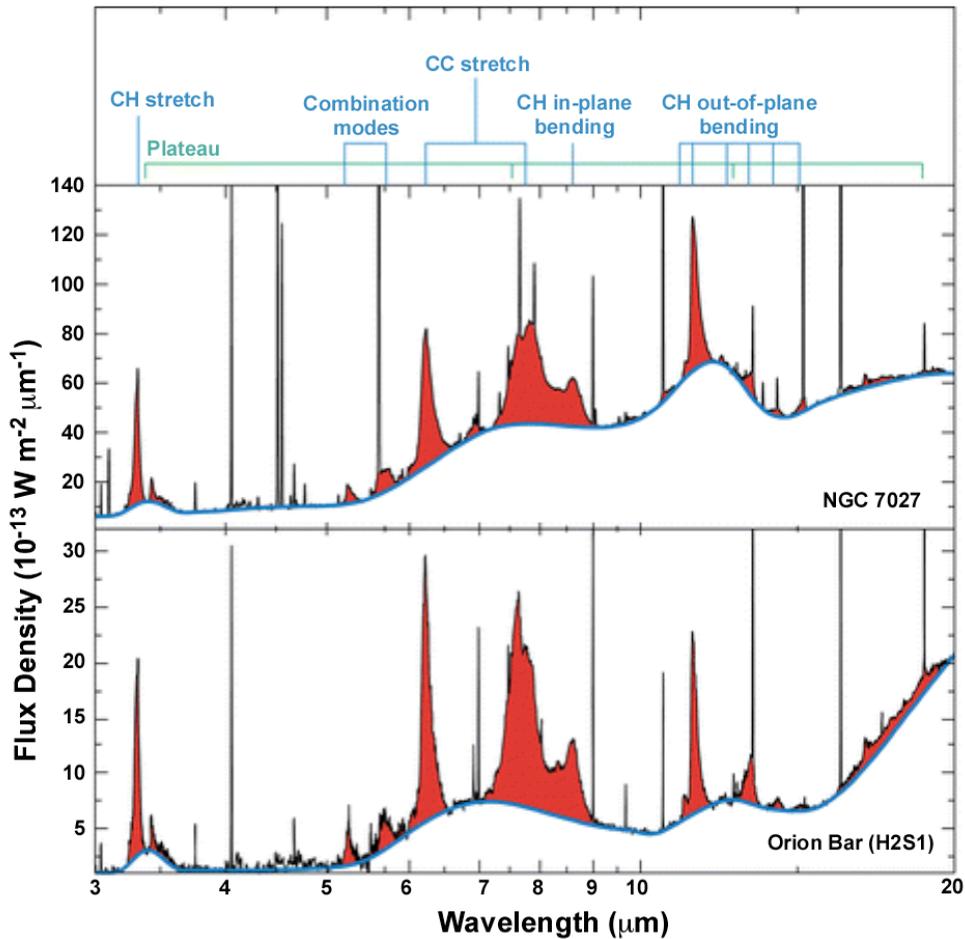

***Figure 3-7.*** *The mid-infrared features attributed to PAHs arising in a planetary nebula and the photodissociation region in the Orion Bar. The PAH features are filled in with red and these features are labeled with the vibrational modes of PAHs at the top (Figure is from Tielens 2008 and adapted from Peeters et al. 2002).*

Far-infrared (Far-IR) spectroscopy of PAH rich objects will provide important constraints on the size and shape of PAH molecules and clusters by measuring the lowest lying vibrational modes corresponding to 'drumhead' modes of PAHs. If the interstellar PAH family is wide and diverse and hundreds of PAH species are contributing to the mid-IR emission spectrum of the interstellar medium, then the far-IR modes — which are molecule specific — of any individual PAH may be too weak to be detectable against the dust continuum. However, there are indications that under the harsh conditions of interstellar space the interstellar PAH family is weeded down until only the most stable structures survive (Tielens 2008). These most likely center on the highly compact PAHs, circumcoronene ($C_{54}H_{18}$) and circumcircumcoronene ($C_{96}H_{24}$) which — because of the superaromaticity associated with their compact structure — are exceedingly stable.

Thus, it seems plausible that we may detect a series of strong far-IR bands in PAH rich sources and constrain the size and shape of these PAHs. Moreover, we could inventory potential variation and evolution of PAHs using far-IR spectroscopy.

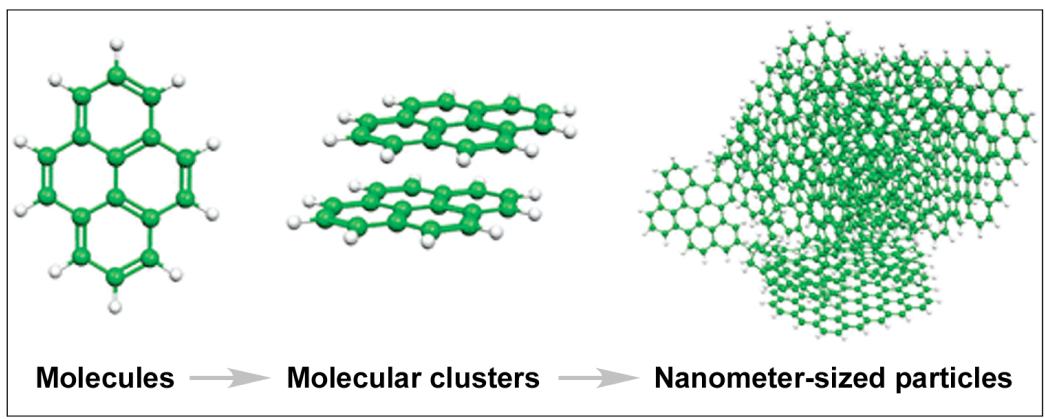

**Molecules** → **Molecular clusters** → **Nanometer-sized particles**

*Figure 3-8.* Schematic chemical structure diagrams of PAH molecules, molecular clusters and nanometer-sized particles. The carbon atoms appear as green spheres and the white spheres, which are attached to the benzene rings, may be an H atom, or something more complex, e.g. $CH_3$. Figure is from Tielens (2008).

The potential utility of far-IR spectroscopy of PAHs is illustrated in Figure 3-9 which shows the calculated spectra of an aggregate of 20 PAHs for conditions relevant for the post-AGB object, the Red Rectangle. While the mid-IR bands of these species line up at the fundamental CC and CH stretching and bending modes, the far-IR is riddled by a large number of specific transitions in individual molecules (Figure 3-9). Positive identification will require detection of not only the lowest frequency mode, but also higher frequency modes throughout the full range at the expected relative strength in agreement with laboratory-measured intrinsic strength and the calculated emission temperature of the species in the particular nebula.

Once a potential PAH species is identified with the low resolution broad-band far-IR spectroscopy, higher spectral resolution observations of individual far-IR bands would be pursued as followup. At high spectral resolution, the P and R branches of the far-IR modes will be revealed and we can derive the moments of inertia from the rotational spacing of the P and R branches (Figure 3-10). The moments of inertia reflect the structure and the number of carbon atoms of individual PAHs and thus provide an additional and independent identification tool.

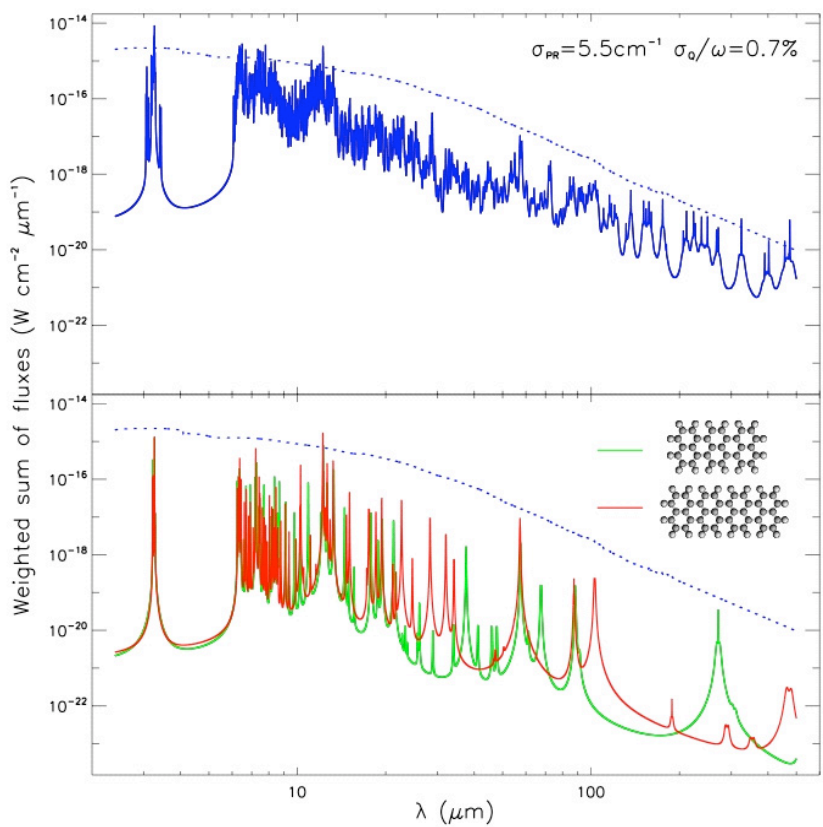

***Figure 3-9.*** *Top: Weighted sum of the spectra of 20 PAHs in their neutral and cationic state, calculated for the excitation conditions of the post-AGB object, the Red Rectangle. The dotted line shows, for comparison, the estimated dust continuum in the same source. Bottom: The predicted spectra of two individual PAHs illustrates that 20-600 μm spectroscopy could lead to unique identification of a PAH molecule. Such a measurement will require a new spectrograph for SOFIA. Figure adapted from Mulas et al. (2006).*

As Figure 3-9 illustrates, the full spectral range from 3-500 μm should be probed and the hitherto unprobed 20-500 μm range is particularly unique for molecular identification purposes. The density of individual bands decreases with wavelength and the range 100-200 μm is of prime importance. We propose a pilot program that uses FIFI LS to cover this spectral range. Sensitivity is generally not an issue since most of these sources are bright in the IR. However, line-to-continuum is of some concern, with predicted ratios for the strongest bands in the long wavelength range of ~0.1. **T**hus ideally, an advanced spectrograph that can cover the full relevant spectral range from 3 to 500 μm is required to control systematic uncertainties. Such a spectrograph would be complex and massive, and would require a flying platform that only SOFIA can provide. Studies of the P and R branches of the decoupled lowest vibrational modes can be accomplished with FIFI LS because they require medium spectral resolution of at least R ~$10^3$ (Figure 3-10).

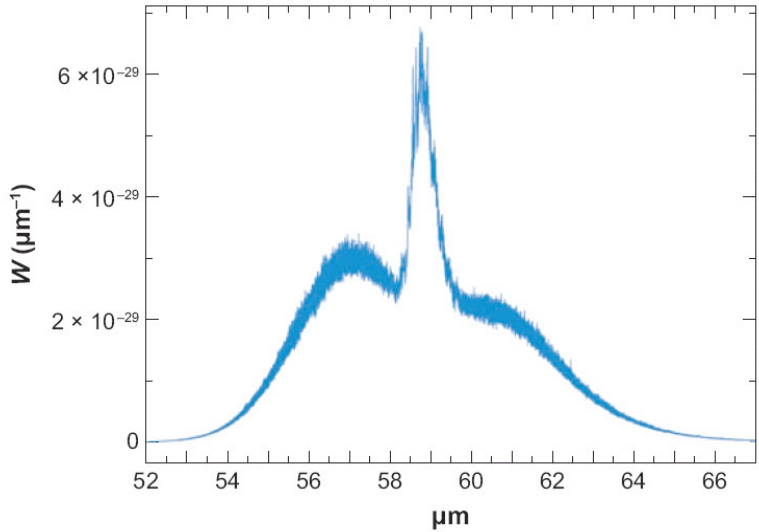

***Figure 3-10.*** *The lowest vibrational modes of PAHs show characteristic P-Q-R branch structure that can be used to uniquely identify the molecule. For example, the PAH molecule naphthalene shown here can be spectrally resolved by FIFI LS.*

The full program would sample a wide range of objects that probe the full lifecycle of interstellar PAHs including post-AGB objects (e.g., the Red Rectangle, IRAS 22272, GL 618), planetary nebula (e.g., NGC 7027, BD 30 3639, IRAS 21280), PDRs in the ISM (e.g., NGC 2023, IC 63, the Orion Bar), molecular cloud surfaces (e.g., ρ Oph, Horsehead nebula), and regions of star and planet formation (e.g., Herbig Ae and T Tauri stars). A comprehensive program spanning the full range of physical and chemical conditions is outlined in the ISM observing plan table in Appendix B.

### 3.5 Deuterium in the ISM: Constraints from HD

The interstellar deuterium (D) abundance provides a direct measure of the cosmic history of nucleosynthesis. All the D in the universe was originally formed during the Big Bang. Observations of the angular distribution of the temperature fluctuations in the cosmic microwave background radiation, in combination with new data on the large-scale structure of the Universe (e.g., Spergel et al. 2003), have constrained the baryon abundance to better than 3%. The resulting Big Bang nucleosynthesis predicted D-abundance is accurate to 5%: D/H= 2.6 ± 0.15 x $10^{-5}$, in good agreement with the measured D-abundance in a few high redshift quasar absorption line systems. Subsequently, D is readily converted into $^3$He,

*With GREAT on SOFIA, the main molecular reservoir of deuterium, HD, can be studied at the individual molecular cloud level.*

$^4$He and heavier elements by nuclear reactions in stars in a process commonly called astration (Epstein et al. 1976). Thus, as material is cycled through stars, D is only destroyed. This steady, monotonic decrease in the D-abundance is counteracted by infall of less astrated material — D-rich and metal-poor — from the intergalactic medium and from small galaxies captured by the Milky Way (Tosi 1988a). Thus, the reigning paradigm has it that the evolution of the D-abundance is simple: the D-abundance is a probe of the fraction of interstellar gas, which has never been cycled through stars (Steigman and Tosi 1992).

Recently, FUSE observations have challenged this picture by revealing that the gas phase D-abundance in the diffuse interstellar medium of the Milky Way is highly variable by factors of 2 to 4 depending on location in the Milky Way which is difficult to account for by the simple scenario (Linsky et al. 2006; Hoopes et al. 2003, Sembach et al. 2004; Savage et al. 2007). Indeed, these FUSE observations of atomic hydrogen and deuterium in the Milky Way indicate that deuterium is preferentially deposited onto dust grains, a mechanism only poorly understood at this time (Draine et al., 2006; Linsky et al. 2006), but reminiscent of the molecular fractionation seen in cold gas. The challenge for all observations of deuterium bearing species is to understand the relation between the measurements and the real isotopic abundances.

In molecular clouds, deuterated molecular hydrogen, HD, is the dominant reservoir of deuterium. HD has been observed with the ISO and Spitzer spectrographs (Figure 3-11, Wright et al. 1999; Neufeld et al. 2006; Caux et al. 2002) through a number of rotational transitions in warm gas associated with strong shocks. However, analysis of these transitions in terms of the HD abundance is hampered by the uncertain excitation conditions in these regions, and the poor spectral resolution. With GREAT on SOFIA, the ground-state transition of HD (at 112 $\mu$m, 2.675 THz) will become uniquely accessible at sub-km/s resolution, in emission in the warm gas associated with photodissociation regions and interstellar shocks, and in absorption toward bright background sources.

The ultimate question for deuterium studies in the ISM is to understand the rate of stellar astration and Galactic infall over cosmic times. To advance this study the SOFIA/GREAT program will make crucial contributions in several areas:

- How does the chemistry, and fractionation, in the ISM affect measurements of deuterium?
- How does the gas and dust characteristics of the gas affect the D gas-phase abundance?

The 100 μm channel on GREAT is uniquely suited to study the gas phase HD abundance in molecular clouds through the J=1-0 transition at 112 μm. Absorption studies against bright background sources is of particular interest since at the low temperatures of molecular clouds, essentially all HD will be in its ground rotational state and this line can be directly translated into a velocity resolved absorbing column of HD molecules. Corresponding hydrogen column densities can be derived from absorption measurements of CO isotopomers. The ratio of the HD to $H_2$ column densities will provide an estimate of the D/H ratio. HD observations of a variety of galactic environments will provide key insight into how chemistry, fractionation, and dust depletion affect the D/H abundance. With the results from these HD observations, the systematic uncertainties in the D/H abundances can be better understood providing insight into galactic chemical evolution. The ISM observing plan table (see Appendix B), demonstrates the feasibility of this proposed plan.

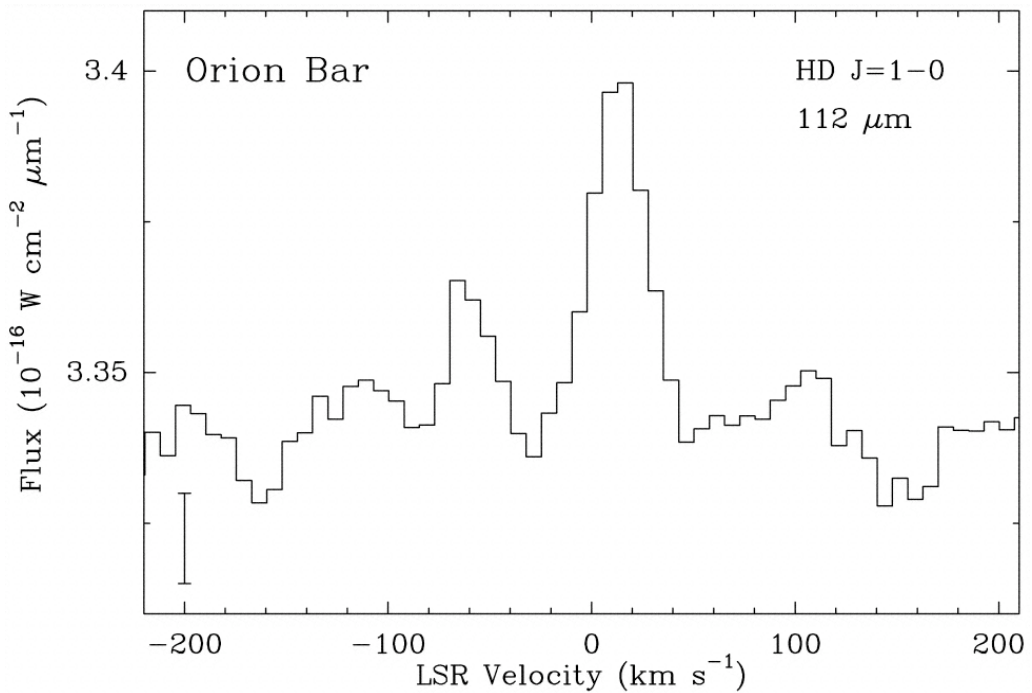

***Figure 3-11.*** *ISO detection of the HD 112 μm line in the Orion Bar from Wright et al. (1999) using the LWS Fabry-Perot spectrometer. The inferred HD column density is $2.9 \times 10^{17}$ cm$^{-2}$. The line is unresolved.*

# Chapter 4

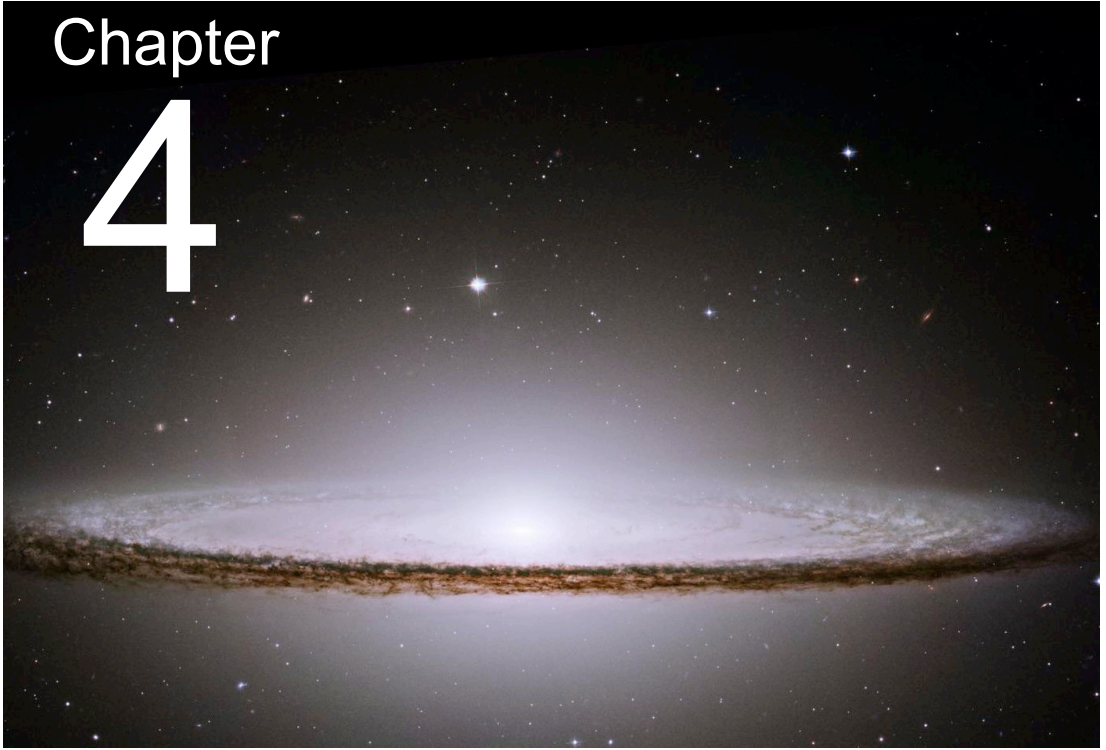

## Galaxies and the Galactic Center

## 4.1 Overview

The center of our Galaxy is a few hundred times closer than even the nearest active galaxies, enabling uniquely detailed studies of phenomena found only in galactic nuclei, including extremely warm dense clouds, strong magnetic fields, and a black hole. Infrared observations, which first opened the window of visibility of the Galactic Center, are particularly valuable for investigating the many questions about this complex region of stars, gas, dust, and the black hole. Of particular interest are the heating source for the molecular clouds in the central molecular zone, and the dynamical and energetic consequences of the strong magnetic fields found in the Galactic Center. Are the clouds heated by stellar photons, X-rays from the nucleus, or shocks due to cloud-cloud collisions or internal supersonic motions? If the clouds are heated by cloud collisions, will the angular momentum loss lead to streaming towards the black hole? Does the magnetic field govern the collection of molecular gas into the central few parsecs of the Galaxy and perhaps guide or direct such streaming motion? The Galactic Center provides a template by which more distant galactic nuclei may be understood, and SOFIA will be a crucial platform for studying the processes and phenomena occurring therein.

A key challenge of extragalactic astrophysics is understanding the star-formation history of external galaxies, and the investigation of the interstellar medium in galaxies is essential to achieving this goal. Far-IR fine structure lines are excellent probes of both the physical conditions of the interstellar medium and the stellar radiation field properties within galaxies. SOFIA observations of these and other lines at unprecedented spatial resolution will enable studies of the variations in the conditions in the interstellar medium in nearby galaxies as a function of location and distance from nuclei and spiral arms.

Typically the brightest far-IR line is the [CII] 158 $\mu$m line, and detection of this line in distant galaxies (at redshifts beyond 1.1) is one of the primary science goals for ALMA. SOFIA is uniquely poised to detect [CII] from galaxies in the redshift range between ~0.3 and 1.1 — from the time of the peak in star formation activity in the Universe to nearly the present day, thereby constraining the strength and spatial extent of the starbursts. Observations of [CII] with SOFIA should illuminate the question of how and why star formation transitioned from the galaxy-wide star burst mode seen at z > 1 to the rather quiescent and localized mode seen in present day spiral galaxies.

## 4.2 The Galactic Center: Warm Clouds and Strong Magnetic Fields

The 300-pc diameter central molecular zone (CMZ) of our Galaxy, with ~$4 \times 10^7$ $M_\odot$ of warm molecular gas and dust responding to the deep potential well of the nuclear stellar bulge and the $4 \times 10^6$ $M_\odot$ galactic black hole, represents the most massive concentration of dense gas in the Galaxy and is an essential template for understanding the phenomenology of galactic nuclei in general.

*The abundant energy of the Galactic Center emerges almost entirely in the IR because of high visual extinction. SOFIA is the most versatile and sensitive platform for observations of the Galactic Center covering the entire IR spectral region.*

At a distance of only 8 kpc, vastly closer than even the nearest active galactic nucleus or starburst nucleus, the Galactic Center (GC) offers us critical, spatially resolved information about how the activity of a galactic nucleus is produced by the interactions of the multitude of contributors: the central black hole, massive stars, dense clouds, strong magnetic fields, intense X-ray emission, and other forms of high-energy radiation. Because of the 20 – 30 magnitudes of visual extinction to the GC, its abundant energy emerges almost entirely in the IR, and is well suited for study by SOFIA (Figure 4-1). Here, we focus on two important, unsolved questions about the Galactic Center interstellar medium that SOFIA is uniquely positioned to address.

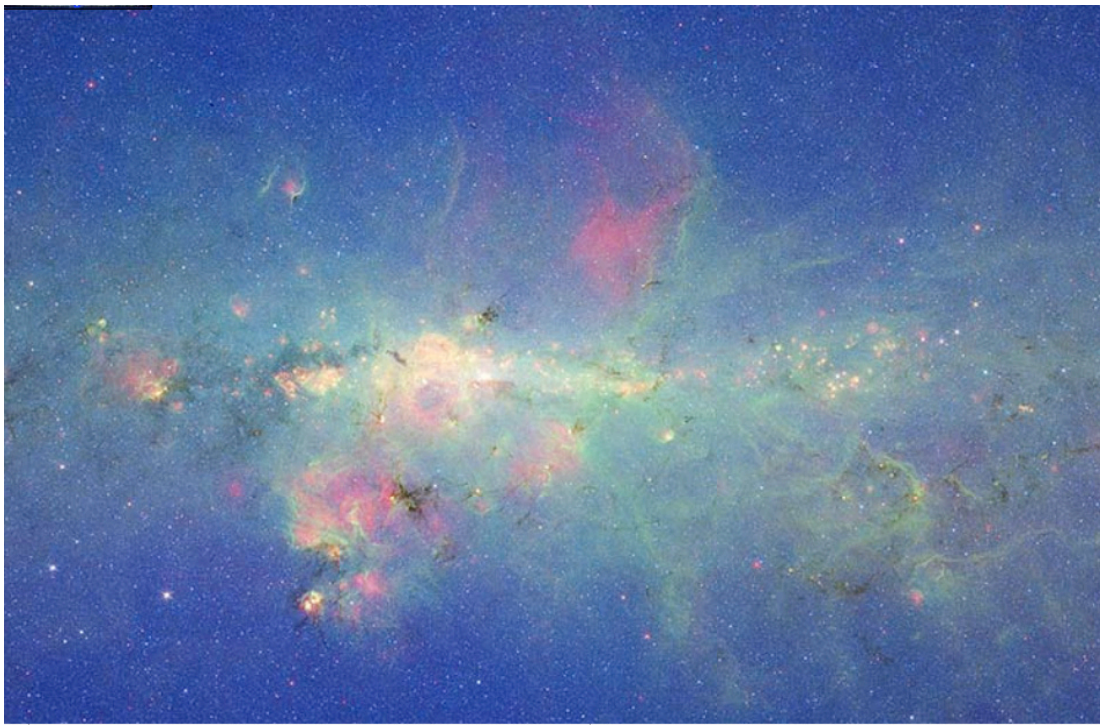

***Figure 4-1.*** *Infrared view of the inner 500 parsecs of the Galaxy, from the legacy programs of the Spitzer Space Telescope. Blue & green: measurements at 3.6 & 8.0 μm with the IRAC camera; red: 24 μm image from the MIPS camera. The CMZ occupies the central 300 parsecs of the Galactic plane.*

***What is the dominant heating source for clouds in the CMZ?*** The average temperature in GC clouds is about 80K, but can reach 500K, and about 1/3 of the total gas column is at a temperature of 150 – 200K. These temperatures are far higher than that of the molecular gas in the disk and require a very substantial heating source, the nature of which has long been a mystery (e.g., Morris et al. 1983); currently, the leading candidates are shocks and X-rays (Rodríguez-Fernández et al. 2004), while the dissipation of hydromagnetic waves is also a possible contributor. The answer to this question is essential to understanding both the unusual chemistry and the star formation in galactic nuclei. The high gas temperature affects the balance between gas-phase and grain surface chemistry, and helps overcome the potential barrier inhibiting many gas-phase reactions. Furthermore, some of the proposed heating mechanisms give rise to local temperature spikes that can have dramatic effects on the chemistry, by speeding up reactions and promoting the evaporation of volatile molecules from grains. In addition, because the Jeans mass increases with temperature, the high gas temperatures found in the GC may skew the stellar initial mass function toward higher masses. Hence, galactic nuclei may tend to favor the formation of high mass stars compared to cooler regions throughout the galaxy.

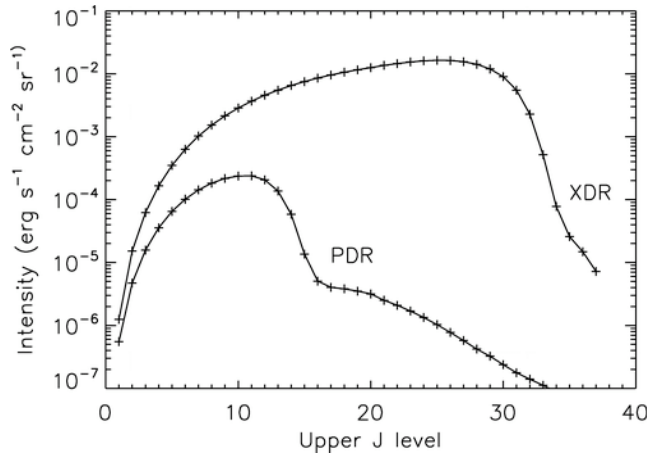

***Figure 4-2.*** *CO line intensities as a function of the upper J level, for a solar metallicity model of density $10^5 cm^{-3}$, and a constant impinging energy flux (from Spaans & Meijerink 2008). The stellar spectrum assumed for the PDR illumination is a 30,000 K blackbody.*

Shock heating can best be investigated with lines of [OI]. SOFIA will measure the 63 and 146 μm lines of [OI] and permit a comparison of their strengths to that of the [CII] 158 μm line. While the [OI] lines can be produced in both shocks and photodissociation regions (PDRs, see Chapter 3), the [CII] line strength is much less affected by shocks; observations of the two [OI] lines are needed to control for density. Previous observations of the [OI] lines with ISO in a number of isolated locations have been interpreted in terms of C-shocks with pre-shock densities in the range $10^3 - 10^5$ cm$^{-3}$, and shock velocities of 10 – 20 km s$^{-1}$ (Rodríguez-Fernández et al. 2004; J-shocks are ruled out by the ISO results). SOFIA's unique contribution to observations of the important [OI] 63 $\mu$m line results from the GREAT instrument having sufficient velocity resolution to discriminate between the multiple velocity components that are typically seen along a given line of sight through the CMZ, and spread out over a few hundred km s$^{-1}$. Many shocks can be spatially resolved, since they result from large-scale phenomena such as cloud collisions, or are mediated by the galactic bar potential. In those cases, a given shock structure will typically be confined to a single velocity component, and both the spatial and spectral resolution of that component is essential. Internal shocks associated with the strong, supersonic (and probably magnetohydrodynamic) turbulence in GC clouds will not be spatially resolved with any instrument, but the shock heating hypothesis can be tested by seeking a correlation between the turbulent linewidth and the [OI]/[CII] shock diagnostic.

X-ray heating of GC clouds has been downplayed by Rodríguez-Fernández et al. (2004) because the currently measured X-ray fluxes are not sufficient to counteract molecular cooling at the high temperatures and densities of GC clouds. However, X-rays up to 2 keV are absorbed by Galactic dust, so only the tip of the X-ray iceberg can be measured. Furthermore, the neutral iron fluorescence line at 6.4 keV suggests that a burst of X-rays many orders of magnitude brighter than anything currently observed in the GC occurred several hundred years ago, so the heating from such an X-ray flash is still propagating through the CMZ. Thus, the X-ray heating hypothesis warrants renewed investigation.

X-ray dominated regions (XDRs) have most recently been investigated by Meijerink et al. (2007), who predicted intensities for atomic fine structure lines and CO lines in both XDRs (illuminated by a nonthermal AGN spectrum) and PDRs (illuminated by a starburst spectrum). The results of this work show that the line intensity ratios [SiII] 34.8 $\mu$m/[CII] 158 $\mu$m and [FeII] 26 $\mu$m/[CII] 158 $\mu$m, as well as the intensities of the high-J CO lines, provide powerful diagnostics. For example, at typical GC cloud densities, the J=16-15 CO line is about three orders of magnitude stronger in XDRs than in PDRs (Fig. 2.2, from Spaans & Meijerink 2008); these lines are accessible to the CASIMIR (1.15 THz) and GREAT (1.84 THz) instruments, respectively. The [FeII] 26 $\mu$m line can be observed with EXES, while a grism on FORCAST will permit observations of the [SiII] 34.8 $\mu$m line, and GREAT will provide access to the [CII] 158 $\mu$m line. All these lines are relatively strong (too strong, in fact, to be observed with Spitzer), so extensive spatial mapping will be possible with all of these instruments, and will allow one to identify the presence and extent of any XDRs in the region.

**What is the geometry of the magnetic fields in the clouds of the CMZ?** Determining the orientation of the strong magnetic fields in the Galaxy's CMZ is critical to elucidating the dynamical and energetic consequences of this field. As indicated above, the internal motions, and probably much of the heating of GC clouds, are governed by hydromagnetic waves. One of the most significant discoveries with the Kuiper Airborne Observatory (KAO) was that magnetic field geometries in molecular clouds — particularly in the GC — could be determined by measuring the polarization of thermal far-IR emission from magnetically aligned dust grains (e.g., Chuss et al. 2003). While similar studies have recently been carried out in the submillimeter, the best wavelength range in which to study warm GC clouds is the far-IR, where the polarization is the strongest. Determinations of the magnetic field orientations in GC clouds have been made in only a few clouds so far, with the results that the emitting grains are well aligned (presumably by a strong field), the field directions in clouds are remarkably uniform, and the field is predominantly parallel to the Galactic plane (Figure 4-3). This has been attributed to the shear of molecular clouds in the strong tidal field of the CMZ.

Beyond extending such observations to a greater number of clouds, the next step is to study the spatial fluctuations in the field direction as traced by fluctuations in the polarization angle. Using the rms fluctuations, in concert with the known velocity dispersions of these clouds, one can apply the Chandrasekhar-Fermi (C-F) method (which assumes equipartition between the magnetic energy density of the field fluctuations and the local turbulent velocities associated with the field distortions) to determine the field strengths. However, it will probably be necessary to employ a modern variation of the C-F method, in which observed rms fluctuations in field direction are compared with beam-convolved numerical models of turbulent clouds (e.g., Heitsch et al. 2001). Since Zeeman measures have so far not been very definitive in the GC, largely because of the very broad linewidths, the C-F method appears to be the best way to determine the field strength in clouds. This is critical for our understanding of cloud dynamics, star formation, and the relationship to the strong intercloud magnetic field (Morris 2006).

> *SOFIA is uniquely poised to discriminate among the multiple velocity components of far-IR lines along sight-lines through the Central Molecular Zone, using the high resolution of GREAT, to study shocks and their effects on clouds in the Galactic Center.*

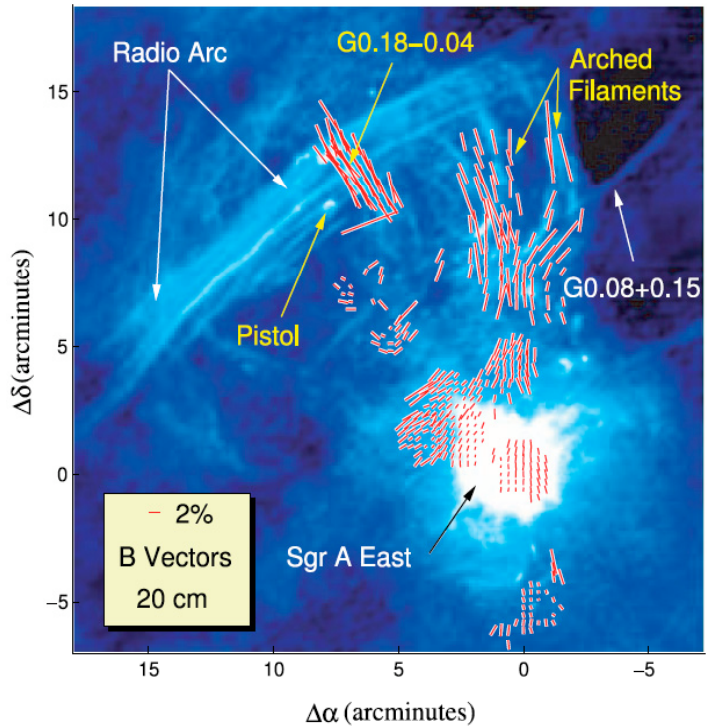

*Figure 4-3.* Magnetic field directions inferred from measurements of the orientation of the polarized E-vectors of thermal emission from magnetically aligned dust grains at a combination of far-IR and submm wavelengths (from Chuss et al. 2003). The underlying VLA radiograph, from Yusef-Zadeh et al. (1984), shows 20 cm emission from a 60x60 pc region. The Galactic plane is oriented at a position angle of about 30° east of north, and the Galactic center is located within the radio-bright Sgr A complex.

*With a polarimeter on the HAWC instrument, SOFIA can measure spatial fluctuations in the magnetic field in the Galactic Center with unprecedented resolution, thus contributing uniquely to outstanding questions of magnetic field geometry in clouds of the Central Molecular Zone.*

The circumnuclear disk surrounding the central black hole at a distance of 1 – 5 pc (0.4 – 2 arcmin in radius) provides a particularly interesting case study (Hildebrand et al. 1993), since it comprises the reservoir of dust and gas from which the central black hole may episodically accrete matter, and from which we stand to learn about activity in low luminosity active galactic nuclei and about circumnuclear disks in general. Information about the geometry and fluctuations of the magnetic field in this structure will be very important for informing models of the dynamical evolution of this disk, and the eventual accretion arising therefrom. Furthermore, such observations can elucidate the connection of the toroidal field in this disk to the vertical (dipole) field that apparently dominates the intercloud medium (Wardle & Königl 1990).

With a polarimeter on the HAWC instrument, the spatial fluctuations of the field direction can be measured in the 53 $\mu$m band with spatial resolution five times better than previously possible. Furthermore, such measurements cannot be made with any other observatory. The spatial resolution is a key issue, because the applicability of the C-F method is maximized when the scale of the fluctuation measurements is matched to the angular scale of the cloud's velocity fluctuations, and previous observations have probably not reached this scale. Because the continuum emission from GC clouds is strong, these polarization measurements could be carried out quickly with SOFIA; we estimate that an entire cloud could be measured at full resolution in ~10 – 15 hours. The GC contains about 10 clouds with sufficiently strong far-IR emissions on which this kind of study could be carried out. A thorough characterization of the full set of such clouds would be important for reaching general conclusions about the magnetic field in the CMZ.

## 4.3 The Interstellar Medium and the Star Formation History of External Galaxies

As mentioned in Chapter 3, by enabling observations of the bright far-IR fine structure lines (e.g., [OI] 63 and 146 $\mu$m, [OIII] 52 and 88 $\mu$m, [NII] 122 and 205 $\mu$m, [NIII] 57 $\mu$m, and [CII] 158 $\mu$m), SOFIA provides a nearly unique opportunity to study the properties of the interstellar medium (ISM) in nearby galaxies and to trace the star formation history of galaxies from the epoch of the peak of the star formation rate per unit volume (z~1) to today (z=0).

### *4.3.1 Mapping the Interstellar Medium in Nearby Galaxies*

Arising in interstellar gas clouds heated by UV photons, the aforementioned lines have distinct advantages as probes of the ISM over those available to optical and near-IR observers. The line emission arises from transitions between levels only a few hundred degrees above the ground state, easily excited at the temperatures characteristic of most of the ISM in galaxies. The lines are almost always optically thin and therefore are important (and often the dominant) coolants for much of the ISM including the Warm Ionized Medium (WIM), Cold Neutral Medium (CNM), Warm Neutral Medium (WNM), and the PDRs on the surfaces of molecular clouds. The level populations (and hence line strengths) are determined primarily via collisional excitation, with critical densities well matched to the densities found in most of the ISM. Hence, the far-IR lines are very sensitive probes of the physical conditions of the ISM, and density and masses of the various ISM components can be estimated from them. The range of elements and ionization levels also allows estimates of the elemental abundances in the ISM. Furthermore, the far-IR lines provide information on the ambient far-UV radiation fields, reflecting both the strength (parameterized in units of the local far-UV interstellar radiation field, $G_o$), and the effective temperature (hardness) of these fields (see Figure 3-3).

Since the UV radiation is generated by stars or accretion disks surrounding active galactic nuclei (AGN), observations of these lines yield information on the stellar population in the galaxy (e.g., the parameters of the Initial Mass Function), and the contribution from accretion onto a black hole to the general radiation field. Finally, and most importantly, because of their long wavelengths, these far-IR fine structure lines are relatively immune to the effects of dust, and therefore provide extinction free probes which are particularly useful for studying both dust enshrouded star formation regions and the heavily obscured regions near galactic nuclei — regions largely inaccessible to optical observers. With observations of these lines, SOFIA will allow investigations of the ISM in nearby galaxies at spatial resolutions matched only by Herschel, at sensitivities nearly equal to or (longward 200 $\mu$m) better than Herschel, and with unrivaled mapping capabilities.

*SOFIA observations of far-IR lines at unprecedented spatial resolution will enable studies of the ISM in nearby galaxies to explore variations with location and distance from nuclei and spiral arms.*

The [CII] 158 $\mu$m line is typically the brightest far-IR line, and is overall one of the strongest lines in the integrated spectrum of a galaxy, with a luminosity on the order of 0.1-1% of the total far-IR continuum luminosity. This line is the dominant coolant of the CNM and an important coolant for the diffuse WIM and for PDRs on the surfaces of moderate density (n ~$10^{2.5} - 10^4$ cm$^{-3}$) molecular clouds. Figure 4-4 presents the best [CII] map taken to date of the galaxy M83; this map was obtained with an imaging spectrometer on the KAO. SOFIA, with FIFI LS, will be able to map this galaxy in all of the far-IR fine structure lines (except [OI] 63 $\mu$m, which is obscured by the Earth's atmosphere at radial velocities between 500 and 1600 km s$^{-1}$) with 3 times higher spatial resolution than the KAO could provide. A map of M83 of this size in the [CII] line with a signal-to-noise ratio of 20 could be obtained in only a few hours. SOFIA observations will easily resolve the spiral arms (which can contain significant amounts of dust) and separate emission from the spiral arms from that of the inter-arm regions. The high spatial resolution will enable detailed studies of the ISM across the arms and allow investigations into the role spiral structures play in the star formation process (e.g., compression of the ISM in spiral density waves, collapse of molecular clouds to form stars) and the interaction between young stars and their natal environment (e.g., disruption of natal clouds by newly formed massive stars).

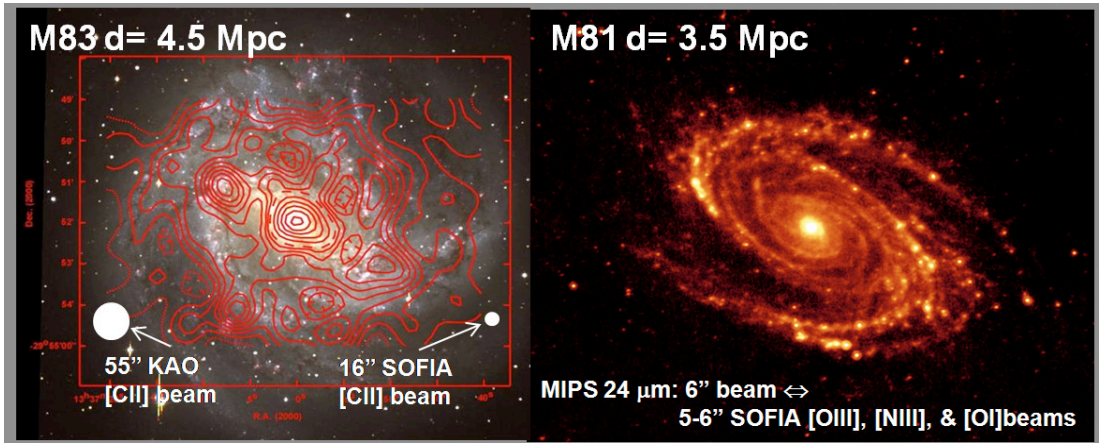

**Figure 4-4.** *(left) KAO [CII] map of M83 (d=4.5 Mpc) (contours, 55" beam) superposed on an optical image (Geis et al., in prep.). (right) MIPS 24 µm (6" beam) continuum image of M81 (d=3.5 Mpc). SOFIA can image nearby galaxies in the [OIII] 52 µm, [NIII] 57 µm, and [OI] 63 µm lines at a spatial resolution comparable to that of the Spitzer 24 µm image.*

### 4.4 Tracing the Universe's Star Formation History with Far-IR Fine-Structure Lines

The peak of star formation activity per unit co-moving volume in the Universe occurred at redshifts between ~1 and 3 (Figure 4-5) when the Universe was between 2 and 6 Gyr old. Since that time, the star formation rate has dropped by a factor of ~30; today, star formation occurs at generally modest rates (few stars/yr/galaxy), although much higher rates are found in local starburst and Ultra-Luminous Infrared Galaxy (ULIRG) systems. In these objects, star formation is characterized by very intense episodes confined to very localized (~100 pc size) regions. In contrast, near the peak of the star formation history of the Universe, galaxies experienced large-scale (several kpc), near-global bursts of star formation at rates of thousands of stars per year (Farrah et al. 2008; Hailey-Dunsheath et al. 2009).

Observations of the [CII] line and far-IR continuum from distant galaxies at redshifts between 0 and 1 permit studies of the evolution of the physical extents and the ages of the starburst episodes with redshift. For moderate density clouds, the [CII]/far-IR continuum luminosity ratio, R, is strongly inversely proportional to the interstellar UV radiation field, $G_o$, for a given density (e.g., see Figure 3-3). Therefore, when combined with an estimate of the density, the ratio R yields a measurement of $G_o$, which in turn provides clues to the nature of the source of the UV radiation. For normal galaxies like the Milky Way, $G_o$ is ~1-100 (R ~$10^{-3}$–$10^{-2}$, Stacey et al. 1991). A moderate value of $G_o$ (~$10^3$, R ~$10^{-3}$) is typically found in star-burst galaxies such as M82, while a very large value of $G_0$ (~$10^4$, R ~$10^{-4}$) indicates the presence of an AGN or very intense, very compact star formation regions with UV fields comparable to those found in Orion (0.4 pc from an O6 star; Stacey et al. 1993). As a starburst episode ages, the value of $G_0$ is expected to drop, and R should likewise rise. Dust in PDRs absorbs most of the far-UV photons generated by the starburst and re-radiates the energy as far-IR continuum. Hence, for distant and unresolved sources the ratio of the observed far-IR continuum intensity to the inferred $G_0$ yields the physical size of the starburst region in $kpc^2$.

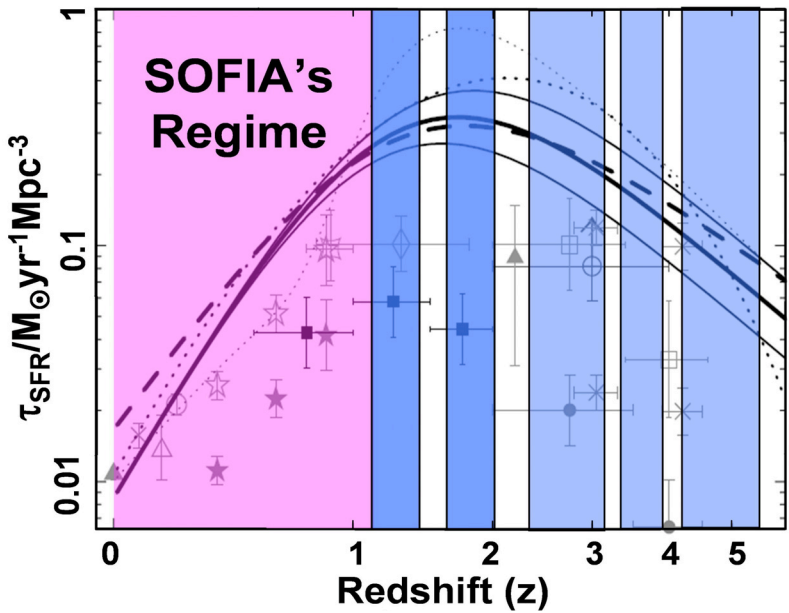

*Figure 4-5.* The co-moving star formation history of the Universe (Smail et al. 2002). The original Madau plot based on optical/UV HDF observations are the filled marks (Madau et al. 1996). Open symbols correct this data for dust extinction (Pettini et al. 1998). These are models based on the SCUBA data. The colored regions mark redshift ranges for the [CII] line available to SOFIA (pink) and to ground based observatories. (blue).

SOFIA can detect the [CII] line from galaxies that are relatively common in the redshift range from 0.25 to 1.25 — most of cosmic history and back to the time when the star formation rate per unit volume in the Universe had peaked. (Figure 4-6, top). (For larger redshifts, the [CII] line shifts out beyond 330 $\mu m$ where the telluric submillimeter windows permit access with much larger ground-based submillimeter telescopes — see Figure 4-5.) The SAFIRE instrument will have the sensitivity to detect galaxies with far infrared luminosities ($L_{far-IR}$) ranging from far below that of the Milky Way at very low redshifts to those with $L_{far-IR}$ ~2-3 times that of the Milky Way at z ~0.25, and on to those with $L_{far-IR}$ up to ~$10^{12}$ $L_\odot$ (ULIRG luminosities) at z ~0.5, and to powerful "hyper"-

LIRGs ($L_{far-IR} > 10^{13}$ $L_\odot$, HLIRGs) at z ~1. At z > 0.3, these capabilities are beyond those of Herschel (Figure 4-6, top), and are matched only by the large ground-based telescopes at z > 1.1. A [CII] survey of galaxies at these redshifts by SOFIA would be critical to understanding the star formation history of the Universe. For example, SOFIA observations will be able to address the question of whether the starbursts at these redshifts are galaxy-wide, or extremely spatially confined and intense.

Figure 4-6 (bottom) shows the detectability (in terms of the total far-IR luminosity of the source) of the [OI] 63 $\mu$m line with PACS/Herschel, and the [OIII] 88 $\mu$m, [NII] 122 and 205 $\mu$m, and [CII] 158 $\mu$m lines with SAFIRE on SOFIA. In the important redshift range from 0.5 to 1, all the lines are nearly equally detectable. Furthermore, any galaxy detectable in [OI] with Herschel/PACS is detectable with SAFIRE in [CII], from the local Universe to redshifts beyond 1. The [OI]/[CII]/far-IR continuum combination strongly constrains PDR models in terms of density and $G_0$, yielding the physical conditions of the clouds and strength and size of the starburst in this crucial epoch of the star formation history of the Universe. The [OIII] line signals the presence of hot massive early (O) type stars while the [NII] lines are important coolants for the diffuse ionized media  As such, they are good proxies for Lyman continuum photon flux, and can be used to estimate the fraction of [CII] flux arising from the ionized gas (Oberst et al. 2006). As Figure 4-6 shows, at (redshifted) wavelengths between 145 and 210 $\mu$m, these far-IR fine structure lines are accessible to both SOFIA and Herschel with roughly equal sensitivity, but between 210 $\mu$m and 330 $\mu$m SOFIA's sensitivity is unparalleled.

> *Because of its sensitivity between 210 μm and 330 μm is unparalleled, SOFIA provides a unique opportunity to study the star formation history of the Universe from redshift z=1 to today (z=0).*

There are many sources known to be above the SOFIA detection thresholds presented in Figure 4-6. For example, Rowan-Robinson et al. (2008) presented a photometric redshift catalogue of over a million galaxies derived from ~33 deg² of the Spitzer/SWIRE survey. This catalogue contains about 40 HLIRGs ($L_{far-IR} > 10^{13}$ $L_\odot$), more than 150 galaxies with $L_{far-IR} > 3\times10^{12}$ $L_\odot$, and many hundreds of ULIRG galaxies.  Nearly all of these are within the SOFIA-[CII] redshift range.  Over the full sky, there are expected to be ~50,000 HLIRG galaxies. Indeed, Elbaz et al. (2004) showed that ~2/3 of the Cosmic Infrared Background (CIB) at 150 $\mu$m arises from LIRGs at z ~0.7 — in the center SOFIA band — and that about 1/6 of this flux arises from galaxies with $L_{far-IR} > 2\times10^{12}$ $L_\odot$!

In summary, SOFIA is a critical platform with which to explore the star formation history of the Universe, from the epoch of the peak of the star formation rate per unit volume to today, including the epoch during which most of the CIB was formed.

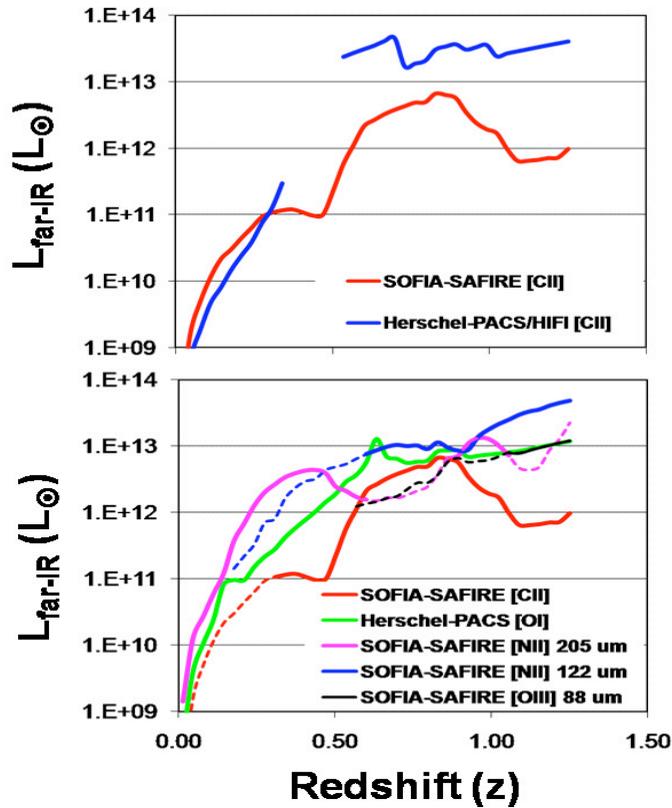

***Figure 4-6.*** *SOFIA-SAFIRE detectability of far-IR lines compared with Herschel-PACS/HIFI as a function of redshift. ULIRG values of R assumed for $L > 10^{12}\ L_\odot$, Milky Way values assumed for lower luminosity systems. Herschel HIFI values assume a line width of 300 km/s. Regions where SOFIA is not uniquely sensitive as a tracer are indicated by dashed lines. For example, at $z < 0.35$, Herschel-PACS is more sensitive for detecting the [CII] line than SOFIA-SAFIRE. Limits are S/N = 5 in 2 hours integration time.*

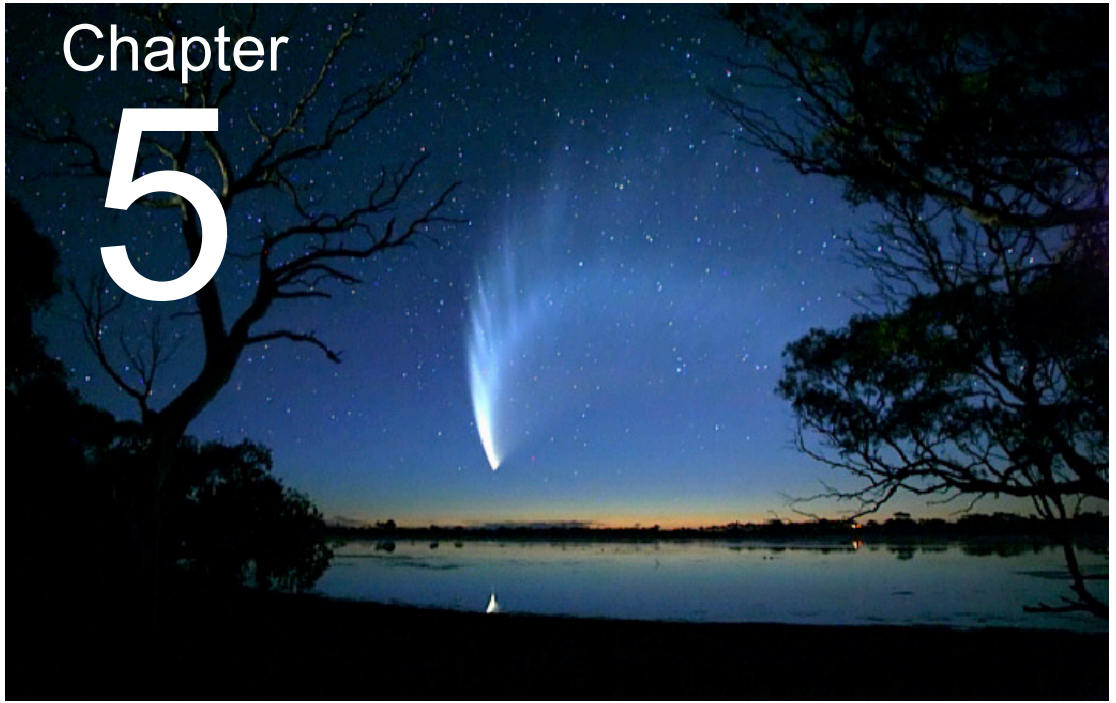

# Chapter 5

# Planetary Science

### 5.1 Overview

> *SOFIA can fill in gaps in our understanding of how Earth, and life on Earth, came to exist.*

Our Solar System contains the Sun, the Earth, Earth's seven sibling planets and the planets' moons, plus several varieties of leftover planet construction material such as comets and asteroids that are sometimes referred to as "primitive bodies."

A general but incomplete picture has emerged over the past few decades regarding how the Earth and the rest of the Solar System formed, bolstered by the discovery of more than 300 extrasolar planets orbiting other suns. It is less well known how the planet formation processes set the stage for life to develop on Earth and perhaps also on other Solar System bodies.

SOFIA's unique capabilities relative to space- or ground-based infrared observatories for investigation of Solar System objects include being able to: (a) observe objects closer to the Sun than Earth's orbit — for example, comets in their most active phases, and the planet Venus, (b) observe stellar occultations and other transient events from optimum locations anywhere on Earth, (c) directly point at bright planets and their inner moons, (d) observe spectral features of water and organic molecules at wavelengths blocked by Earth's atmosphere, and (e) monitor seasonal and episodic changes in slow-orbiting outer planets over decade timescales.

We identify selected topics within which questions of basic importance remain unanswered, or for which objects in need of close study are many and diverse. The few categories of planetary science described here are representative of high-value investigations that are difficult or impossible to do, or to do as well, from any facility other than SOFIA from now until the year 2020 at least.

SOFIA can fill in gaps in our understanding of how Earth, and life on Earth, came to exist by select observations of: (1) primitive bodies containing material little altered since the birth of the planets, including water and organic substances like those that may have rained down on Earth during and after its formation; (2) the giant planets, whose bulk compositions are most like the original raw material from which all the planets formed, and whose properties provide local comparisons for interpreting observations of extrasolar giant planets; (3) Venus, similar to Earth in size, location, and composition but with radically different surface conditions, including evidence (needing confirmation) that Venus boiled away its oceans via a runaway greenhouse effect; and (4) Saturn's moon Titan, a low-temperature organic factory possibly analogous to the pre-biological Earth.

## 5.2 Primitive Bodies

Our view of the formation and evolution of the most primitive Solar System objects has been radically altered in recent years by five developments: (1) comet flyby, sample return, and impact missions; (2) sensitive (microJansky) infrared space-based remote sensing of comets and asteroids at moderate spectral resolution (e.g., by Spitzer); (3) high-dispersion spectroscopic ground-based studies of parent volatiles in comets; (4) the emergence of a new dynamical model for the Solar System's early evolution; and (5) Spitzer spectroscopy of debris disks that are extrasolar analogs to our Solar System's remnant planetesimal belts and disks.

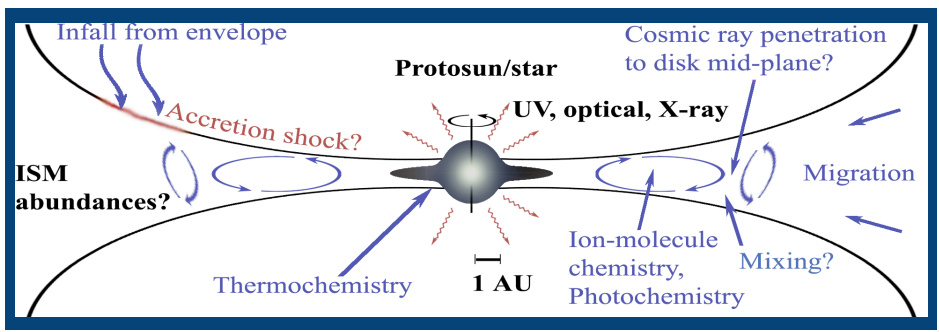

*Figure 5-1.* Schematic diagram of processes affecting material in the proto-planetary nebula during planetary system formation (from Mumma 2004).

***A Mixed-up Solar System.*** The "Nice Model" of Solar System dynamical evolution (named for its city of origin; originally presented by Gomes et al. 2005, reviewed by Morbidelli et al. 2008) infers an episode of giant planet migration and mutual perturbation 600-800 Myr after the Solar System formed that would have destabilized orbits of Kuiper Belt Objects (KBOs) and, possibly also, asteroids. Small bodies would have deluged the inner Solar System, resulting in the Late Heavy Bombardment (LHB) and potentially bringing significant amounts of volatiles to Earth, the other terrestrial planets, and the Moon. Simulations predict that many of the objects scattered inward from the Kuiper Belt joined the outer main Asteroid Belt and Jupiter Trojan swarms. If the model is correct, each of today's planetesimal reservoirs may contain significant numbers of bodies that originally formed, and spent most of their first billion years, as members of the other groups. The Nice Model further implies that analogs to the LHB event could occur in other planetary systems. During the LHB, widespread collisions and fragmentation of remnant planetesimals would have made the Solar System a strong IR emitter, possibly analogous to prominent debris disk systems detected by Spitzer around some stars with ages of a Gyr or more (reviewed by Meyer et al. 2007, but see also Wyatt et al. 2007).

### 5.2.1 Comets

Until recently, it was thought that most Jupiter-family comets formed in the Kuiper Belt (KB) region (> 30 AU) while Oort cloud (OC) comets formed in the giant planet region (5-30 AU), implying these two populations should have different native volatile compositions. However, the "Nice Model" described above predicts considerable mixing, compositional diversity, and overlap between the nearly isotropic (OC) and ecliptic (KB) dynamical populations. SOFIA can provide unique tests of early Solar System formation hypotheses by quantifying the diversity among these comet populations, connecting the mineralogy, formation temperature, and organic content of a large number of comets with their respective dynamical families. Combined with modeling of disk clearing, the comet taxonomy can provide key data for evaluating the possible contribution of water and prebiotic organic chemicals to early Earth by each class of objects.

**Comet Mineralogy.** The finding that a significant fraction of comet silicate particles are crystalline, hinted by ground and space-based spectroscopy and confirmed by Stardust samples from Comet Wild 2, is a surprise and puzzle. Possible origins of these crystalline grains include direct condensation at high temperature in the inner solar nebula coupled with some radial transport mechanism, or annealing in nebular shocks near Jupiter (Figure 5-1).

The magnesium-to-iron (Mg:Fe) ratio of crystalline silicates is diagnostic of their formation environment. Silicates with a high Mg:Fe ratio would be expected to condense from the canonical solar nebula, a relatively high-temperature, low-oxygen, low-water, Fe-reducing environment. Fe can be incorporated into condensing silicate grains, however, if the local oxygen or water content is high. The crystalline silicate Mg and Fe contents of three comets have been measured so far: the Mg-rich comets Hale-Bopp and Wild 2, and the distributed-Mg and -Fe comet Tempel 1 (review by Kelley & Wooden 2009).

With EXES, and FORCAST plus grism, spectra at segments of the mid-IR containing resonant emission peaks of crystalline silicates will be able to discern compositional details, especially the Fe:Mg value (Figure 5-2). Note that most of the diagnostic differences are at wavelengths impossible to observe from the ground. An advanced spectrograph covering all of the mid-IR (5 to 40 $\mu$m) at R ~ 1000 in single observations would allow a more comprehensive study of comet mineralogy; adding such an instrument to SOFIA's complement is also desired for studies of the interstellar medium.

> *SOFIA can provide unique tests of early Solar System formation hypotheses by quantifying the diversity among these comet populations, connecting the mineralogy, formation temperature, and organic content of a large number of comets with their respective dynamical families.*

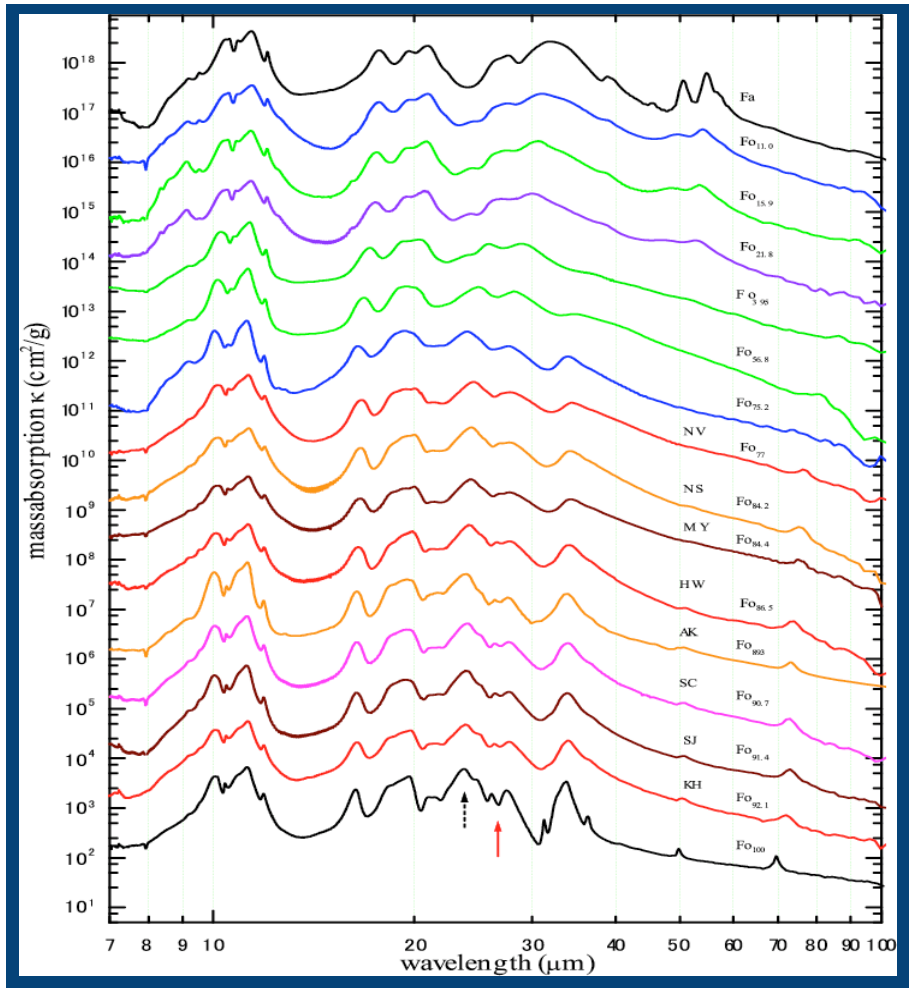

*Figure 5-2.* Measured mass absorption coefficients (equivalent to thermal emission spectra of small particles) for silicates of different Fe/Mg ratio. Top: pure Fe end-member silicate; bottom: pure Mg end-member silicate. Notice that most of the diagnostic features are at mid-IR wavelengths inaccessible from Earth. Plot taken from Koike et al. (2003).

> *SOFIA's ability to observe comets near perihelion when their activity is greatest is critical for all comets, but particularly so for Jupiter-family comets that are less active than dynamically new comets.*

SOFIA's ability to observe comets near perihelion when their activity is greatest (< 2 AU), at solar elongation angles inaccessible to other facilities, is critical for all comets, but particularly so for Jupiter-family comets that are less active than dynamically new comets because of many prior solar passages. SOFIA's high spatial resolution can also play a role in determining the degree of heterogeneity of individual comet nuclei, for example via spectral mapping of comae using the wide field-of-view FORCAST plus grism, or via long-slit spectroscopy, assessing whether or not the grains in jets arising from isolated active areas differ from those of the average coma.

*Water in Comets.* Water is the principal volatile in comets, so quantifying its abundance and other properties is of prime importance to understanding comets. Cometary $H_2O$ was first detected in emission from Comet Halley using the Kuiper Airborne Observatory (KAO) (Mumma et al. 1986). After the KAO was decommissioned, new methods for measuring water from the ground (Mumma et al. 1995) were developed, targeting "hot bands" with lower states not significantly populated in Earth's atmosphere (Dello Russo et al. 2004). However, this is a poor substitute for observing the fundamental bands $\nu_3$ and $\nu_2$ — observable from SOFIA — as these are stronger by a factor of ~100, permitting fainter comets to be observed.

<u>Ortho-to-Para Ratio.</u> The water molecule consists of two distinct nuclear spin species, ortho- and para-$H_2O$. The ortho- to para- abundance ratio (dubbed "OPR") can be related to a nuclear spin temperature ($T_{spin}$), that measures the water formation temperature (Bonev et al. 2007 and references therein) (Figure 5-3). There is a wide range of formation temperatures, extending to > 50 K for Comet Wilson, a dynamically new comet. Moderately high values of $T_{spin}$ (> 40-50 K) are found in both nearly-isotropic and ecliptic comet populations. Moderate spectral resolution Spitzer results for Comet K4 are shown in Figure 5-3 (right panel), along with theoretical predictions of narrow actual (unconvolved) spectral line profiles. The much higher spectral resolving power of EXES on SOFIA will permit robust measurements of individual rotational temperatures for ortho- and para-$H_2O$ isomers, and thus derivation of more accurate nuclear spin temperatures. Moreover, EXES can extend such measurements to the ortho-para-meta spin isomers of $CH_4$ that cannot be sampled from the ground except for comets with rare, very large Doppler shifts.

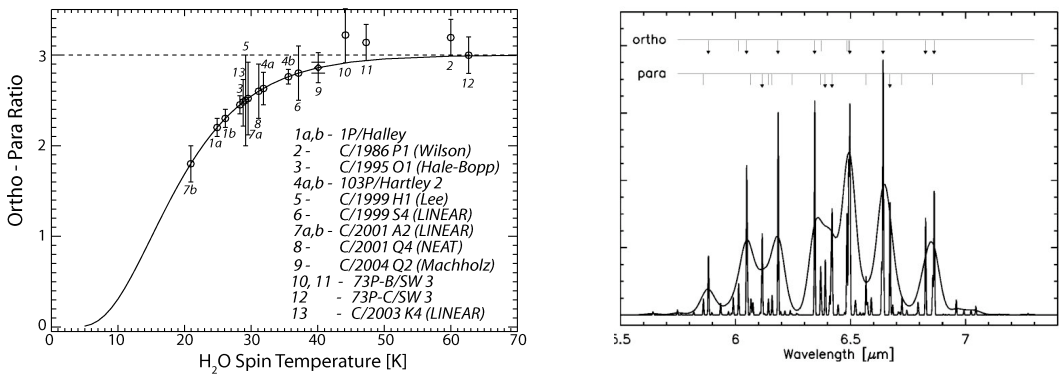

***Figure 5-3.*** *(Left): Ortho-para ratios for $H_2O$ in comets (Bonev et al. 2007) placed on a theoretical curve connecting them to the corresponding formation temperature. (Right): The 6.5 µm $H_2O$ band in comet K4, both fully resolved and also convolved to the resolution of Spitzer (Woodward et al. 2007). Ortho and para lines are indicated. EXES on SOFIA would provide major improvement in the detection limits by resolving lines of each spin isomer, and eliminating spectral confusion from interloping lines.*

D/H Ratio. The water D/H ratio measured in comets Halley, Hyakutake, and Hale-Bopp is enriched by a factor of two relative to terrestrial water, indicating ion-molecule chemistry at temperatures below 30 K. That cometary D/H ratio has been used to argue against supply of Earth's oceans by comets of this type, presumably formed in the Kuiper Belt. However, the ratio $HDO/H_2O$ should be lower in comets formed at higher temperatures near Jupiter and Saturn, or from water diffused or advected outward from the terrestrial planets region, and such bodies could have supplied Earth's water. Measurements of the D/H ratio in future bright comets can test this hypothesis. HDO has a strong fundamental band ($\nu_2$) near 7.1 $\mu$m that is inaccessible from the ground but accessible to EXES on SOFIA. Such comets are usually targets-of-opportunity that could require observations by SOFIA from remote locations and at small solar elongation.

Oxygen Isotopes. For many comets, SOFIA's CASIMIR can access several different lines of $H_2^{18}O$, which is optically thin and provides a measurement of the total water production rate complementary to EXES measurements (Bensch & Bergin 2004). Moreover, CASIMIR can observe the ratio of $^{18}OH$ to $^{16}OH$, providing direct measurements of the $^{18}O/^{16}O$ ratio in the comet formation region (Bergin et al. 2008), yielding an independent measure of comet formation temperatures (Yurimoto et al. 2007).

*Organics in Comets.* Comets are rich in diverse classes of organic species (Bockelée-Morvan et al. 2004). Based on their parent volatile compositions, three comet groups are emerging: organics-enriched, -normal, and -depleted. The discovery of organics-enriched comets that also exhibit very low nuclear spin temperatures suggests that this organic material formed at very low temperatures, perhaps even from remnants of the natal interstellar cloud core. Also, the discovery of a class of comet with drastically depleted organic composition (Mumma et al. 2001; Villanueva et al. 2006) argues that icy planetesimals which formed in the Jupiter-Saturn feeding zones of the protoplanetary disk represent a distinct population the members of which were emplaced in both the Oort Cloud and the Kuiper Belt, as suggested by the Nice Model, and which might have D/H ratios closer to terrestrial values.

For comet observations, SOFIA affords unique advantages over ground-based observatories and spacecraft in terms of duration, mobility, and flexibility. Based on recent experience, we expect that comets suitable for characterization will be discovered at a rate of 1-3 per year, bringing the total prospects to as many as ~60 comets in SOFIA's two decades of operation. This extended time span maximizes our chances of observing a rare comet "new" to the inner Solar System, even if its optimum visibility is in the southern hemisphere.

SOFIA's ability to observe at small solar elongation angles, and during daytime, is particularly important for studies of comets, which are generally brightest when close to the Sun. In contrast, ground-based observations are often limited or impossible when comets are at their most productive phase. Table 5-1 lists two comet apparitions during the first few years of SOFIA's scientific operations. The "Figure-of-Merit" is an estimate of spectral line detectability based on each comet's respective volatile production rate and the comet-Earth-Sun geometry during the indicated perihelion passage.

**Table 5-1. 2011 – 2013 Known Comet Targets**

| COMET | DATE | $Q_1$ $10^{29}$ s$^{-1}$ | $\Delta$ AU | R AU | $\Delta$-dot km/s | Figure-of-Merit |
|---|---|---|---|---|---|---|
| 45P/H-M-P | September 2011 | 0.072 | 0.70 | 0.96 | +38 | 0.12 |
| 2P/Encke | October 2013 | 0.04 | 0.54 | 0.69 | +21 | 0.30 |

$Q_1$ – Maximum release rate of water molecules per second   $\Delta$ – Minimum distance from Earth
R – Distance from Sun at closest approach to Earth   $\Delta$-dot – Velocity relative to Earth at closest approach

SOFIA offers many advantages over current and future facilities for comet studies, and promises to extend the frontiers significantly during the coming two decades. SOFIA will characterize dozens of comets from several different dynamical families, going anywhere on Earth to obtain the most favorable geometry.

*SOFIA's ability to observe at small solar elongation angles, and during daytime, is particularly important for studies of comets, which are generally brightest when close to the Sun.*

### 5.2.2 Trans-Neptunian Objects, Centaurs, and Asteroids

Aside from comets, the tribes of primitive bodies include the asteroids of the inner Solar System, plus groups of outer Solar System dynamical families including Centaurs (10-30 AU), Kuiper Belt Objects (KBOs; 30-40 AU), and even more remote "scattered disk" objects (SDOs). SDOs and KBOs are collectively referred to as Trans-Neptunian Objects or TNOs, and known objects now number over 1000. These classes of primitive bodies are notable in their diversity — meaning that large samples need to be observed to properly characterize typical properties and ranges of properties. SOFIA's contribution to TNO studies comes mainly from its ability to observe stellar occultations.

*Atmospheres of TNOs.* As a planetary body with even a very tenuous atmosphere passes in front of a background star, refraction in that atmosphere, the presence of aerosols or dust particles, and variation in gas temperature with altitude can all be discerned from the light curve. Measurements of stellar occultations by KBOs observed simultaneously with two SOFIA instruments, HIPO and FLITECAM, can probe for atmospheres with surface pressure as small as ~0.1 $\mu$bar, comparable to the atmospheres of Pluto and Triton.

A small object's occultation shadow rarely crosses the Earth at the location of a large ground-based telescope, but SOFIA can be positioned almost anywhere, free from clouds and scintillation noise, with a large telescope and optimized high-speed photometers operating in several colors to give the maximum achievable spatial resolution. It is estimated that SOFIA can capture 30 or more TNO occultations over its lifetime. Moreover, SOFIA can observe occulted and occulting objects a few days before the event to precisely determine trajectory parameters and be able to guarantee flying in the shadow track center when the moment comes.

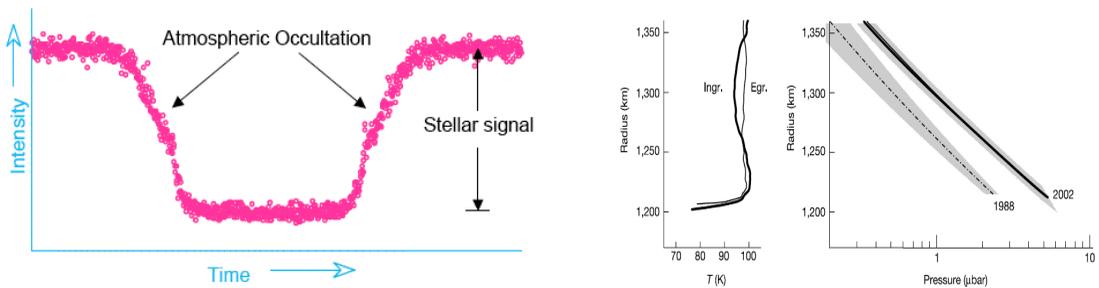

*Figure 5-4.* KAO photometric measurement of a star occulted by Pluto. The gradual transitions on either side of the occultation indicate the presence of the dwarf planet's atmosphere (from Eliot et al. 1989). b: Small irregularities in Pluto's occultation light curves can be analyzed to provide detailed information about atmospheric temperature and pressure profiles (from Sicardy et al. 2003).

> SOFIA's contribution to TNO studies comes mainly from its ability to observe stellar occultations.

SOFIA will greatly expand our knowledge of several compositionally diverse populations of primitive bodies — determining comet mineralogy, water content, and organic content, and also constraining trace atmospheres and densities of dozens of TNOs, Centaurs, and asteroids. SOFIAs unique observations will test recent Solar System formation and evolution theories that predict substantial radial mixing of these dynamical populations.

### 5.3 Extrasolar Planetary Material

With high-resolution spectroscopy over a wide wavelength range, SOFIA can build on Spitzer's legacy of mid-infrared spectroscopy of unresolved material around solar-type stars, connecting our understanding of the composition of primitive bodies in extrasolar systems to our Solar System.

Most, if not all, solar-mass stars are born with circumstellar disks. These "primordial" disks evidently form planetesimals and planets in some cases, and are finally cleared out by photo-evaporation, radiation pressure, and other effects within a few Myr after their formation

Debris disks, in contrast, are features of older systems in which planet formation has finished or nearly finished, but significant amounts of relatively short-lived second generation dust are generated by collisions and sublimation of remnant planetesimals (e.g., reviews by Backman & Paresce 1993, Meyer et al. 2007). The Solar System's Kuiper Belt and Asteroid Belt can be considered old, low-density remnant debris disks, with 99% or more of their original mass lost, continuing to create dust via collisions and comet activity, which in turn are the results of planetary perturbations. SOFIA can target dozens of debris disks discovered photometrically by Spitzer but that were not studied spectroscopically, covering a broad wavelength range at high spectral resolution. Figure 5-2 above illustrates the type of spectra SOFIA can be expected to return regarding Solar System primitive bodies. Figure 5-5 compares mid-IR spectra of two comets, some terrestrial minerals, and various examples of extrasolar circumstellar dust.

> SOFIA's mid- and far-IR spectroscopic capabilities can support detailed mineralogical analyses and comparison of extrasolar debris disks containing dust produced by planetesimal collisions and sublimation, with Solar System primitive bodies.

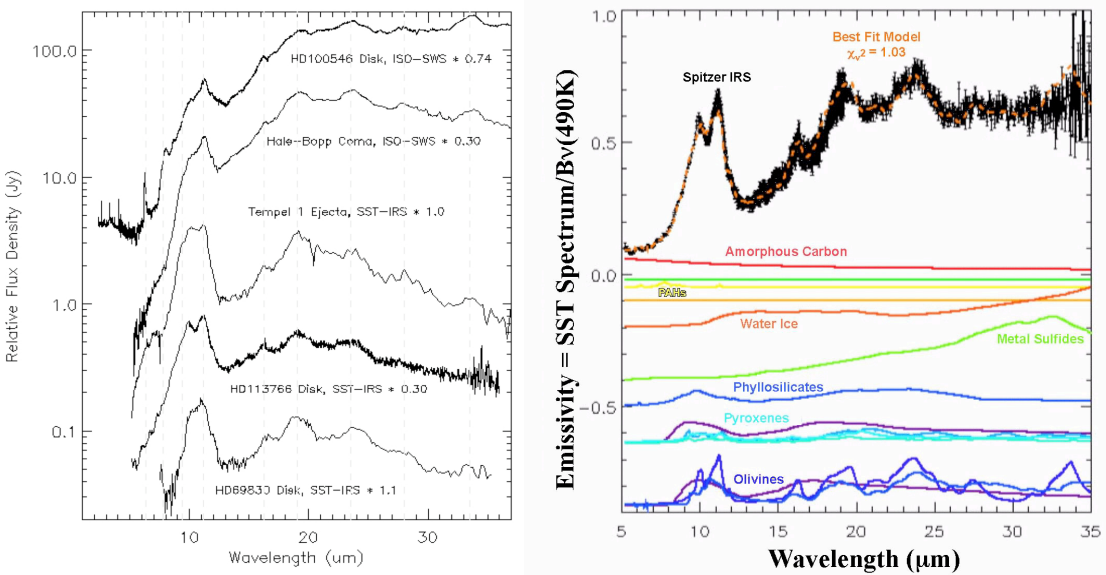

***Figure 5-5.*** *a. Comparison of dust emission spectra from two Solar System comets and a variety of extrasolar protoplanetary or planetary debris disk systems: HD 100546, a Herbig A0V star, a younger version of the Vega system; HD 113766, a 12 Myr-old F5 main sequence star; and HD 69830, a nearby mature K0V star, an older version of epsilon Eridani. Note the similar emission features at common wavelengths in the spectra, despite the highly diverse source environments. b. Detailed mineralogical analysis of the Spitzer spectrum of one of the circumstellar dust disks in panel (a), HD 113766. The star is encircled by at least a Mars-mass of terrestrial planet construction material within its "habitable" (liquid water-temperature) zone, plus two belts at greater radii dominated by water ice-laden dust. (Both panels after Lisse et al. 2008.)*

SOFIA's 1st-generation spectrographs can observe segments of the mid-IR containing important debris disk mineralogy diagnostics. As with solar system miner-alogy, there is need for an advanced SOFIA instrument capable of covering the entire mid-IR (5 to 40 $\mu$m) at resolution R ~ 1000 in a single observation.

SOFIA will build on Spitzer's discoveries, adding significantly to the number of spectroscopically well-studied debris disks, while extending measurements to longer wavelengths. SOFIA's mid- and far-IR spectroscopic capabilities can support detailed mineralogical analyses and comparison of extrasolar debris disks containing dust produced by planetesimal collisions and sublimation, with Solar System primitive bodies.

## 5.4 Giant Planets

The properties of our Solar System's giant planets, including cloud formation and atmospheric dynamics, serve as ground truth for interpreting necessarily cruder observations of the ever-growing number of extrasolar planets.

### 5.4.1 Global Studies — Bulk Composition and Dynamics

*SOFIA can target dozens of debris disks discovered photometrically by Spitzer but that were not studied spectroscopically.*

Observations of outer planet SEDs across the far-IR and sub-mm range of their peak thermal emission are key to determining their bulk compositions, thermal structures, and internal fluid dynamics. Just as for stars, the departure of a gas planet's emitted thermal spectrum from a perfect blackbody can be used to determine the temperature versus optical depth in the atmosphere. Surprisingly, the giant planets' SEDs remain incompletely determined in spite of multiple spacecraft investigations and decades of ground-based observations (Atreya et al. 1999; 2003).

SOFIA can contribute to the vexing question of the H/He ratio in the giant planets, a problem that remains unsettled even after several spacecraft flybys and the Galileo entry probe into Jupiter's atmosphere. Observations with FIFI LS (42-210 $\mu$m), calibrated by FORCAST and HAWC broadband photometry between 5 and 215 $\mu$m, will give complete SEDs from which temperatures may be deduced from the upper troposphere down to several bars of pressure. Complete and consistent coverage of this spectral range allows determination of the ortho-para hydrogen ratio and H-He abundance.

Neptune's SED, and by implication its atmospheric structure, have varied over the last two decades, perhaps as a seasonal response (Figure 5-6). Major episodic changes of Neptune's near-infrared spectrum have also been seen (Joyce et al. 1977). Observations by SOFIA over the next two decades could resolve this puzzle.

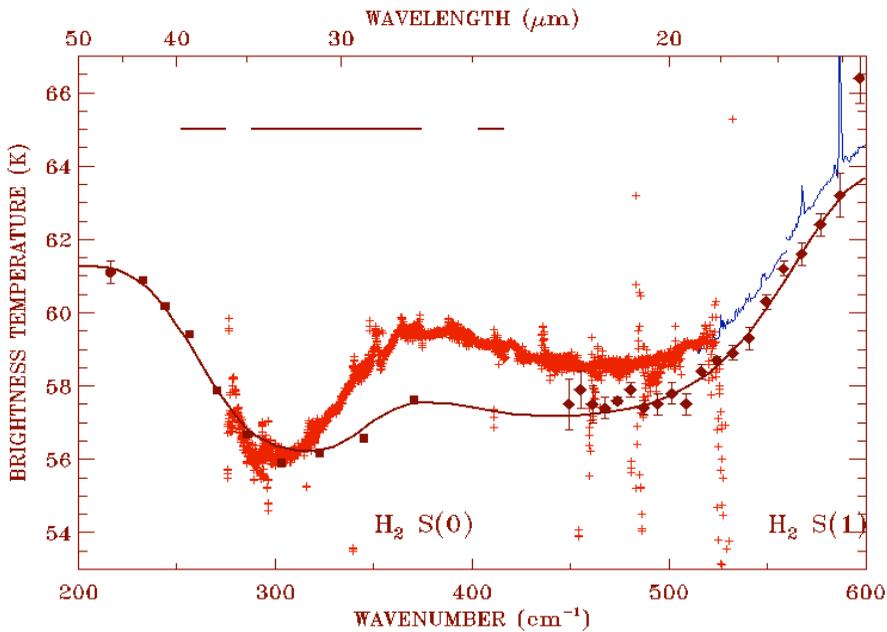

***Figure 5-6.*** *Spectrum of Neptune in the difficult and important mid-IR region unique to SOFIA. The locations of the emission cores of the broad $H_2$ collision-induced S(0) and S(1) rotational lines are also indicated. Spitzer IRS LH spectral data are red crosses, and SH spectra are the blue lines. A model fitting ground-based data from the 1980s (Orton et al. 1987, diamonds) plus ISO LWS (filled circle) and SWS (filled boxes) data are also shown. Spectral ranges covered by SOFIA FORCAST's 38.0, 30.0, and 24.4 µm broadband filters are shown schematically at the upper left.*

### 5.4.2 Atmospheric Chemistry

SOFIA's high-resolution spectrometers (CASIMIR, GREAT and EXES) enable investigation of the global chemical inventory of all the gas giants, with especially good spatial resolution for Jupiter and Saturn. SOFIA's GREAT spectrometer can provide far better spectral resolution on all gas giants in the 60-180 µm range than *Herschel*'s PACS, and CASIMIR will provide unique observations of Jupiter and Saturn at 200-600 µm. The 120-200 µm range will be covered for Uranus and Neptune only by SOFIA's FIFI LS and SAFIRE. While *Herschel*'s HIFI is able to cover the 200-600 µm region at very high spectral resolution, it will be focused on water, so SOFIA observations of other molecular species and wavelength ranges will be unique. FORCAST plus grism will provide much higher spectral resolution than *Herschel* through the mid-infrared. EXES will provide high-spectral resolution capabilities over selected spectral regions between 5 and 28 µm. SOFIA can thus improve on *Herschel*'s incomplete wavelength and species coverage, providing systematic observation of all four giant planets across most of their thermal SEDs at high spectral resolution.

Although we know that complex photochemistry takes place in the atmospheres of the giant planets, the relevant wavelengths are mostly inaccessible from Earth. Among the key opportunities in SOFIA's accessible spectral range for investigations of organic chemistry would be detection and measurement of spectral lines from species such as benzene ($C_6H_6$) and propane ($C_3H_8$) in Uranus and Neptune which are otherwise buried among acetylene ($C_2H_2$) lines, and the verification or identification of isotopes of species such as diacetylene ($C_4H_2$) and methyl acetylene / propyne ($C_3H_4$) (see section 5.6 - Titan: a Pre-biological Organic Laboratory on page 5-17). Lines in the 50-500 $\mu$m range can be detected from known constituents such as $H_2O$, CO, $NH_3$, $PH_3$ and HD, together with possible so-far undetected lines of HF, HBr, HI, InBr, $AsF_3$, $BgH_3$, HCP and $H_2Se$ (Bezard et al. 1986). Only upper limits to several of these species have been obtained for the atmospheres of Jupiter (Fouchet et al. 2004) and Saturn (Teanby et al. 2006). Several of these species (phosphorus compounds in particular) have been proposed as potentially important opacity sources, perhaps capable of resolving the global spectrum paradoxes mentioned above, and others are tracers of non-equilibrium processes, perhaps manifestations of upwelling flows or even micrometeoroid bombardment. SOFIA's spectroscopic sensitivity in the relevant wavelength ranges will be better than ISO's, with which these searches were done originally.

*SOFIA's unique capabilities of wavelength coverage, high spatial resolution, and long duration will open new windows of understanding of the giant planets through studies of their atmospheric compositions and structures, and temporal variability, both seasonal and secular.*

### 5.4.3 Spatial and Temporal Variations

Long-term spatially-resolved monitoring of the para- *vs.* ortho-$H_2$ ratio with the help of FORCAST imaging observations at 24.5, 30, and 38 $\mu$m (Figure 5-7) would enable an assessment of spatial/seasonal variability in atmospheric structure, horizontal and vertical winds, and heat flow over the giant planets' disks. SOFIA's longevity means that variation in atmospheric properties of the outer planets can be observed over decades. This will be particularly useful for Saturn, Uranus, and Neptune which all undergo substantial changes in Earth and Sun-facing geometries. Uranus has no measurable internal heat escaping, unlike the other three giant planets.

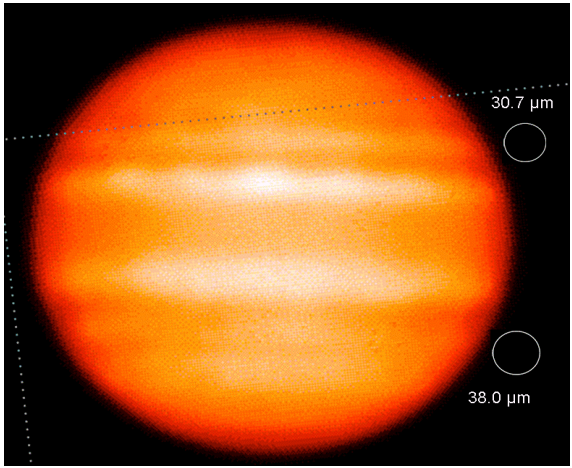

*Figure 5-7.* 24.5 µm image of Jupiter from the NASA IRTF (2008 Aug. 8), showing variability of thermal emission across the disk. Diffraction-limited resolution for FORCAST 30.0 and 38.0 µm images are indicated schematically. These are sufficiently small to resolve major banded structure and large features. Jupiter's diameter viewed from Earth ranges from 35 to 42 arc seconds.

Is that puzzling property a long-term characteristic of the planet, perhaps connected to the catastrophe that turned Uranus on its side, or will decade-scale seasonal atmospheric changes "uncork" heat flow? SOFIA is best suited to find out.

SOFIA will observe all four giant planets across their full bolometric spectrum, giving broad SED coverage that can solve outstanding questions regarding atmospheric structure (temperature, bulk composition, opacity, vertical upwelling). Moreover SOFIA's unparalleled spectral coverage and resolution, generally exceeding any other facility, can discover and map many key molecules spatially and (via modeling of line profiles) vertically.

SOFIA's unique capabilities of wavelength coverage, high spatial resolution, and long duration will open new windows of understanding of the giant planets through studies of their atmospheric compositions and structures, and temporal variability, both seasonal and secular.

### 5.5 Venus: Earth's Neglected Sibling

Venus, the planet most similar to Earth in bulk composition and location, has an atmosphere with chemistry and dynamics that are poorly understood. Investigating the current characteristics and history of Venus's atmosphere bears directly on understanding Earth's corresponding properties.

Venus may have initially had as much water as Earth, but atmospheric D/H ratios suggest that it subsequently lost its oceans to a fierce runaway greenhouse effect. The hydrogen was probably lost to space, but the fate of the oxygen remains unknown. Understanding Venus's initial conditions, and how this loss occurred, can provide crucial clues for unraveling the formation and evolution of Earth and other terrestrial planets. Insight can also be gained regarding the width of the "Continuously Habitable Zone" around Sun-like stars — that is, at what distances from a primary star would an originally Earth-like "water world" remain Earth-like, or evolve into a Venus? Furthermore, the hot, dry, acidic atmo-sphere of Venus may be a good analog for the atmospheres of "super-Earth" exoplanets orbiting close to their parent stars.

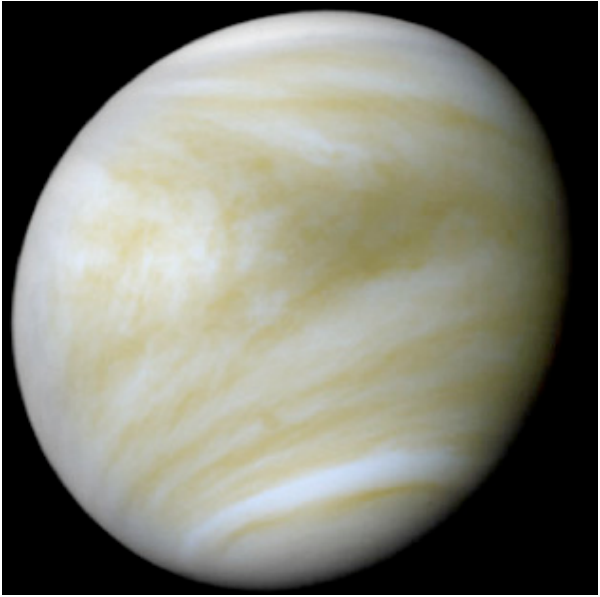

*Figure 5-8.* Pioneer Venus UV images of Venus's $H_2SO_4$ cloud layers were used to reveal temporal and spatial variations, and track wind speeds.

Venus's slow rotation leaves it in an unusual atmospheric dynamics regime, including a puzzling "super-rotating" mesosphere circling the planet in only 4 Earth days, with substantial dayside-nightside energy transport overlain by a symmetric sub-solar to anti-solar flow at high altitudes (Figure 5-8). Chemical models predict that UV insolation should decompose Venus's current predominantly $CO_2$ atmosphere into a mixture with substantial abundances of CO and $O_2$, which is not observed.

One area in which observations of Venus have been lacking, where SOFIA can play the discovery-level role of a spacecraft, is high-resolution spectroscopy. With the failure of the high-resolution (1.2 cm$^{-1}$) mid-infrared Planetary Fourier Spectrometer (PFS) aboard the otherwise highly successful Venus Express, the last spectroscopic observations of Venus at middle- to far-infrared wavelengths by a spacecraft were by a lower resolution (5 cm$^{-1}$) instrument on Venera-15 in the 1980s. The last attempts at high-resolution spectroscopy of Venus from the KAO (Aumann & Orton 1979; Aumann et al. 1982) did not have sufficiently high spectral resolution to detect individual molecular signatures. Venus was not observable by ISO and will not be observable by Herschel because of solar elongation limitations. In contrast, Venus can be observed by SOFIA for as much as six months around its maximum eastern and western elongations. Hence, SOFIA will become the facility of choice for observations in spectral regions unavailable from the Earth at least until another spacecraft with a mid- to far-infrared high-resolution spectrometer visits Venus — and none is currently approved to do so.

*Atmospheric structure.* Pioneer Venus and Venus Express mapped the high-level $H_2SO_4$ cloud layer in the UV, using stable spatial variations in the clouds to derive wind speeds and constrain the mesospheric super-rotation, many aspects of which are still not understood. SOFIA's EXES spectrometer can follow up by observing the $v_2$ $CO_2$ band at wavelengths of 11-13 $\mu$m and 16-18 $\mu$m, outside the half-power points of the corresponding telluric band, sampling the temperature and wind speed profiles in 5 km vertical increments (determined by the spectral resolution) through Venus's entire middle atmosphere (50-90 km altitudes). The 70-90 km portion, which has never been studied, lies between the observable cloud tops and the higher altitude regions accessible via observations of CO rotational lines at microwave wavelengths. With its good spatial resolution of about 2 arc seconds at 18 $\mu$m, SOFIA data can be used to study diurnal latitudinal and longitudinal variations in atmospheric structure and infer wind structure. Very high-resolution heterodyne spectroscopy with CASIMIR, which provides good altitude resolution, can extend ground-based observations of $^{12}$CO and $^{13}$CO to stronger lines which sample higher altitudes and lower abundances. The 90-100 km altitude range that can be studied in this way is the transition region between Venus's two different atmospheric dynamics regimes. This altitude is also the location of a controversial possible temporally variable warm layer (Bertaux et al. 2007; Clancy *et al.* 2008) that may result from localized heating correlated with intense infrared airglow in this altitude range (Crisp et al. 1996; Ohtsuki et al. 2005; Bailey et al. 2008).

> *SOFIA observations of Venus will address critical and unsolved questions about the planet's atmospheric structure, dynamics and variable composition, using different wavelengths to probe all levels of the atmosphere.*

***Chemical and isotopic composition.*** Key goals for spectroscopy of Venus include measurement of a number of molecules of interest to atmospheric chemistry analyses that have absorption and emission lines throughout the mid- and far-IR spectral regions accessible only to SOFIA. Some of these molecules may act as catalysts, controlling chemical pathways, atmospheric evolution, and atmospheric escape. Chlorine, for instance, is expected to be an important catalyst. HCl is known to be present in Venus's atmosphere, and modeling suggests a range of chlorine oxides should be formed from photo-dissociation of HCl. These chlorine oxides may help answer the question of why $CO_2$ appears to be even more stable against ultraviolet photo-dissociation on Venus than it is on Mars. As mentioned above, it remains a major puzzle why the chief constituent of Venus's atmosphere does not break down under solar UV irradiation into a mixture of mostly CO and $O_2$ (Mills & Allen 2007).

Isotopic ratios in key species such as H and O can help constrain how Venus's oceans were lost. The most important molecules for atmospheric escape are $H_2O$ and HDO, the latter of which is hundreds of times more abundant in Venus's than in Earth's atmosphere. Deuterated species (e.g., DCl) readily observable by SOFIA provide diagnostics of atmospheric chemical pathways and also an independent measurement of the D/H ratio. DCl is potentially observable by Venus Express, but the only DCl lines that it can observe at its wavelengths of observation are blended with strong $CO_2$ lines.

SO and $SO_2$ abundances, critical to understanding the cycle which maintains Venus's mesospheric sulfuric acid haze, show diurnal, temporal, and vertical variations in sparse ground-based 200-300 GHz observations (Sandor et al. 2008). In addition, Venus Express measurements show that thermospheric temperatures and $SO_2$ abundances at the cloud tops have changed dramatically since Pioneer Venus (Bertaux et al. 2007; 2008). Venus Express also found large abundances of $SO_2$ at 100-110 km altitude (Bertaux et al. 2008) which greatly exceed both model predictions and upper limits from ground-based observations (Sandor et al. 2008). Both sets of Venus Express observations suggest strong vertical mixing; if confirmed, such a discovery would be important for understanding the atmospheric sulfur cycles that maintain the clouds of Venus. SOFIA (CASIMIR, GREAT) can observe vertical profiles and temporal variations of these important molecules, improving on ground-based observations by accessing stronger lines of SO and $SO_2$ which provide more sensitive probes. SOFIA's long mission duration is important for continued surveillance of processes that might have decade timescales.

By circumstance, Earth's sister planet has never been thoroughly explored with broadband, high-resolution spectroscopy, leaving fundamental questions about the atmospheric composition and evolution. SOFIA can play the role of a Venus-focused spacecraft with the potential for discovery-level science regarding atmospheric dynamics and the atmospheric chemical network, via its ability to map lateral, vertical and temporal variations in composition, temperature and associated wind structure, above, within and below the visible haze layer.

### 5.6 Titan: a Pre-biological Organic Laboratory

Titan has long been a target of central interest from the standpoint of organic chemical evolution. Its low abundance of atmospheric $H_2$ (~0.1%) allows chemical pathways to proceed to great complexity that cannot be reached in the atmospheres of the gas giant planets. In spite of many new results obtained by the Cassini orbiter and Huygens probe, there are a number of missing key measurements that SOFIA is in the best position to make. SOFIA has several advantages over Herschel, Spitzer and ground-based telescopes for Titan studies: SOFIA's wavelength range is broader and captures more molecules, especially at short IR wavelengths and outside terrestrial atmospheric windows (*e.g.,* CO, $CH_4$, $C_2N_2$, $C_4N_2$, $HC_3N$), it allows for the detection of more lines per molecule, and it has a much longer mission duration.

During a 20-year span the Saturn system undergoes more than half of a seasonal cycle (northern solstice to southern solstice). SOFIA can observe Titan's full seasonal and latitudinal variability, connecting Cassini observations (and Voyager 1980, ISO 1997) with possible future missions. Cassini has taught us that the Saturn system is temporally and seasonally variable. Herschel will observe Titan, but will focus on water, and is limited in wavelength coverage and duration.

***Atmospheric chemistry.*** The main novel results that SOFIA can contribute to Titan science are breakthroughs in atmospheric composition and chemistry from high-spectral resolution far-infrared (GREAT and CASIMIR) and mid-infrared (EXES) spectroscopy. Of particular importance in the sub-millimeter spectral range are nitriles and heavy-C hydrocarbons, which Cassini's mass spectrometer INMS has shown to exist in the ionosphere up to its mass limit of 6 or 7 carbon atoms. Observations of Titan with ISO spectral resolution of a few times $10^3$ allowed the first detections of $H_2O$ and $C_6H_6$ (Coustenis et al. 1998; 2003). Benzene has also been confirmed in Cassini/CIRS data, albeit with low S/N ratio that doesn't allow precise determination of its abundance (Coustenis et al. 2007). Thus, SOFIA's ability to achieve R ~$10^4$ – $10^7$ at comparable sensitivity can be expected to allow detection of new molecular and isotopic species.

Lines of key molecules including $CH_4$, CO, and HCN — the starting links in the organic chemical evolution chain — are observable by SOFIA in the sub-millimeter range not covered by ISO, and recorded but with poor S/N by Cassini. SOFIA's broad wavelength range will allow these molecules to be sampled in different resolved lines, e.g., at 191 $\mu$m (1569 GHz), constraining the thermal profile and vertical distribution in novel altitude regions. Several hours' integration will be sufficient to assess the rotational lines of $CH_4$ with GREAT and CASIMIR. The outcome of these complementary studies will be the exploration of a rich spectral region as yet unexploited at Titan. Also, studies of isotope ratios in C, N, and O, bearing on the formation of Saturn, Titan, and the Solar System, can be done uniquely with this high spectral resolution and wavelength coverage.

Simulations of the signatures of heavy nitriles ($CH_3CN$, etc.) indicate that they can be detected for the first time in the 250-550 $\mu$m (1200-550 GHz) range with CASIMIR and hence provide information on the maximum degree of complexity achieved in Titan's organic chemistry.

The far-infrared and sub-millimeter wavelength coverage of the SOFIA instruments enable the search for and quantitative studies of organic and other molecules in Titan's rich and evolving atmosphere beyond the capabilities of spacecraft and ground-based observatories.

> *The far-infrared and sub-millimeter wavelength coverage of the SOFIA instruments enable the search for and quantitative studies of organic and other molecules in Titan's rich and evolving atmosphere beyond the capabilities of spacecraft and ground-based observatories.*

In the mid-IR range, complementing ground-based Cassini and ISO observations, EXES can be used to search for complex hydrocarbons and nitriles in Titan's stratosphere, benefiting from the lack of atmospheric interference by telluric $H_2O$ and $CO_2$. This gives access to windows unattainable from the ground in which organics predicted by models and laboratory experiments, such as $C_6H_2$, $C_8H_2$, $C_5H_2$, $C_4H_4$, $CH_2CHCN$, $CH_3CH_2CN$, and many others, remain to be seen (Coustenis et al. 2003 and references therein) (Figure 5-9).

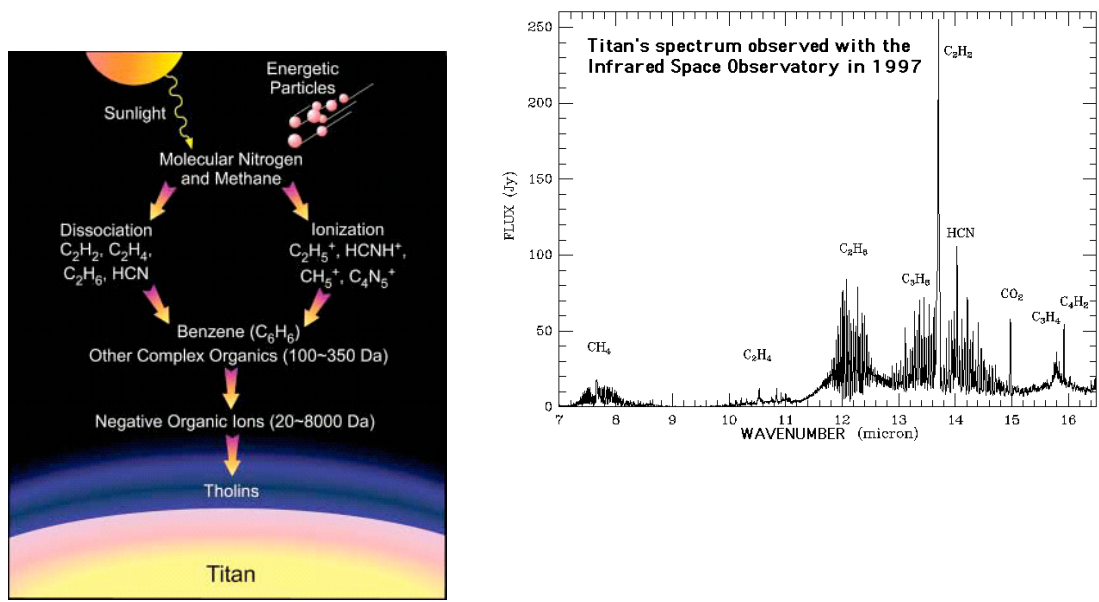

***Figure 5-9.*** *(left) Complex organic chemistry hypothesized to occur in Titan's atmosphere (courtesy NASA/JPL). (right) Data from ISO SWS with resolution R ~1600-2000 (from Coustenis et al. 2003). SOFIA's EXES instrument would be able to search for pathway-critical species such as $CH_3$ (16.5 μm), $C_6H_2$ (16.1 μm) and Crotonitrile (13.7 μm) predicted by models.*

Because of its higher sensitivity and spectral resolution, SOFIA can greatly extend and enhance existing ISO and Cassini observations (which have lower spectral resolution) and also Herschel observations that will be limited in wavelength range and time. Thanks to its long operational lifetime, SOFIA may be a bridge to future spacecraft exploration of the Saturn system. Major atmospheric constituents such as $CH_4$, CO and HCN can be studied and monitored. High molecular weight hydrocarbons and nitriles only hinted at by Cassini observations, or seen only in the laboratory, may be observed directly, and their globally averaged vertical distributions inferred. SOFIA's long mission lifetime will also contribute to monitoring secular and seasonal atmospheric variations, including the methane "monsoon" cycle, over a large fraction of Saturn's 29-year orbital period.

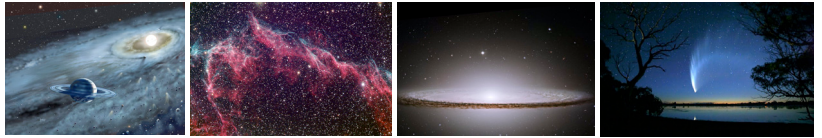

# APPENDIX A — Acronyms and Terminology

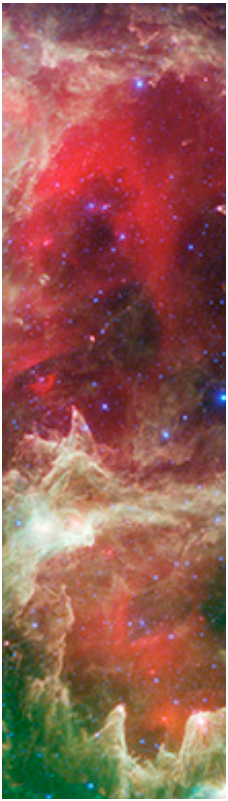

**Table A-1. Acronyms and Terminology**

| Acronym/Term | Description |
|---:|---|
| Accretion Disk | The whirling disk of gas that forms around a compact object such as a white dwarf, neutron star or black hole as matter is drawn in. |
| AGB | Asymptotic Giant Branch stars |
| AGN | Active Galactic Nuclei. A class of galaxies which spew massive amounts of energy from their centers, far more than ordinary galaxies. Black holes may lie at the center of these galaxies. |
| AKARI | Japanese satellite dedicated to infrared astronomy from the Institute of Space and Astronautical Science (ISAS) of the Japanese Aerospace Exploration Agency (JAXA). Formerly ASTRO-F. |
| ALMA | Atacama Large Millimeter/submillimeter Array located in Chile. |
| ASTRO-F | See AKARI. |
| AU | Astronomical Unit roughly equal to the mean distance between the Earth and the Sun. It is approximately 150 million kilometers (93 million miles). |
| BIB | Blocked Impurity Band detector |
| CARMA | Combined Array for Research in Millimeter-wave Astronomy |
| CASIMIR | CAltech Submillimeter Interstellar Medium Investigations Receiver |
| C-F | Chandrasekhar-Fermi method for evaluation of the mean magnetic field strength in molecular clouds. |
| CIB | Cosmic Infrared Background |
| CMZ | Central Molecular Zone (of the Milky Way galaxy) |
| CNM | Cold Neutral Medium |
| D | Interstellar Deuterium |
| DAOF | Dryden Aircraft Operations Facility. Flight operations will be conducted out of NASA Dryden Flight Research Center's Aircraft Operations Facility in Palmdale, CA. |
| DLR | Deutsches Zentrum für Luft-und Raumfahrt |
| DSI | Deutsches SOFIA Institut |
| ES | Early Science |
| EXES | Echelon-Cross-Echelle Spectrograph |
| FFI | Fine Field Imager (see FPI). |
| FIFI LS | Field Imaging Far-Infrared Line Spectrometer |
| FIR | Far InfraRed |
| FLITECAM | First Light Infrared Test Experiment CAmera |
| FORCAST | Faint Object InfraRed CAmera |
| FOV | Field of View. SOFIA's 8 arcminute diameter field of view (FOV) allows use of very large format detector arrays. |
| FPI | Focal Plane Imager. An optical focal plane guiding system. Independent of the FPI there are two other optical imaging and guiding cameras available: a Wide Field Imager (WFI) and Fine Field Imager (FFI). Both of these cameras are attached to the front ring of the telescope. |
| FSIs | Facility Class Science Instruments: HAWC, FORCAST, and FLITECAM. |

**Table A-1. Acronyms and Terminology**

| Acronym/Term | Description |
|---:|---|
| G | Interstellar UV radiation field |
| GAIA | Satellite instrument designed to produce a three-dimensional map of the stars in the Milky Way galaxy. Named for the Greek Earth goddess, Gaia. |
| GC | Galactic Center |
| GI | General Investigator |
| GLIMPSE | Galactic Legacy Infrared Mid-Plane Survey Extraordinaire (Spitzer) |
| GREAT | German Receiver for Astronomy at Terahertz Frequencies |
| HAWC | High-resolution Airborne Wideband Camera |
| HD | Deuterated molecular hydrogen |
| HDO | Hydrogen Deuterium Oxide |
| HIFI | Heterodyne Instrument for the Far Infrared (Herschel). A high-resolution spectrograph that operates in the range of 480 to 1250 GHz in five bands and 1410 to 1910 GHz in two additional bands. |
| HIPO | High-speed Imaging Photometer for Occulation |
| IPHAS | Isaac Newton Telescope Photometric Hydrogen-alpha Survey of the northern galactic plane |
| IR | Infrared |
| IRAC | Infrared Array Camera (Spitzer) |
| IRAS | InfraRed Astronomical Satellite (NASA Explorer mission) |
| IRDC | Infrared-dark Clouds |
| ISM | Interstellar Medium |
| ISO | Infrared Space Observatory (European Space Agency) |
| JWST | James Webb Space Telescope |
| KAO | Kuiper Airborne Observatory |
| KBOs | Kuiper Belt Objects |
| KID | Kinetic Induction Detector Spectrometer (future instrument) |
| LHB | Late Heavy Bombardment |
| ly | Light Year. The distance light travels in a vacuum in one Julian year. It is equal to just under ten trillion kilometers. |
| MDCF | Minimum Detectable Continuum Flux |
| MIPS | Multiband Imaging Photometer for Spitzer. Provides the Spitzer Space Telescope with capabilities for imaging and photometry in broad spectral bands centered nominally at 24, 70, and 160 $\mu$m. |
| OC | Oort Cloud. A widely accepted theory that the Sun is surrounded by a distant cloud of comet matter called the Oort Cloud, bits of which are occasionally hurled into the solar system as comets. |
| ODIN | Swedish dual disciplinary (astrophysics and atmospheric science) spacecraft |
| PACS | Photodetector Array Camera and Spectrometer (Herschel). A bolometer array photometer and a photoconductor array imaging spectrometer operating at a wavelength range between 60 and 210 $\mu$m. |
| PAHs | Polycyclic Aromatic Hydrocarbons |

**Table A-1. Acronyms and Terminology**

| Acronym/Term | Description |
|---:|---|
| pc | Parsec (parallax of one arcsecond). The distance at which 1 AU length would subtend one second of arc. This is equal to approximately 3.26 light-years. |
| PDRs | Photodissociation Regions |
| PI | Principal Investigator |
| R | Spectral resolution defined as $\lambda/\Delta\lambda$. |
| RSGs | Red Supergiants |
| SAFIRE | Submillimeter And Far InfraRed Experiment |
| SDOs | Scattered Disk Objects |
| SEDs | Spectral Energy Distributions |
| SFR | Star Formation Rate |
| SIS | Superconductor-Insulator-Superconductor |
| SMA | Submillimeter Array |
| SMO | Science Mission Operations. The US FSIs will be maintained and operated by the science staff of the SMO. |
| SOFIA | Stratospheric Observatory For Infrared Astronomy |
| SPUD | Silicon Pop-Up Detector (HAWC) |
| SSC | SOFIA Science Center |
| SWAS | Submillimeter Wave Astronomy Satellite |
| TA | Telescope Assembly (SOFIA) |
| TEXES | Texas Echelon Cross Echelle Spectrograph high-resolution mid-infrared spectrograph |
| TNOs | Trans-Neptunian Objects. Scattered disk and Kuiper Belt Objects are collectively referred to as Trans-Neptunian Objects. |
| UKIDSS | UKIRT Infrared Deep Sky Survey |
| ULIRG | Ultra-Luminous Infrared Galaxy |
| USRA | Universities Space Research Association |
| VISTA | Visible and Infrared Survey Telescope for Astronomy, a 4-m class wide field survey telescope for the southern hemisphere, equipped with a near infrared camera. |
| VPHAS+ | Very Large Telescope / Survey Telescope Photometric Hydrogen-alpha Survey of the Southern Galactic Plane |
| WFI | Wide Field Imager (see FPI) |
| WIM | Warm Ionized Medium |
| WNM | Warm Neutral Medium |
| XDRs | X-ray Dominated Regions |

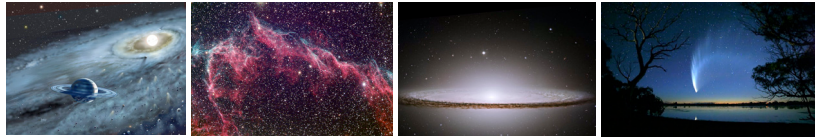

**APPENDIX B**     Additional Tables

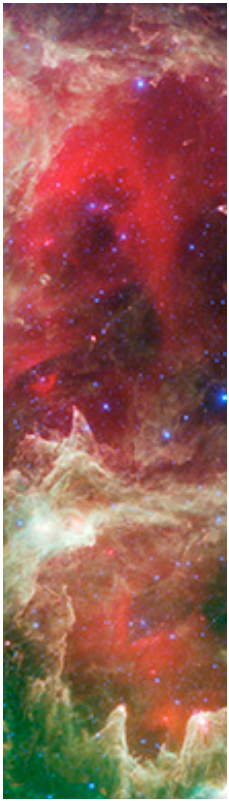

## B.1 Observing Plan Feasibility

The following tables give a sample of some of the observing parameters for a selected subset of the science to be carried out in the four science themes. In most cases these are based on the estimated sensitivities of the present instruments given in Figure 1-9 and Figure 1-10 of the Introduction. In a few cases an extrapolation was made to a future upgrade or instrument, as noted in the comment field.

Each scientific subtheme also identifies the number of targets that would need to be observed to adequately address the science described in the body of the text. The total time on target includes overheads and chopper efficiency when appropriate. The estimated sensitivity is 10 $\sigma$, and comes from previous missions such as KAO, IRAS, ISO and Spitzer, corrected for beam size and spectral resolution.

## B.2 The Formation of Stars and Planets

| Subtheme | Wavelength range ($\mu m$) | Instrument used | Sensitivity required $10\sigma$ | Spectral resolution ($\lambda/\Delta\lambda$) | Spatial resolution (arcsec) | Number of targets | Total time on targets | Comments |
|---|---|---|---|---|---|---|---|---|
| Massive Star Formation | 6 - 216 | FORCAST - HAWC | Few Jy (38$\mu$m) and few 10s Jy (100$\mu$m) | N/A | 2 - 20 | 20 | 15 hrs | SED of massive cores |
| Proto-Planetary Disks | 6 - 655 | FORCAST - FIFILS – HAWC - SAFIRE | 2Jy (10$\mu$m), 4Jy(100$\mu$m), 1Jy(450$\mu$m) | N/A | 6 - 60 | 40 | 60 hrs | SED of Herbig stars |
| Chemistry in Disks: $H_2O$ gas phase | several lines 6.6 - 7.1 | EXES | $6\times10^{-18}$ W/m$^2$ | $10^5$ | 3 | 20 | 20 hrs | Detection of water in disks. |
| Chemistry in Disks: $H_2O$ -ice band | broad feature 43 - 46 | FIFI-LS | 10 Jy (43.8$\mu$m) (Needs 15 settings) | N/A | 5 | 20 | 50 hrs | Modeling challenging due to the strong atmospheric absorption in this band. |
| Astrochemistry [OI] | 63 and 145 | GREAT | $3\times10^{-16}$ and $1\times10^{-16}$ W/m$^2$ | $10^6$ | 6 and 14 | 20 | 10 hrs | Looking for the Oxygen in cold clouds. |
| Astrochemistry Water in Absorption | several lines 5.7 - 6.2 | EXES | $2\times10^{-17}$ W/m$^2$ with 50$\sigma$ | $10^5$ | 3 | 10 | 40 hrs | Survey of bright protostars |
| Astrochemistry Water in shocks | 250 | CASIMIR | $3\times10^{-17}$ W/m$^2$ | $10^5$ | 25 | 20 | 10 hrs | p-$H_2^{18}O$ |

## B.3 The Interstellar Medium of the Milky Way

| Subtheme | Wavelength range (μm) | Instrument used | Sensitivity required 10 σ | Spectral resolution (λ/Δλ) | Spatial resolution (arcsec) | Number of targets | Total time on targets | Comments |
|---|---|---|---|---|---|---|---|---|
| Massive star Feedback: PDR astrophysics | 5 - 20 | FORCAST GRISM | 10 Jy | 200 - 1000 | 2 | 15 | 50 hrs | PAHs, 10'x10' maps, some from Spitzer |
| | 50 - 200 | HAWC | 10 Jy | NA | 5, 9, 16, 22 | 20 | 20 hrs | continuum, 10'x10' maps |
| | 40 - 210 | FIFI-LS | $5 \times 10^{-18}$ W/m² to $10^{-13}$ W/m² | 2000 | 12, 24 | 20 | 80 hrs | lines, 10'x10' maps |
| | 63 - 600 | GREAT/CASIMIR | | $10^5$ | 6 - 60 | 20 | 20 hrs | line flux for OI (63, 146 μm), CII, CO, points |
| Origins of Dust: Evolved Stars | 5 - 40 | FORCAST GRISM | 1 Jy | 200-1000 | | 100 | 50 hrs | compact or point sources |
| | 40 - 100 | FIFI-LS | 20 Jy | 2000 | N/A | 20 | 100 hrs | full spectral coverage for FIFI-LS |
| | 40 - 100 | New instrument | 1 Jy | 200 | | 100 | 50 hrs | new full spectral coverage instrument is more efficient |
| Identification of PAHs | 5 - 40 | FORCAST GRISM | 100 Jy | 200 - 1000 | Matched to beam size at longest wavelengths ~24 | 20 | 10 hrs | Much better done with a new instrument for improved systematics |
| | 40 - 210 | FIFI-LS | 100 Jy | 2000 | | 20 | 100 hrs | |
| | 60 - 600 | GREAT/CASIMIR | $5 \times 10^{-15}$ W/m² | $10^5$ | | 10 | 20 hrs | |
| | 5 - 600 | New Instrument | 100 Jy | 200 | | 50 | 100 hrs | |
| Deuterium: HD abundance | 112 | GREAT | $5 \times 10^{-17}$ W/m² | $10^5$ | 11 | 10 | 40 hrs | single line, single pointings |

## B.4 Galaxies and the Galactic Center

| Subtheme | Wavelength range (μm) | Instrument used | Sensitivity required 10 σ | Spectral resolution (λ/Δλ) | Spatial resolution (arcsec) | Number of targets | Total time on targets | Comments |
|---|---|---|---|---|---|---|---|---|
| Galactic Center | 53 | HAWC Polarimeter | 160 mJy | NA | 5.3 | 40 | 20 hrs | Polarimetry mode is not in current baseline |
| | 63 to 163 | GREAT | $1 \times 10^{-15}$ W/m$^2$ | 30000 | 6 to 20 | 120 | 16 hrs | [OI], [CII], CO(16-15) lines |
| | 260 | CASIMIR | $2 \times 10^{-16}$ W/m$^2$ | 30000 | 26 | 120 | 4 hrs | CO(10-9) line |
| | 26 | EXES high res | $8 \times 10^{-17}$ W/m$^2$ | 30000 | 3 | 120 | 8 hrs | [FeII] 26 um line |
| | 34.8 | FORCAST Grisms | $2 \times 10^{-16}$ W/m$^2$ | 800 | 3.4 | 30 | 15 hrs | [SiII] 34.8 um line |
| Nearby Galaxies | 52 | FIFI-LS | $6 \times 10^{-17}$ W/m$^2$ | 2000 | 7 | 3 | 12 hrs | Nucleus and bar/Spiral arm interfaces of M83 |
| | 57 | FIFI-LS | $6 \times 10^{-17}$ W/m$^2$ | 2000 | 7 | 3 | 12 hrs | Nucleus and bar/Spiral arm interfaces of M83 |
| | 63 | FIFI-LS | $7 \times 10^{-17}$ W/m$^2$ | 2000 | 7 | 10 | 20 hrs | Cut across NGC 6946 |
| | 88 | FIFI-LS | $7 \times 10^{-17}$ W/m$^2$ | 2000 | 7 | 10 | 20 hrs | Cut across M83 |
| | 122 | FIFI-LS | $7 \times 10^{-17}$ W/m$^2$ | 2000 | 14 | 5 | 10 hrs | Cut across M83 |
| | 146 | FIFI-LS | $6 \times 10^{-17}$ W/m$^2$ | 2000 | 14 | 5 | 5 hrs | Cut across M83 |
| | 158 | FIFI-LS | $1 \times 10^{-16}$ W/m$^2$ | 2000 | 14 | 25 | 4 hrs | Maps all of M83 |
| | 205 | FIFI-LS | $4 \times 10^{-17}$ W/m$^2$ | 2000 | 14 | 5 | 10 hrs | Cut across M83 |
| Evolution of Galaxies | 145-450 | SAFIRE | 2 to $8 \times 10^{-18}$ W/m$^2$ | 1000 | Point source | 20 | 80 hrs | 20 targets for first sample, many thousands available for study |

## B.5 Planetary Science

| Subtheme | Wavelength range (µm) | Instrument used | Sensitivity required 10 σ | Spectral resolution (λ/Δλ) | Spatial resolution (arcsec) | Number of targets in 5 years | Total time on targets | Comments |
|---|---|---|---|---|---|---|---|---|
| Comets | 1 - 5 | FLITECAM+grism | $1\times10^{-18}$ W/m$^2$ @ 3 µm | 2,000 | 2 | 5 | 10 hrs | 0.01 line/cont, S/N>100, 900s (1) |
| | 5 - 30 | EXES | $3\times10^{-17}$ W/m$^2$ @ 20 µm | 10,000 | 2 | 5 | 10 hrs | 0.01 line/cont, S/N=30, 900s (1) |
| | 5 - 40 | FORCAST+grism | $3\times10^{-16}$ W/m$^2$ @ 30 µm | 200 | 5 | 5 | 10 hrs | 0.01 line/cont, S/N=10, 900s (1) |
| | 150 - 600 | CASIMIR | $5\times10^{-18}$ W/m$^2$ @ 180 µm | $10^5$ | 20 | 5 | 10 hrs | line S/N=10, 900s (2) |
| Asteroids | 1 - 5 | FLITECAM+grism | 2 mJy @ 3 µm reflected | 2,000 | 2 | 25 | 50 hrs | 100 objects d>60km |
| | 5 - 40 | FORCAST+grism | 200 mJy @ 30 µm thermal | 200 | 5 | 100 | 50 hrs | 1000 objects d>20km |
| TNOs | 0.3 - 0.9, 1 - 5 | HIPO+FLITECAM | V = +17 background star | 10 | 2 | 10 | 200 hrs | 2 Hz, open filt S/N > 10 |
| Giant Planets | 5 - 30 | EXES | 2 Jy = $3\times10^{-17}$ W/m$^2$ @ 20 µm | 10,000 | 2 | 4 planets x 5 | 20 hrs | 0.3 line/cont, S/N=10, 9000s |
| | 5 - 40 | FORCAST+grism | 40 Jy @ 30 µm | 200 | 2 | 4 planets x 5 | 5 hrs | (easy) (3) |
| | 50 - 250 | HAWC | 400 Jy @ 100 µm | 10 | 10 | 4 planets x 5 | 5 hrs | (easy) (3) |
| | 40 - 200 | FIFI-LS | 400 Jy @ 100 µm | 2,000 | 10 | 4 planets x 5 | 20 hrs | 0.01 line/cont, S/N=10, 9000s (3) |
| | 60 - 200 | GREAT | 400 Jy @ 100 µm | $10^5$ | 10 | 4 planets x 5 | 20 hrs | 0.5 line/cont, S/N=10, 9000s (3) |
| | 150 - 600 | CASIMIR | 200 Jy @ 200 µm | $10^5$ | 20 | 4 planets x 5 | 20 hrs | 0.1 line/cont, S/N=10, 9000s (3) |
| Titan | 5 - 30 | EXES | 40 Jy = $5\times10^{-18}$ W/m$^2$ @ 25 µm | 10,000 | 2 | 1x5 | 15 hrs | 0.02 line/cont, S/N=10, 9000s |
| | 150 - 600 | CASIMIR | 40 Jy = $1\times10^{-17}$ W/m$^2$ @ 200 µm | $10^5$ | 20 | 1x5 | 15 hrs | 0.5 line/cont, S/N=10, 9000s |
| Venus | 5 - 30 | EXES | $3\times10^{-17}$ W/m$^2$ @ 18 µm | 10,000 | 2 | 7 elongs | 5 hrs | continuum S/N=10, 900s (4) |
| | 60 - 200 | GREAT | $3\times10^{-17}$ W/m$^2$ @ 100 µm | $10^6$ | 10 | 7 elongs | 5 hrs | 0.01 line/cont, S/N=10, 900s (5) |
| | 150 - 600 | CASIMIR | $3\times10^{-18}$ W/m$^2$ @ 200 µm | $10^6$ | 20 | 7 elongs | 5 hrs | 0.01 line/cont, S/N=60, 900s |

(1) comets Borrelly, Faye, Schaumasse  (2) comet T1-McNaught-Hartley, at H$_2$O(18) 180 um line  (3) flux from Neptune  (4) at CO$_2$ 18 um line  (5) equal to Venus Express PFS performance

## B.6 Selected Spectral Lines Referenced in the Text

| Species | Line Position $\lambda$ ($\mu$m) | Atmospheric Transmission* (+/- 0.01) |
|---|---|---|
| $H_2O$ | 6.08 | 0.92 |
| HDO ($v_2$) | 7.13 | 0.99 |
| $C_4H_5N$ | 13.74 | 0.88 |
| $C_6H_2$ ($v_{11}$) | 16.1 | 0.12 |
| $CH_3$ | 16.5 | 0.86 |
| [FeII] | 25.99 | 0.26 |
| [SiII] | 34.82 | 0.98 |
| [OIII] | 51.81 | 0.94 |
| [NIII] | 57.33 | 0.85 |
| [OI] | 63.18 | 0.59 |
| [OIII] | 88.36 | 0.94 |
| HD (J=1-0) | 112.07 | 0.55 |
| OH | 119.23 | 0.93 |
| OH | 119.44 | 0.91 |
| [OI] | 145.53 | 0.77 |
| [CII] | 157.68 | 0.84 |
| CO (16-15) | 162.81 | 0.88 |
| $CH_4$ | 191.1 | 0.93 |
| CO (13-12) | 200.27 | 0.92 |
| [NII] | 205.18 | 0.94 |
| p-$H_2$(18)O | 301.40 | 0.68 |
| p-$H_2$(18)O | 402.23 | 0.66 |

*Flight Altitude = 41,000 ft., zenith water vapor overburden = 7.1 $\mu$m, elevation angle = 40 deg.

**APPENDIX C** References

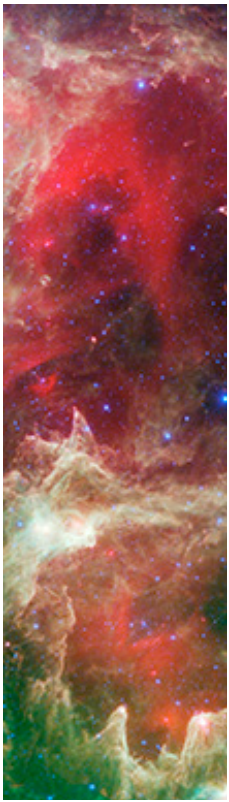

## C.1 Chapter 1 – Introduction

Gardner, J. P., Mather, J. C., Clampin, M., et al., 2006. The James Webb Space Telescope. Space Science Reviews 123, 485.

Gehrz, R.D., Roellig, T.L., Werner, M.W., et al., 2007. The NASA Spitzer Space Telescope. Rev. Sci. Instrum 78, (011302). Original

Pilbratt, G. L., 2003. Herschel Space Observatory mission overview. Proc. SPIE 4850, 586.

Werner, M.W., Roellig, T.L., Low, F.J., et al., 2004. The Spitzer Space Telescope Mission. Ap. J. Supp. Series 154, 1-9.

Young, E. et al., 2008. The 20/20 Vision of SOFIA: Scientific and Technical Opportunities in the next 20 years. http://www.sofia.usra.edu/Science/07Dec_SOFIA_Vision/SOFIA_2020Vision_white_paper_final.pdf

## C.2 Chapter 2 – The Formation of Stars and Planets

Boss, A.P., 2008. Flux-limited Diffusion Approximation Models of Giant Planet Formation by Disk Instability. Ap. J. 677, 607.

Boogert, A. C. A., Tielens, A. G. G. M., Ceccarelli, C., Boonman, A. M. S., van Dishoeck, E. F., Keane, J. V., Whittet, D. C. B., de Graauw, T., 2000. Infrared observations of hot gas and cold ice toward the low mass protostar Elias 29. A&A 360, 683.

Bonnell, I. A., Bate, M. R., Clarke, C. J., Pringle, J. E., 1997. Accretion and the stellar mass spectrum in small clusters. MNRAS 285, 201.

Boonman, A. M. S., Doty, S. D., van Dishoeck, E. F., Bergin, E. A., Melnick, G. J., Wright, C. M., Stark, R., 2003. Modeling gas-phase $H_2O$ between 5 m and 540 m toward massive protostars. A&A 406, 937.

Bouwman, J., Henning, Th., Hillenbrand, L. A., et al., 2008. The Formation and Evolution of Planetary Systems: Grain Growth and Chemical Processing of Dust in T Tauri Systems. Ap. J. 683, 479.

Carr, J. S., Evans, N. J., II, Lacy, J. H., Zhou, S., 1995. Observation of Infrared and Radio Lines of Molecules toward GL 2591 and Comparison to Physical and Chemical Models. Ap. J. 450, 667.

Carr, J. S., Najita, J. R., 2008. Organic Molecules and Water in the Planet Formation Region of Young Circumstellar Disks. Science 319, 1504.

Ciesla, F. J, Cuzzi, J.N., 2006. The evolution of the water distribution in a viscous protoplanetary disk. Icarus 181, 178.

Crapsi, A., van Dishoeck, E. F., Hogerheijde, M. R., Pontoppidan, K. M., Dullemond, C. P., 2008. Characterizing the nature of embedded young stellar objects through silicate, ice and millimeter observations. A&A 486, 245.

Crutcher, R.M., Hakobian, N., Troland, T. H., 2008. Testing Magnetic Star Formation Theory. arXiv:0807.2862.

de Wit, W. J., Testi, L., Palla, F., Vanzi, L., Zinnecker, H., 2004. The Origin of Massive O-type Field Stars. I. A Search for Clusters. A&A 425, 937.

de Wit, W. J., Testi, L., Palla, F., Zinnecker, H., 2005. The Origin of Massive O-type Field Stars: II. Field O Stars as Runaways. A&A 437, 247.

Doty, S. D., Neufeld, D. A., 1997. Models for Dense Molecular Cloud Cores. Ap. J. 489, 122.

Egan, M. P., Shipman, R. F., Price, S. D., Carey, S. J., Clark, F. O., Cohen, M., 1998. A Population of Cold Cores in the Galactic Plane. Ap. J. Letters 494, L199.

Evans, N. J., II, 1999. Physical Conditions in Regions of Star Formation. ARA&A 37, 311.

Franklin, J., Snell, R. L., Kaufman, M. J., Melnick, G. J., Neufeld, D. A., Hollenbach, D. J., Bergin, E. A., 2008. SWAS Observations of Water in Molecular Outflows. Ap. J. 674, 1015.

Goldsmith, P.~F., et al., 2000. $O_2$ in Interstellar Molecular Clouds. Ap. J. Letters 539, L123.

Grün, E., et al., 2001. Broadband infrared photometry of comet Hale-Bopp with ISOPHOT. A&A 377, 1098.

Hillenbrand, L. A., Strom, S. E., Vrba, F. J., Keene, J., 1992. Herbig Ae/Be stars - Intermediate-mass stars surrounded by massive circumstellar accretion disks. Ap. J. 397, 613.

Kamp, I., Dullemond, C. P., 2004. The Gas Temperature in the Surface Layers of Protoplanetary Disks. Ap. J. 615, 991.

Kandori, R., et al., 2007. Near-Infrared Imaging Polarimetry of the Star-Forming Region NGC 2024. PASJ 59, 487.

Knez, C., Lacy, J. H., Evans, N. J., II, Boonman, A. M. S., van Dishoeck, E. F, 2003. Mid-Infrared Absorption Spectroscopy towards NGC 7538 IRS 1. SFChem 2002: Chemistry as a Diagnostic of Star Formation, 325.

Li, W., Evans, N. J., II, Jaffe, D. T., van Dishoeck, E. F., Thi, W. F., 2002. Photon-dominated Regions in Low-Ultraviolet Fields: A Study of the Peripheral Region of L1204/S140. Ap. J. 568, 242.

Mannings, V., Sargent, A. I., 1997. A High-Resolution Study of Gas and Dust around Young Intermediate-Mass Stars: Evidence for Circumstellar Disks in Herbig AE Systems. Ap. J. 490, 792.

McKee, C. F., Tan, J. C., 2003. The Formation of Massive Stars from Turbulent Cores. Ap. J. 585, 850.

Meeus, G., Waters, L. B. F. M., Bouwman, J., van den Ancker, M. E., Waelkens, C., Malfait, K., 2001. ISO Spectroscopy of Circumstellar Dust in 14 Herbig Ae/Be Systems: Toward an Understanding of Dust Processing. Astr. Ap. 365, 476.

Meijerink, R., Poelman, D. R., Spaans, M., Tielens, A. G. G. M., Glassgold, A. E., 2008. Rotational Line Emission from Water in Protoplanetary Disks. arXiv:0810.1769.

Meyer, D. M., Jura, M., Cardelli, J. A., 1998. The Definitive Abundance of Interstellar Oxygen. Ap. J. 493, 222.

Mouschovias, T. C., Tassis, K., 2008. Recent OH Zeeman Observations: Do they Really Contradict the Ambipolar-Diffusion Theory of Star Formation? arXiv:0807.4571.

Natta, A., Testi, L., Neri, R., Shepherd, D. S., Wilner, D. J., 2004. A search for evolved dust in Herbig Ae stars. A&A 416, 179.

Neufeld, D. A., et al., 2000. Observations of Interstellar Water Vapor in Outflow Regions. Ap. J. Letters 539, L107.

Nisini, B., 2003. Mid and Far Infrared Observations of Protostellar Jets. Ap. J. 287, 207.

Pillai, T., Wyrowski, F., Menten, K.M., Krügel, E., 2006. High mass star formation in the infrared dark cloud G11.11-0.12. A&A 447, 929.

Richter, M. J., Lacy, J. H., Jaffe, D. T., Greathouse, T. K., Hemenwaya, M. K., 2000. EXES: A Progress Report on the Development of a High-resolution, Mid-infrared Grating Spectrograph for SOFIA. Proc. of SPIE Vol. 4014, ed. R. Melugin, H. Roeser (SPIE), 53.

Robitaille, T. P., Whitney, B. A., Indebetouw, R., Wood, K., Denzmore, P., 2006. Interpreting Spectral Energy Distributions from Young Stellar Objects. I. A Grid of 200,000 YSO Model SEDs. Ap. J. 167, 256.

Robitaille, T.P., Whitney, B.A., Indebetouw, R., Wood, K., 2007. Interpreting Spectral Energy Distributions from Young Stellar Objects. II. Fitting Observed SEDs Using a Large Grid of Precomputed Models. Ap. J. 169, 328.

Salyk, C., Pontoppidan, K. M., Blake, G. A., Lahuis, F., van Dishoeck, E. F., Evans, N. J., II 2008. $H_2O$ and OH Gas in the Terrestrial Planet-forming Zones of Protoplanetary Disks. Ap. J. Letters 676, L49.

Simon, R., Rathborne, J. M., Shah, R. Y., Jackson, J. M., Chambers, E. T., 2006. The Characterization and Galactic Distribution of Infrared Dark Clouds. Ap. J. 653, 1325.

Smith, R. G., Sellgren, K., Tokunaga, A. T., 1989. Absorption features in the 3 micron spectra of protostars. Ap. J. 344, 413.

Troland, T. H., Crutcher, R. M., 2008. Magnetic Fields in Dark Cloud Cores: Arecibo OH Zeeman Observations. Ap. J. 680, 457.

van Boekel, R., Min, M., Waters, L. B. F. M., de Koter, A., Dominik, C., van den Ancker, M. E., Bouwman, J., 2005. A 10 m spectroscopic survey of Herbig Ae star disks: Grain growth and crystallization. A&A 437, 189.

van Boekel, R., 2007. Water worlds in the making. Nature 447, 31.

van den Ancker, M. E., et al., 2000. ISO spectroscopy of circumstellar dust in the Herbig Ae systems AB Aur and HD 163296. A&A 357, 325.

van der Tak, F.F.S., 2005. The chemistry of high-mass star formation. Massive Star Birth: A Crossroads of Astrophysics 227, 70.

Voshchinnikov, N. V., Il'in, V. B., Henning, Th., Dubkova, D. N., 2006. Dust extinction and absorption: the challenge of porous grains. A&A 445, 167.

## C.3 Chapter 3 – The Interstellar Medium of the Milky Way

Benjamin, R. A., et al., 2003. GLIMPSE. I. An SIRTF Legacy Project to Map the Inner Galaxy. PASP 115, 953B.

Bernard, J-P, et al., 2008. Spitzer Sage Survey of the Large Magellanic Cloud. III. Star Formation and ~1000 New Candidate Young Stellar Objects. Ap. J. 136, 919.

Bernstein, M., et al., 2002. Side Group Addition to the Polycyclic Aromatic Hydrocarbon Coronene by Ultraviolet Photolysis in Cosmic Ice Analogs. Ap. J. 576, 1115B.

Blum, R., et al., 2006. Spitzer SAGE Survey of the Large Magellanic Cloud. II. Evolved Stars and Infrared Color-Magnitude Diagrams. Ap. J. 132, 2034.

Boersma, C., Bouwman, J., Lahuis, F., van Kerckhoven, C., Tielens, A.G.G.M., Waters, L.B.F.M., Henning, T., 2008. The characteristics of the IR emission features in the spectra of Herbig Ae stars: evidence for chemical evolution. A&A 484, 241.

Calzetti, D., et al., 2007. The Calibration of Mid-Infrared Star Formation Rate Indicators. Ap. J. 666, 870.

Cami, J., 2001. Molecular gas and dust around evolved stars. PhD thesis, University of Amsterdam.

Carey, S., et al., 2005. MIPSGAL: A Survey of the Inner Galactic Plane at 24 and 70 microns, Survey Strategy and Early Results. AAS 207, 6333C.

Caux, E., Ceccarelli, C., Pagani, L., Maret, S., Castets, A., Pardo, J. R., 2002. HD 112 mu m in absorption and extreme CO depletion in a cold molecular cloud. A&A 383, L9.

Cherchneff, I., Barker, J.R., Tielens, A.G.G.M., 1992. Polycyclic Aromatic Hydrocarbon Formation in Carbon-rich Stellar Envelopes: Erratum. Ap. J. 413, 445.

Draine, B. T., 2006. Can Dust Explain Variations in the D/H Ratio? ASPC 348, 58.

Draine, B. T., et al., 2007. Dust Masses, PAH Abundances, and Starlight Intensities in the SINGS Galaxy Sample. Ap. J. 663, 866.

Drew, J., et al., 2005. IPHAS: Surveying the North Galactic Plane in H$\alpha$. ING Newsletter issue 9, 3.

Epstein, R.LI., 1976. The origin of deuterium. Nature 263, 5574.

Flagey, N., Boulanger, F., Verstraete, L., Miville Deschenes, M.A., Noriega Crespo, A., Reach, W.T., 2006. Spitzer/IRAC and ISOCAM/CVF insights on the origin of the near to mid-IR Galactic diffuse emission. A&A 453, 969.

Frenklach, M. Feigelson, E.D., 1989. Formation of polycyclic aromatic hydrocarbons in circumstellar envelopes. Ap. J. 341, 372.

Galliano, F., Madden, S., Tielens, A.G.G.M., Peeters, E., Jones, A.P., 2008. Variations of the Mid-IR Aromatic Features inside and among Galaxies. Ap. J. 679, 310.

Goto, M., Kobayashi, N., Terada, H., Tokunaga, A.T., 2002. Imaging and Spatially Resolved Spectroscopy of AFGL 2688 in the Thermal Infrared Region. Ap. J. 572, 276.

Hollenbach, D. J., Tielens, A. G. G. M., 1999. Photodissociation regions in the interstellar medium of galaxies. Reviews of Modern Physics 71, 173.

Hony, S., Van Kerckhoven, C., Peeters, E., Tielens, A. G. G. M., Hudgins, D. M., Allamandola, L. J., 2001. The CH out-of-plane bending modes of PAH molecules in astrophysical environments. A&A 370, 1030.

Hony, S., Waters, L.B.F.M., Tielens, A.G.G.M., 2002. The carrier of the "30'' mu m emission feature in evolved stars. A simple model using magnesium sulfide. A&A 390, 533.

Hoopes, J.L. et al., 2006. Deuterium toward Two Milky Way Disk Stars: Probing Extended Sight Lines with the Far Ultraviolet Spectroscopic Explorer. Ap. J. 586, 1094.

Joblin, C, Szczerba, R., Berne, O., Szyszka, C., 2008. Carriers of the mid-IR emission bands in PNe reanalysed. Evidence of a link between circumstellar and interstellar aromatic dust. A&A 490, 189.

Jochims, H.W., Ruhl, E., Baumgartel, H., Tobita, S., Leach, S., 1994. Size effects on dissociation rates of polycyclic aromatic hydrocarbon cations: Laboratory studies and astophysical implications. Ap. J. 420, 307.

Jones A.P., Tielens, A.G.G.M., Hollenbach, D.J., 1996. Grain Shattering in Shocks: The Interstellar Grain Size Distribution. Ap. J. 469, 740.

Kaufman. M.J., Wolfire, M.G., Hollenbach, D.J., Luhman, M.L., 1999. Far-Infrared and Submillimeter Emission from Galactic and Extragalactic Photodissociation Regions. Ap. J. 527, 795.

Lawrence, A., et al., 2007. The UKIRT Infrared Deep Sky Survey (UKIDSS). MNRAS 379, 1599.

Le Page, V., Snow, T., Bierbaum, V.M., 2001. Hydrogenation and Charge States of PAHS in Diffuse Clouds. I. Development of a Model. Ap. J. 132, 233.

Linsky, J.L., et al., 2006. What Is the Total Deuterium Abundance in the Local Galactic Disk? Ap. J. 647, 1106.

Meixner, M., et al., 2006. Spitzer Survey of the Large Magellanic Cloud: Surveying the Agents of a Galaxy's Evolution (SAGE). I. Overview and Initial Results. Ap. J. 132, 2268.

Micelotta, E., Jones, A.P., Tielens, A.G.G.M., 2009. in preparation.

Molster, F.J., Waters, L.B.F.M., Tielens, A.G.G.M., Koike, C., Chihara, H., 2002. Crystalline silicate dust around evolved stars. III. A correlations study of crystalline silicate features. A&A 382, 241.

Molster, F.J., Waters, L.B.F.M., Tielens, A.G.G.M., Barlow, M.J., 2002. Crystalline silicate dust around evolved stars. I. The sample stars. A&A, 382, 184.

Mulas, G., Malloci, Joblin, C., Toublanc, D., 2006. A general model for the identification of specific PAHs in the far-IR. A&A 460, 93.

Neufeld, D. A., et al., 2006. Spitzer Observations of Hydrogen Deuteride. Ap. J. Letters 647, L33.

Peeters, E., Hony, S., van Kerckhoven, C., Tielens, A.G.G.M., Allamandola, L.J., Hudgins, D.M., Bauschlicher, C.W., 2002. The rich 6 to 9 vec mu m spectrum of interstellar PAHs. A&A 390, 1089.

Pino, T., Cao, A.T., Carpentier, Y., Dartois, E., d'Hendecourt, L., Brechignac, Ph., 2008. The 6.2 m band position in laboratory and astrophysical spectra: a tracer of the aliphatic to aromatic evolution of interstellar carbonaceous dust. A&A 490, 665.

Rapacioli, M., Joblin, C., Boissel, P., 2005. Spectroscopy of polycyclic aromatic hydrocarbons and very small grains in photodissociation regions. A&A 429, 193.

Romano, D., Tosi, M., Chiappini, C., Matteucci, F., 2006. Deuterium astration in the local disc and beyond. MNRAS 369, 295.

Savage et al., 2007. The Abundance of Deuterium in the Warm Neutral Medium of the Lower Galactic Halo. Ap. J. 659, 1222.

Sellgren, K., Uchida, K. I., Werner, M. W., 2007. The 15-20 m Spitzer Spectra of Interstellar Emission Features in NGC 7023. Ap. J. 659, 1338.

Sembach et al., 2004. The Deuterium-to-Hydrogen Ratio in a Low-Metallicity Cloud Falling onto the Milky Way. Ap. J. 150, 587.

Sloan, G. C., Jura, M., Duley, W. W., Kraemer, K. E., Bernard-Salas, J., Forrest, W. J., Sargent, B., Li A., Barry, D. J., Bohac, C. JK., Watson, D. M., Houck, J. R., 2007. The Unusual Hydrocarbon Emission from the Early Carbon Star HD 100764: The Connection between Aromatics and Aliphatics. Ap. J. 664, 1144.

Smith, J. D. T., et al., 2007. The Mid-Infrared Spectrum of Star-forming Galaxies: Global Properties of Polycyclic Aromatic Hydrocarbon Emission. Ap. J. 656, 770.

Spergel, D.N., 2003. First-Year Wilkinson Microwave Anisotropy Probe (WMAP) Observations: Determination of Cosmological Parameters. Ap. J. 148, 175.

Steigman, G., Romano, D., Tosi, M., 2007. Connecting the primordial and Galactic deuterium abundances. MNRAS 378, 576.

Sylvester, R.J., Kemper, F., Barlow, M.J., de Jong, T., Waters, L.B.F.M., Tielens, A.G.G.M., Omont, A., 1999. 2.4-197 mu m spectroscopy of OH/IR stars: the IR characteristics of circumstellar dust in O-rich environments. A&A 352, 587.

Tosi, M., 1988. The effect of metal-rich infall on galactic chemical evolution. A&A 197, 47.

Tielens, A.G.G.M., 2008. Interstellar Polycyclic Aromatic Hydrocarbon Molecules. ARAA 46, 289.

Whitelock, P. A., Feast, M.W., van Leeuwen, F., 2008. AGB variables and the Mira period-luminosity relation. MNRAS 386, 313.

Wolfire, M.G., Tielens, A.G.G.M., Hollenbach, D.J., 1990. Physical conditions in photodissociation regions - Application to galactic nuclei. Ap. J. 358, 116.

Wright, C.M., van Dishoeck, E. F., Cox, P., Sidher, S., Kessler, M.F., 1999. Infrared Space Observatory-Long Wavelength Spectrometer Detection of the 112 Micron HD J = 1 --> 0 Line toward the Orion Bar. Ap. J. Letters 515, L29.

## C.4 Chapter 4 – Galaxies and the Galactic Center

Chuss, D.T., Davidson, J.A., Dotson, J.L., Dowell, C.D., Hildebrand, R.H., Novak, G., Vaillancourt, J.E., 2003. Magnetic Fields in Cool Clouds within the Central 50 Parsecs of the Galaxy. Ap. J. 599, 1128.

Elbaz, D. C. J. Cesarsky, C.J., Chanial, P., Aussel, H., Franceschini, A. Fadda1, D., Chary, R.R., 2002. The bulk of the cosmic infrared background resolved by ISOCAM. A&A 384, 848.

Farrah, D., Lonsdale, C. J., Weedman, D. W., Spoon, H. W. W., Rowan-Robinson, M., Polletta, M., Oliver, S., Houck, J. R., & Smith, H. E., 2008. The Nature of Star Formation in Distant Ultraluminous Infrared Galaxies Selected in a Remarkably Narrow Redshift Range. Ap. J. 677, 957.

Geis et al., 2009. in preparation.

Hailey-Dunsheath et al., 2009. in preparation.

Heitsch, F., Mac Low, M.-M., Klessen, R.S., 2001. Gravitational Collapse in Turbulent Molecular Clouds. II. Magnetohydrodynamical Turbulence. Ap. J. 547, 280.

Hildebrand, R.H., Davidson, J.A., Dotson, J., Figer, D.F., Novak, G., Platt, S.R., Tao, L., 1993. Polarization of the Thermal Emission from the Dust Ring at the Center of the Galaxy. Ap. J. 417, 565.

Kaufman, M.J., Wolfire, M.G., Hollenbach, D.J., Luhman, M.L., 1999. Far-Infrared and Submillimeter Emission from Galactic and Extragalactic Photodissociation Regions. Ap. J. 527, 795.

Madau, P., Ferguson, H. C., Dickinson, M. E., Giavalisco, M. Steidel, C. C., Fruchter, A., 1996. High-redshift galaxies in the Hubble Deep Field: Color selection and star formation history to z~4. MNRAS 283, 1388.

Meijerink, R., Spaans, M., Israel, F.P., 2007. Diagnostics of irradiated dense gas in galaxy nuclei. II. A grid of XDR and PDR models. A&A 461, 793.

Morris, M., 2006. The Galactic Center Magnetosphere. *J. Phys.: Conf. Ser*. 54, eds: R. Schödel, G.C. Bower, M.P. Muno, S. Nayakshin & T. Ott, pp. 1-9.

Morris, M., Polish, N., Zuckerman, B., Kaifu, N., 1983. The Temperature of Molecular Gas in the Galactic Center Region. Ap. J. 88, 1228.

Oberst, T. E., Parshley, S. C., Stacey, G. J., Nikola T, Löhr, A., Harnett, J. I., Tothill, N. F. H., Lane, A. P. Stark, A. A., Tucker, C. E., 2006. Detection of the 205 m [N II] Line from the Carina Nebula. Ap. J. 652, L125.

Rowan-Robinson, M., et al., 2008. Photometric redshifts in the SWIRE Survey. MNRAS 386, 697.

Pettini, M., Kellogg, M., Steidel, C.C., Dickenson, M., Adelberger, K.L., Giavalisco, M., 1998. Infrared Observations of Nebular Emission Lines from Galaxies at z ≈ 3. Ap. J. 508, 539.

Rodríguez-Fernández, N.J., Martín-Pintado, J., Fuente, A., Wilson, T.L., 2004. ISO observations of the Galactic center interstellar medium. Neutral gas and dust. A&A 427, 217.

Smail, I., Ivison, R. J., Blain, A. W., Kneib, J.-P., 2002. The nature of faint submillimetre-selected galaxies. MNRAS 331, 495.

Spaans, M., Meijerink, R., 2008. On the Detection of High-Redshift Black Holes with ALMA through CO and $H_2$ Emission. Ap. J. 678, L5.

Stacey, G. J., Geis, N., Genzel, R., Lugten, J. B, Poglitsch, A. Sternberg, A., Townes, C.H., 1991. The 158 microns Forbidden C II Line – A Measure of Global Star Formation Activity in Galaxies. Ap. J. 373, 423.

Stacey, G. J., Jaffe, D. T., Geis, N., Genzel, R., Harris, A. I., Poglitsch, A., Townes, C. H., 1993. 158 micron Forbidden C II Mapping of the Orion Molecular Cloud. Ap. J. 404, 219.

Wardle, M., Königl, A., 1990. A Model for the Magnetic Field in the Molecular Disk at the Galactic Center. Ap. J. 362, 120.

Yusef-Zadeh, F., Morris, M., Chance, D., 1984. Large, highly organized radio structures near the galactic centre. Nature 310, 557.

## C.5 Chapter 5 – Planetary Science

Atreya, S., Wong, M.H., Owen, T.C., Mahaffy, P.R., Niemann, H.B., de Pater, I., Drossart, P., Encrenaz, T., 1999. A comparison of the atmospheres of Jupiter and Saturn: deep atmospheric composition, cloud structure, vertical mixing, and origin. Planet. & Sp. Sci. 47, 1243-1262.

Atreya, S., Mahaffy, P.R., Niemann, H.B., Wong, M.H., Owen, T.C., 2003. Composition and origin of the atmosphere of Jupiter-an update, and implications for the extrasolar giant planets. Planet. & Sp. Sci. 51, 105-112.

Aumann, H., Orton, G., 1979. The 12- to 20-micron spectrum of Venus - Implications for temperature and cloud structure. Icarus 38, 251-266.

Aumann, H., Martonchik, J.V., Orton, G.S., 1982. Airborne spectroscopy and spacecraft radiometry of Venus in the far infrared. Icarus 49, 227-243.

Backman, D., Paresce, F., 1993. Main-sequence stars with circumstellar solid material - The Vega phenomenon. In: Levy, E.H., Lunine, J.L. (Eds.), Protostars and Planets III, University of Arizona Press, Tucson, pp. 1253-1304.

Bailey, J. Meadows, V.S., Chamberlain, S., Crisp, D., 2008. The temperature of the Venus mesosphere from $O_2$ (a$\Delta$g1) airglow observations. Icarus 197, 247-259.

Bensch, F., Bergin, E., 2004. The pure rotational line emission of ortho-water vapor in comets. I. radiative transfer model. Ap. J. 615, 531-544.

Bergin, E., Blake, G., Goldsmith, P., Harris, A., Melnick, G., Zmuidzinas, J., 2008. SOFIA Science Steering Committee (SSSC) Design Reference Mission Study (DRM).

Bertaux, J.-L., and 51 colleges, 2007. A warm layer in Venus' cryosphere and high-altitude measurements of HF, HCl, $H_2O$ and HDO. Nature 450, 646-649.

Bertaux, J-L., Montmessin, F., Marcq, E., 2008. Horizontal and Vertical Distribution of $SO_2$ in the Clouds from SPICAV UV Spectrometer. American Geophysical Union Fall Meeting 2008, abstract #P22A -05.

Bezard, B., Gautier, Marten, A., 1986. Detectability of HD and non-equilibrium species in the upper atmospheres of the giant planets from their submillimeter spectrum. A&A 161, 387-402.

Bockelée-Morvan, D., Crovisier, J., Mumma, M.J., Weaver, H.A., 2004. The composition of cometary volatiles. In: M. C. Festou, M.C., Keller, H.U., Weaver, H.A. (Eds.), Comets II, University of Arizona Press, Tucson, pp. 391-423.

Bonev, B.P. Mumma, M.J., Villanueva, G.L., Disanti, M.A., Ellis, R.S., Magee-Sauer, K., Dello Russo, N., 2007. A Search for Variation in the H2O Ortho-Para Ratio and Rotational Temperature in the Inner Coma of Comet C/2004 Q2 (Machholz). Ap. J. Letters 661, L97-L100.

Clancy, R.T., Sandor, B.J., Moriarty-Schieven, G.H., 2008. Venus upper atmospheric CO, temperature, and winds across the afternoon/evening terminator from June 2007 JCMT sub-millimeter line observations. Planet. & Space Sci. 56, 1344-1354.

Coustenis, A., Salama, A., Lellouch, E., Encrenaz, Th., Bjoraker, G., Samuelson, R.E., de Graauw, Th., Feuchtgruber, H., Kessler, M F., 1998. Evidence for water vapor in Titan's atmosphere from ISO/SWS data. A&A 336, L85-L89.

Coustenis, A., Salama, A., Schulz, B., Ott, S., Lellouch, E., Encrenaz, Th., Gautier, D., Feuchtgruber, H. 2003. Titan's atmosphere from ISO mid-infrared spectroscopy. Icarus 161, 383-403.

Coustenis, A., and 24 co-authors, 2007. The composition of Titan's stratosphere from Cassini/CIRS mid-infrared spectra. Icarus 189, 35-62.

Crisp, D., Meadows, V.S., Bezard, B., de Bergh, C., Maillard, J.-P., and Mills, F.P., 1996. Ground-Based Near-Infrared Observations of the Venus Night Side: 1.27mm $O_2$('D) Airglow from the Venus Upper Atmosphere. JGR 101, 4577-4593.

Dello Russo, N., DiSanti, M.A., Magee-Sauer, K., Gibb, E.L., Mumma, M.J., Barber, R.J., Tennyson, J., 2004. Water production and release in Comet 153P/Ikeya-Zhang (C/2002 C1): accurate rotational temperature retrievals from hot-band lines near 2.9-µm. Icarus 168, 186-200.

Elliot, J.L., Dunham, E.W., Bosh, A.S., Slivan, S.M., Young, L.A., Wasserman, L.H., Millis, R.L., 1989. Pluto's atmosphere. Icarus 77, 148-170.

Fouchet, T., Orton, G., Irwin, P.G.J., Calcutt, S.B., Nixon, C.A., 2004. Upper limits on hydrogen halides in Jupiter from Cassini/CIRS observations. Icarus 170, 237-241.

Gomes, R., Levison, H.F., Tsiganis, K., Morbidelli, A., 2005. Origin of the cataclysmic Late Heavy Bombardment period of the terrestrial planets. Nature 435, 466-469.

Griffith, C., Hall, J.L., Geballe, T.R., 2000. Detection of daily clouds on Titan. Science 290, 509-513.

Joyce, R.R., Pilcher, C.B., Cruikshank, D.P., Morrison, D., 1977. Evidence for weather on Neptune. Ap. J. 214, 657.

Kelley, M.S., Wooden, D.H., 2009. The composition of dust in Jupiter-family comets as inferred from infrared spectroscopy. Accepted for publication in Planet and Space Sci., arXiv:0811.3939.

Koike, C., Chihara, H., Tsuchiyama, A., Suto, H., Sogawa, H., Okuda, H., 2003. Compositional dependence of infrared absorption spectra of crystalline silicate. II. Natural and synthetic olivines. A&A 399, 1101-1107.

Lisse, C.M., Chen, C.H., Wyatt, M.C., Morlok, A., 2008. Circumstellar dust created by terrestrial planet formation in HD 113766. Ap. J. 673, 1106-1122.

Meyer, M.R., Backman, D.E., Weinberger, A.J., Wyatt, M.C., 2007. Evolution of circumstellar disks around normal stars: placing our solar system in context. In: Reipurth, B., Jewitt, D., Kiel, K. (Eds.), Protostars and Planets V, University of Arizona Press, Tucson, pp. 573-588.

Mills, F., Allen, M., 2007. A review of selected issues concerning the chemistry in Venus' middle atmosphere. Planet. Space Sci. 55, 1729-1740.

Morbidelli, A., Levison, H.F., Gomes, R., 2008. The dynamical structure of the Kuiper Belt and its primordial origin. In: Barucci, M.A., Boehnhardt, H., Cruikshank, D.P., Morbidelli, A. (Eds.), The Solar System Beyond Neptune, University of Arizona Press, Tucson, pp. 275-292.

Mumma, M.J., Weaver, H.A., Larson, H.P., Davis, D.S., Williams, M., 1986. Detection of water vapor in Halley's comet. Science 232, 1523-1528.

Mumma, M.J., DiSanti, M.A., Tokunaga, A., Roettger, E.E., 1995. Ground-based detection of water in comet Shoemaker-Levy 1992 XIX; probing cometary parent molecules by hot-band fluorescence. Bull. Amer. Astron. Soc. 27, 1144.

Mumma, M.J., Dello Russo, N., DiSanti, M.A., Magee-Sauer, K., Novak, R.E., Brittain, S., Rettig, T., McLean, I.S., Reuter, D.C., Xu, Li-H., 2001. Organic composition of C/1999 S4 (LINEAR): A comet formed near Jupiter? Science 292, 1334-1339.

Mumma, M.J., 2004. Private communication, 2008.

Ohtsuki, S., Iwagami, N., Sagawa, H., Kasaba, Y., Ueno, M., Imamura, T., 2005. Ground-based observation of the Venus 1.27-ìm O2 airglow. Adv. Space Res. 36, 2038-2042.

Orton, G.S., Baines, K.H., Bergstralh, J.T., Brown, R.H., Caldwell, J., Tokunaga, A.T., 1987. Infrared radiometry of Uranus and Neptune at 21 and 32 microns. Icarus 69, 230-238.

Sandor, B.J., Clancy, T., Moriarty-Schieven, G.H., Mills, F.P., 2008. Diurnal and altitude behavior of $SO_2$ and SO in the Venus mesosphere. American Astronomical Society, DPS meeting #40, abstract #62.06.

Sicardy, B., and 40 colleagues, 2003. Large changes in Pluto's atmosphere as revealed by recent stellar occultations. Nature 424, 168-170.

Teanby, N.A., Fletcher, L.N., Irwin, P.G.J., Fouchet., T., Orton, G.S., 2006. New upper limits for hydrogen halides on Saturn derived from Cassini-CIRS data. Icarus 185, 466-475.

Villanueva, G.L., Bonev, B.P., Magee-Sauer, K., DiSanti, M.A., Salyk, C., Blake, G.A., Mumma, M.J., 2006. The volatile composition of the split ecliptic comet 73P/Schwassmann-Wachmann 3: A comparison of fragments C and B. Ap. J. Letters 650, L87-L90.

Woodward, C.E., Kelley, M.S., Bockelée-Morvan, D., Gehrz, R.D., 2007. Water in comet C/2003 K4 (LINEAR) with Spitzer. Ap. J. 671, 1065-1074.

Wyatt, M.C., Smith, R., Greaves, J.S., Beichman, C.A., Bryden, G., Lisse, C.M. 2007. Transience of Hot Dust Around Sun-Like Stars. Ap. J. 658, 569-583.

Yurimoto, H., Kuramoto, K., Krot, A.N., Scott, E.R.D., Cuzzi, J.N., Thiemens, M.H., Lyons, J.R., 2007. Origin and evolution of oxygen-isotopic compositions of the solar system. In: Reipurth, B., Jewitt, D., Kiel, K. (Eds.), Protostars and Planets V, University of Arizona Press, Tucson, pp. 849-862.

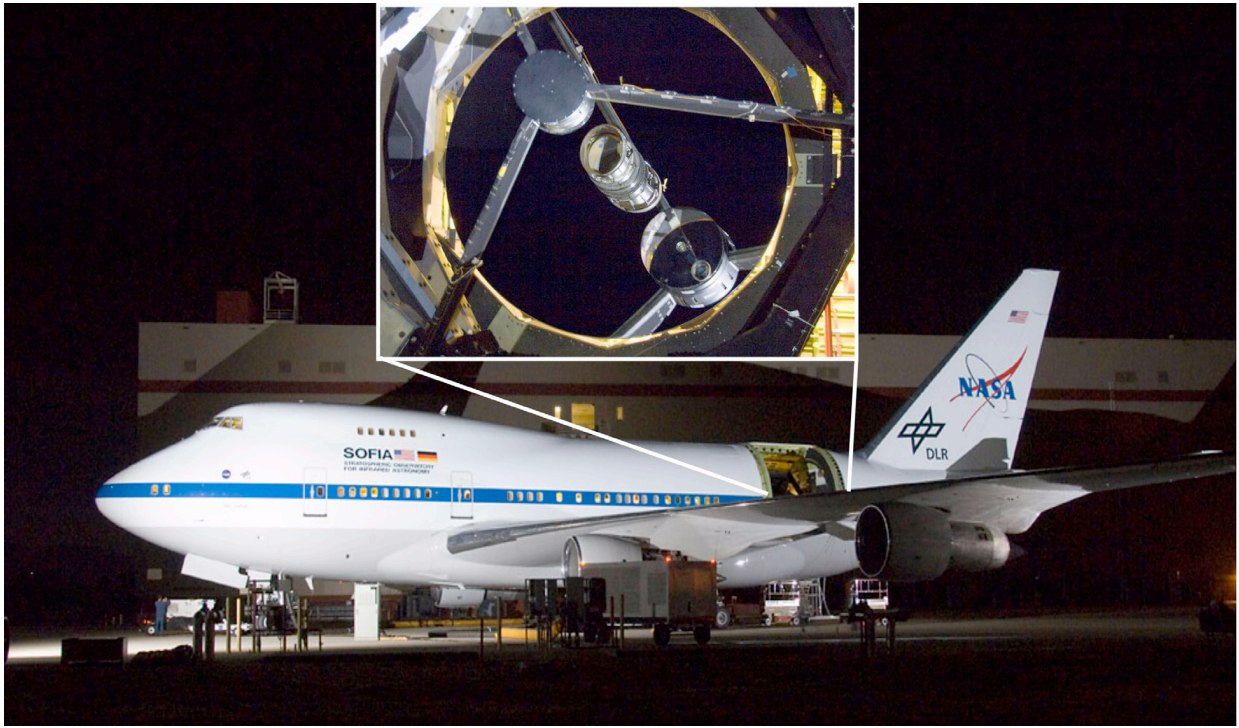

*The SOFIA flying observatory and telescope (inset)*

**SOFIA Program Office/Platform Project Office**
NASA Dryden Flight Research Center
P.O. Box 273
Edwards, CA 93523
(661) 276-2065

**SOFIA Science Project Office**
NASA Ames Research Center
Mailstop 211-1
Moffett Field, CA 94035
(650) 604-0999

**Universities Space Research Association (USRA)**
SOFIA Science Center
NASA Ames Research Center
Mailstop 211-3
Moffett Field, CA  94035
(650) 604-0949

**Deutsches Zentrum für Luft- und Raumfahrt (DLR)**
Königswinterer Str. 522-524
53227 Bonn-Oberkassel, Germany
+49 228 447-0

**Deutsches SOFIA Institut (DSI)**
Pfaffenwaldring 31
70569 Stuttgart, Germany
+49 711/685-62375

**For more information:**

http://www.nasa.gov/mission_pages/SOFIA/
or
http://www.sofia.usra.edu

*The Science Vision for the*

**Stratospheric Observatory for Infrared Astronomy**

NASA Ames Research Center
Moffett Field, CA

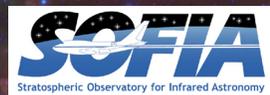
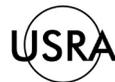

*SOFIA Science Mission Operations are conducted jointly by the Universities Space Research Association and the Deutsches SOFIA Institut on behalf of NASA and DLR.*